\documentclass{SciPost}
\usepackage{graphicx} 

\usepackage{amsmath,amssymb,amsfonts,xspace}
\usepackage{dsfont}
\usepackage[Symbol]{upgreek}

\numberwithin{equation}{section}

\def\det{{\rm det}\, }

\def\Re{\,{\rm Re}\, }

\def\mod{\,{\rm mod}\,}

\def\({\left(}
\def\){\right)}
\def\[{\left[}
\def\]{\right]}

\newcommand{\de}{\mathrm{d}}

\newcommand{\I}{\mathrm{i}}

\newcommand{\cL}{\mathcal{L}}

\newcommand{\p}{\partial}

\newcommand{\cF}{\mathcal{F}}

\newcommand{\cA}{\mathcal{A}}

\newcommand{\cS}{\mathcal{S}}
\newcommand{\cG}{\mathcal{G}}
\newcommand{\cH}{\mathcal{H}}
\newcommand{\cI}{\mathcal{I}}

\newcommand{\cM}{\mathcal{M}}

\newcommand{\cD}{\mathcal{D}}
\newcommand{\cE}{\mathcal{E}}

\newcommand{\cP}{\mathcal{P}}
\newcommand{\cR}{\mathcal{R}}

\DeclareSymbolFont{AMSa}{U}{msa}{m}{n}
\DeclareSymbolFont{AMSb}{U}{msb}{m}{n}
\DeclareMathSymbol{\fieldR}{\mathalpha}{AMSb}{"52}
\newcommand{\ar}[2]{{\textstyle \big[{#1\atop #2}\big]}}
\newcommand{\colvec}[2][.8]{\scalebox{#1}{\renewcommand{\arraystretch}{.8}$\begin{matrix}#2\end{matrix}$}}

\newcommand{\mf}{\mathfrak}

\newcommand{\kahler}{{K\"ahler}\xspace}

\newcommand{\cZ}{\mathcal{Z}}
\newcommand{\cO}{\mathcal{O}}

\renewcommand{\Re}{{\rm Re}}

\newcommand{\nn}{\nonumber}

\newcommand{\IR}{\mathds{R}}

\newcommand{\IZ}{\mathds{Z}}

\newcommand{\Tr}{{\rm Tr}}

\newcommand{\RV}{R}
\def\trF{F_{\rm tr}\hspace{-1.7mm}}
\def\ttrF{\tilde F_{\rm tr}\hspace{-1.7mm}}

\newcommand{\CQ}{{\widetilde{Q}}}
\newcommand{\sLambda}{I \hspace{-1mm}I}

\newcommand{\Part}[1]{\Gamma_{\hspace{-0.4mm}\Lambda_{\scalebox{0.6}{$#1$}}}}
\newcommand{\sPart}[1]{\Gamma_{\hspace{-0.4mm}\sLambda_{\scalebox{0.6}{$#1$}}}}
\newcommand{\tPart}[1]{\Gamma_{\hspace{-0.4mm}\tilde\Lambda_{\scalebox{0.6}{$#1$}}}}
\newcommand{\PartD}[1]{\Gamma_{\hspace{-0.4mm}\Lambda^*_{\scalebox{0.6}{$#1$}}}}
\newcommand{\tPartD}[1]{\Gamma_{\hspace{-0.4mm}\tilde \Lambda^*_{\scalebox{0.6}{$#1$}}}}
\newcommand{\ShiftPart}[2]{\Gamma_{\hspace{-0.4mm}\Lambda_{\scalebox{0.6}{$#1$}}+#2}}

\def\bea{\begin{eqnarray}}
\def\eea{\end{eqnarray}}
\def\be{\begin{equation}}
\def\ee{\end{equation}}
\def\ba{\begin{align}}
\def\ea{\end{align}}
\def\bse{\begin{subequations}}
\def\ese{\end{subequations}}

\def\RN{{\rm R.N.}}
\def\hb{\hat b}

\renewcommand{\_}{\hspace{0.05em}}

\begin{document}

\begin{center}{\Large \textbf{
Four-derivative couplings and BPS dyons \\ in heterotic CHL orbifolds
}}\end{center}

\begin{center}
G. Bossard\textsuperscript{1}, 
C. Cosnier-Horeau\textsuperscript{1,2,3}, 
B. Pioline\textsuperscript{2,3,4*}
\end{center}

\begin{center}
{\bf 1} Centre 
de Physique Th\'eorique, Ecole Polytechnique, Universit\'e Paris-Saclay, \\
91128 Palaiseau Cedex, France
\\
{\bf 2} Sorbonne Universit\'es, UPMC Universit\'e Paris 6, UMR 7589, F-75005 Paris, France
\\
{\bf 3} Laboratoire de Physique Th\'eorique et Hautes
Energies, CNRS UMR 7589, \\
Universit\'e Pierre et Marie Curie,
4 place Jussieu, 75252 Paris cedex 05, France
\\
{\bf 4} Theoretical Physics Department, CERN,
Case C01600, CH-1211 Geneva 23, Switzerland

* boris.pioline@cern.ch
\end{center}

\begin{center}
\today
\\[2mm]
CERN-TH-2017-028, CPHT-RR004.022017, arXiv:1702.01926v3
\end{center}


\section*{Abstract}
{\bf 
Three-dimensional string models with half-maximal supersymmetry are believed to be invariant
under a large U-duality group which unifies the S and T dualities in four dimensions. We propose
an exact, U-duality invariant formula for four-derivative scalar couplings of the 
form $F(\Phi) (\nabla\Phi)^4$ in a class of string vacua known as  CHL $\IZ_N$ heterotic orbifolds with $N$ prime, generalizing our previous work which dealt with the case of heterotic string on $T^6$.
We derive the Ward identities that $F(\Phi)$ must satisfy, and check that our formula obeys  them. We analyze the weak coupling expansion of $F(\Phi)$, and show that it
reproduces the correct tree-level and one-loop contributions, plus an infinite series of 
non-perturbative contributions. Similarly, the large radius
expansion reproduces the exact $F^4$ coupling in four dimensions, including both 
supersymmetric invariants, plus infinite series of instanton corrections from half-BPS dyons
winding around the large circle, and from Taub-NUT instantons. The summation measure
for dyonic instantons agrees with the helicity supertrace  for half-BPS dyons in 4 dimensions in
all charge sectors. In the process we clarify several subtleties about CHL models in $D=4$ and $D=3$, in particular we obtain the  exact helicity supertraces for 1/2-BPS dyonic states in all duality orbits.
}

\vspace{10pt}
\noindent\rule{\textwidth}{1pt}
\tableofcontents\thispagestyle{fancy}
\noindent\rule{\textwidth}{1pt}
\vspace{10pt}

\section{Introduction}

In the absence of a first principle non-pertubative formulation of superstring theory, the study of string vacua with extended supersymmetry continues to be one of the few sources of insight into the strong coupling regime. By exploiting invariance under U-dualities, which the full quantum theory is believed to enjoy \cite{Font:1990gx,Sen:1994fa,Hull:1994ys,Witten:1995ex}, as well as supersymmetric Ward identities, it is often possible to determine certain couplings in the low energy effective action exactly, for all values of the moduli (as demonstrated by \cite{Green:1997tv} and numerous subsequent works). 
The expansion of these couplings near boundaries of
the moduli space, corresponding to cusps of the U-duality group, then reveals, beyond power-like terms computable in perturbation theory, infinite series of exponentially suppressed corrections interpreted as semi-classical contributions in the putative string field theory. A particularly interesting
class of examples is that of BPS saturated couplings in three-dimensional string vacua: in the limit where a circle in the internal space decompactifies, these couplings receive exponentially suppressed contributions from BPS states in four dimensions, along with further suppressed contributions from Taub-NUT instantons. These couplings can therefore be viewed as BPS black hole partitions, which encode the exact degeneracies (or more precisely, helicity supertraces) of BPS black hole micro-states \cite{Gunaydin:2005mx,Bossard:2016zdx,Bossard:2016hgy}. 

In the recent letter  \cite{Bossard:2016zdx}, we investigated the $F(\Phi) (\nabla\Phi)^4$ and $
G(\Phi) \nabla^2 (\nabla\Phi)^4$ couplings in the low energy effective action of three-dimensional string vacua with 16 supercharges, focussing on the simplest example of such vacua, namely heterotic strings compactified on a torus $T^7$, or equivalently, type II strings compactified on $K3\times T^3$. Based on the known perturbative contributions to these couplings, we conjectured
exact formulae for the coefficients $F(\Phi)$ and $G(\Phi)$ for all values of the moduli $\Phi$,
which satisfy the requisite supersymmetric Ward identities and are manifestly invariant under U-duality.
In the limit where one circle inside $T^7$ decompactifies, we claimed that these formulae reproduce the correct helicity supertraces for 1/2-BPS and 1/4-BPS states with primitive charges, for all values of the moduli $\phi$ in four dimensions. 

The goal of the present work is to demonstrate these claims
in the case of the $(\nabla\Phi)^4$ coupling,\footnote{An analysis of the 
$\nabla^2 (\nabla\Phi)^4$ couplings will appear in a separate publication.} 
revisiting the analysis in \cite{Obers:2000ta}, and extend our conjecture to a class of  string vacua with 16 supercharges known as CHL orbifolds \cite{Chaudhuri:1995fk}, restricting to  $\IZ_N$ orbifolds with $N$ prime for simplicity. 

In Section \ref{sec_rev}, after reviewing relevant aspects of heterotic CHL vacua with 16 supercharges in four and three dimensions, we state the helicity supertraces of 1/2-BPS dyons 
with arbitrary charge in four dimensions (referring to Appendix \ref{sec_F4CHL} for the derivation of
the perturbative BPS spectrum), and determine the precise form of the U-duality group $G_3(\IZ)$ in three dimensions, consistent with S-duality and T-duality in four dimensions.
We then propose a manifestly U-duality invariant formula  \eqref{f4exact}
for the coefficient $F_{abcd}(\Phi)$ of the $(\nabla\Phi)^4$ couplings, obtained 
by covariantizing the known one-loop contribution under $G_3(\IZ)$, extending the proposal 
in \cite{Bossard:2016zdx} for the maximal rank case ($N=1$). 

In Section \ref{sec_susy}, using superspace arguments we establish the supersymmetric
Ward identities \eqref{wardf3} which constrain the coupling $F_{abcd}(\Phi)$, and show that the proposal \eqref{f4exact} satisfies these relations.

In Section \ref{pertLimit},  we analyze  \eqref{f4exact} in
the limit where $g_3\to 0$, and show that it reproduces the known tree-level and one-loop contributions in heterotic
perturbation theory, plus an infinite series of NS5-brane, Kaluza--Klein monopole and H-monopole instanton corrections.

In Section \ref{decompLimit}, we similarly  analyze \eqref{f4exact} in the large radius limit $R\to\infty$,  and show that it reproduces the known $F^4$ and $\cR^2$ couplings in $D=4$, along with  an infinite series of exponentially suppressed corrections of order $e^{-R \cM(Q,P)}$ with $Q$ and $P$ collinear,
weighted by the helicity supertrace $\Omega_4(Q,P)$, and further exponentially suppressed corrections from Taub-NUT monopoles. 

In most computations, we allow for lattices of arbitrary 
signature $(p,q)$, before specifying to the most relevant case $(p,q)=(2k,8)$ at the end. 
Details of some computations are relegated to 
Appendices. 
The one-loop vacuum amplitude for heterotic CHL models, from which the perturbative BPS
spectrum, $F^4$ and $(\nabla\Phi)^4$ couplings are easily read off, is constructed in 
Appendix \S\ref{sec_F4CHL}. In \S\ref{eqDiffDecomp} we decompose the Ward
identity on all Fourier modes in the degeneration limit $O(p,q)\to O(p-1,q-1)$, and show that
all Fourier coefficients are uniquely determined up to a moduli-independent summation measure.
In \S\ref{DecompPolynomials} and \S\ref{tensEinsensteinSeries} we collect some notations
which arise in the analysis of \S\ref{pertLimit} and \S\ref{decompLimit}. In Appendix \S\ref{sec_AFP}
we obtain a Poincar\'e series representation of the relevant genus-one modular integrals, and use the same method to construct Eisenstein series for $O(p,q,\IZ)$.

\section{Dualities, BPS spectrum and $(\nabla\Phi)^4$ couplings in CHL vacua\label{sec_rev}}

In this section, we recall relevant aspects of heterotic CHL vacua with 16 supercharges in four and three dimensions, restricting to the case of  $\IZ_N$ orbifolds with $N$ prime for simplicity.  While 
most of the results are well known, we pay special attention to the quantization conditions for the 
electromagnetic charges of 4D dyons, and to the precise form of the U-duality groups in $D=4$ and $D=3$. Finally, we state our proposal for the non-perturbative $(\nabla\Phi)^4$ coupling, which 
is the focus of the remainder of this work.

\subsection{Moduli space and 1/2-BPS dyons in $D=4$}
Recall that in four-dimensional string vacua with 16 supercharges, the moduli space is 
locally a product 
\be
\label{defM4}
\cM_4 = \left[ \frac{SL(2,\IR)}{SO(2)} \times  G_{r-6,6} \right] \slash G_4(\IZ)\ ,
\ee
where $G_{p,q}\equiv O(p,q)/[O(p)\times O(q)]$ denotes the orthogonal Grassmannian of positive $q$-planes in a fiducial vector space $\IR^{p,q}$ of signature $(p,q)$ (a real symmetric space of dimension $pq$),  $r$ is the rank of the Abelian gauge group, and $G_4(\IZ)$ is an arithmetic subgroup of $SL(2,\IR)\times O(r-6,6,\IR)$. 
In heterotic string theory compactified on a torus $T^6$, the first factor 
is parametrized by the 
axiodilaton $S=b+2\pi\I/g_4^2$, where $b$ is the scalar dual to the Kalb-Ramond two-form, 
while the second factor, with $r=28$, is the Narain 
moduli space  \cite{Narain:1985jj}. The U-duality group $G_4(\IZ)$ is then the 
product of the S-duality group
$SL(2,\IZ)$, acting on $S$ by fractional linear transformations $S\mapsto \tfrac{aS+b}{cS+d}$ \cite{Font:1990gx,Sen:1994fa}, 
and of the T-duality group
$O(22,6,\IZ)$, which is the automorphism group of the  even self-dual Narain lattice 
$\Lambda_{22,6}=E_8\oplus E_8 \oplus \sLambda_{6,6}$, where $E_8$
denotes the root lattice of $E_8$ and $\sLambda_{d,d}$ denotes $d$ copies of the standard hyperbolic lattice
$\sLambda_{1,1}$. The effective action is singular on real codimension-$6$ loci where the projection $Q_R$ of a vector
$Q\in \Lambda_{22,6}$ with norm $Q^2=2$ on the negative 6-plane parametrized by $G_{r-6,6}$ vanishes, corresponding to points of gauge symmetry enhancement. 
The same moduli space \eqref{defM4} arises in type IIA string compactified
on $K3\times T^2$, where the first factor parametrizes the \kahler modulus of $T^2$, while
the second factor parametrizes the axiodilaton, the complex modulus of $T^2$, the $K3$ moduli and the holonomies of the RR gauge fields on $T^2\times K3$. These two string vacua are in fact related by heterotic/type II duality \cite{Kachru:1995wm}, which in particular turns S-duality into a geometrical symmetry.

Vacua with lower values of $r$ can be constructed as freely acting orbifolds of the maximal rank
model with 
$r=28$ \cite{Chaudhuri:1995fk,Chaudhuri:1995dj,Chaudhuri:1995bf,Schwarz:1995bj}. On the heterotic side, one mods out by a $\IZ_N$ rotation of the heterotic lattice
$\Lambda_{22,6}$ at values of the Narain moduli where
such a symmetry exists, combined with an order $N$ shift along one circle inside $T^{6}$. 
This projection removes $28-r$ of the gauge fields in 4 dimensions, along with their scalar partners.
On the type II side, one can similarly mod out by a symplectic automorphism of order $N$ on K3,
combined with an order $N$ shift on $T^2$. It is convenient to label this action by the data $\{m(a), a|N\}$ and the associated cycle shape $\prod_{a|N}a^{m(a)}$ such that $\sum_{a|N} a m(a)=24$, corresponding to the cycle
decomposition of the $\IZ_N$ action on the even homology lattice $H_{\rm even}(K3)\sim \IZ^{24}$.
For simplicity we shall restrict ourselves to CHL  orbifolds with $N$ prime and cycle shape 
$1^k N^k$ with $k=24/(N+1)$. In this case, one can decompose 
$\Lambda_{22,6}=\Lambda_{Nk,8-k}\oplus \sLambda_{1,1} \oplus \sLambda_{k-3,k-3}$,
such that  the $\IZ_N$ action acts on the first term by a $\IZ_N$ rotation, on the second term
by an order $N$ shift, leaving $\sLambda_{k-3,k-3}$ invariant (see \S\ref{ZNPartitionFunction}
for details on this construction). We denote by $\Lambda_{k,8-k}$ the quotient of $\Lambda_{Nk,8-k}$  under the $\IZ_N$ rotation (see Table \ref{TableauCHL}). 
The U-duality group $G_4(\IZ)$ includes $\Gamma_1(N) \times \widetilde{O}(r-6,6,\IZ)$,
where $\Gamma_1(N)$ is the congruence subgroup of $SL(2,\IZ)$ corresponding to matrices $\big(\colvec[0.7]{a&b\\c&d}\big)$ with $c=0 \mod N, a=d=1\mod N$, and $\widetilde{O}(r-6,6,\IZ)$ 
is the restricted automorphism group of the lattice 
\be
\label{LambdaCHL}
\Lambda_{r-6,6} = \Lambda_{k,8-k}\oplus \sLambda_{1,1}[N] \oplus \sLambda_{k-3,k-3} \ ,
\ee
{\it i.e.} the subgroup of the automorphism group of $\Lambda_{r-6,6}$
which acts trivially on the discriminant group $\Lambda^*_{r-6,6} / \Lambda_{r-6,6}$. Here and below, for any lattice $\Lambda$, we denote by $\Lambda[\alpha]$ the same lattice with a quadratic form rescaled by a factor $\alpha$ (which is equivalent to rescaling the lattice vectors 
by $\sqrt{\alpha}$). Note that the lattice \eqref{LambdaCHL} is still even, {\it i.e.} $Q^2\in 2\IZ$ for 
$Q\in \Lambda_{r-6,6} $, but it is no longer unimodular, rather it is a lattice of level $N$, in the sense
that $Q^2\in 2\IZ/N$ for any $Q\in \Lambda^*_{r-6,6}$. Singularities now occur on codimension-$q$
loci where $Q_R^2=0$ for a norm 2 vector $Q \in \Lambda_{r-6,6}$, or for a norm $2/N$ vector 
$Q\in \Lambda^*_{r-6,6}$.

\begin{table}\begin{center}
$\begin{array}{|c|c|c|c|c|c|c|}
\hline
\,\,N\,\, &\, {\rm Cycle \,\, Shape}\, & k & \quad r\quad  & 
\Lambda_{k,8-k} &\,\,\Lambda_m \cong\Lambda_e^* \,&\,\,|\Lambda_m^*/\Lambda_m|\,\\
\hline
1 & 1^{24} &  12  & 28 &  & \,E_8 \oplus E_8\oplus\sLambda_{6,6} & \,1\\\hline
2 & 1^82^8 & 8 & 20 & E_8[2]   &  \, E_8[2]\oplus\sLambda_{1,1}[2] \oplus \sLambda_{5,5}  & \,2^{10} \\\hline
3 & 1^63^6 & 6 & 16 &   D_6[3] \oplus D_2[-1] & \, A_2 \oplus A_2\oplus\sLambda_{3,3}[3] \oplus \sLambda_{3,3}  & \,3^8\\\hline
5 & 1^45^4  & 4 & 12 &  D_4[5] \oplus D_4[-1] & \,\sLambda_{3,3}[5] \oplus \sLambda_{3,3} & \,5^6\\
\hline
7 & 1^37^3  &  3 & 10 & D_3[7] \oplus D_5[-1] & \, {\scriptsize \big[\colvec[0.7]{ -4 & -1 \\ -1 & -2 }\big]}\oplus\sLambda_{2,2}[7]  \oplus \sLambda_{2,2} & \,7^{5}\\
\hline 
\end{array}$
\caption{\label{TableauCHL}
A class of $\IZ_N$ CHL orbifolds. Here $k=24/(N+1)$ is the weight of the cusp form whose inverse counts 1/2 BPS states, $r=2k+4$ is the rank of the gauge group  and $\Lambda_m$
is the lattice of magnetic charges in four dimensions. The discriminant group $\Lambda_m^*/\Lambda_m$ is isomorphic to $\IZ_N^{k+2}$. Agreement between the lattice $\Lambda_m$ listed
here and $\Lambda_{r-6,6}$ defined in \eqref{LambdaCHL} follows from the lattice isomorphisms \eqref{latiso}.}
\end{center}
\end{table}

While the U-duality group $G_4(\IZ)$ must certainly include $\Gamma_1(N) \times \widetilde{O}(r-6,6,\IZ)$, it may actually be larger. Moreover, special BPS observables may well be invariant under an even larger group. In particular the four-derivative couplings in $D=4$ turn out to be invariant under the action of the larger duality group $\Gamma_0(N) \times O(r-6,6,\IZ)$, where $\Gamma_0(N)$
is the subgroup  of matrices $\big(\colvec[0.7]{a&b\\c&d}\big)$ with $c=0 \mod N$
and $O(r-6,6,\IZ)$ is the full automorphism group of the lattice $\Lambda_{r-6,6}$. 
For example, the exact $\cR^2$ coupling in the low-energy effective action is given by \cite{Harvey:1996ir,Gregori:1997hi,David:2006ud}
\be
\label{fR2}
- \frac{1}{(8\pi)^2} \int \de^4 x\, \sqrt{-g} \log( S_2^{\; k} |\Delta_k(S)|^2)  ( \mathcal{R}_{\mu\nu\rho\sigma}\mathcal{R}^{\mu\nu\rho\sigma}-4 \mathcal{R}_{\mu\nu} \mathcal{R}^{\mu\nu} + \mathcal{R}^2) 
\ee
where $\Delta_k$ is the unique cusp form of weight $k$ under $\Gamma_0(N)$, nowhere vanishing
except at the cusps $\I\infty$ and $0$,
\be
\label{defg}
\Delta_k(\tau) = \eta^k(\tau)\, \eta^k(N\tau)\ .
\ee
In the weak coupling limit $S_2\to\infty$, the expansion
\be
\label{R2weak}
- \log( S_2^{\; k} |\Delta_k(S)|^2)  = 4\pi S_2 - k\log S_2 + k\sum_{m=1}^{\infty}
\left( \sum_{d|m}  d + \sum_{Nd|m} Nd \right) \frac{q_S^m+\bar q_S^m}{m} 
\ee
with $q_S=e^{2\pi\I S}$
reveals, beyond the expected tree-level contribution and logarithmic 
mixing with the non-local part of the effective action, an infinite series of exponentially suppressed corrections ascribed to NS5-branes wrapped on $T^6$ \cite{Harvey:1996ir}. While not all 
 $\Gamma_0(N) \times O(r-6,6,\IZ)$ transformations are expected to be U-dualities of the full theory but only of the  BPS sector, for brevity we shall refer to them respectively as S- and T-dualities. 

In \cite{Persson:2015jka} it was observed that the coupling \eqref{fR2} is in fact invariant under the larger group $\widehat{\Gamma}_0(N)$, obtained by adjoining to  $\Gamma_0(N)$ the Fricke involution,  which acts on modular forms of weight $k$ under $\Gamma_0(N)$ via $f_k(\tau)\mapsto \hat f_k(\tau)=(-\I\tau\sqrt{N})^{-k} f_k(-1/(N\tau))$. Based on a detailed study of geometric dualities in the type II dual description,  
it was  conjectured\footnote{More generally, Fricke S-duality is conjectured to hold whenever 
the cycle shape satisfies the balancing condition $m(a)=m(N/a)$ for all $a|N$. \cite{Persson:2015jka}}
 that the full U-duality group in $D=4$ 
also includes the so-called Fricke S-duality,  which acts 
on the first factor in \eqref{defM4} by the Fricke  involution $S\mapsto -1/(NS)$, 
accompanied by a suitable action of $O(r-6,6,\IR)$ on the second factor.
Additional evidence for the existence of Fricke S-duality
comes from the spectrum of BPS states, to which we now turn.

Point-like particles in $D=4$ carry electric and magnetic charges $(Q,P)\in \Lambda_{em}$ under the $r$ Maxwell fields, where
\be
\Lambda_{em}=\Lambda_e\oplus \Lambda_m \ ,\quad 
\Lambda_m= \Lambda_{r-6,6} = \Lambda_e^* \ .
\ee
The lattice $\Lambda_m$ is tabulated in the sixth column of Table \ref{TableauCHL}, taken from
 \cite{Persson:2015jka}. It agrees with the result \eqref{LambdaCHL} upon making use of the 
lattice isomorphisms \eqref{latiso}. In view of the remarks below \eqref{LambdaCHL}, one has,  
for any $(Q,P)\in\Lambda_{em}$, 
\be
\label{Q2P2PQ}
Q^2 \in \frac{2}{N} \IZ\ ,\quad P^2 \in 2\IZ\ ,\quad P\cdot Q\in \IZ\ .
\ee
The last property in particular ensures that the Dirac-Schwinger-Zwanziger pairing $Q \cdot P' - Q' \cdot P$ is integer. Moreover, it was observed in \cite{Persson:2015jka} that the lattice $\Lambda_{m}$ is in fact $N$-modular, {\it {\it i.e.}} it satisfies 
\be
\label{Nmodular}
\Lambda_{m}^* \simeq \Lambda_{m}[1/N]\ .
\ee
In other words, there exists an $O(r-6,6,\IR)$ matrix $\sigma$ such that $\sqrt{N} \sigma$ maps the lattice $\Lambda_{m}$ into itself and such that 
\be
\label{Nmodular2}
\Lambda_{m}^* = \frac{\sigma}{\sqrt{N}}  \Lambda_m \quad ( \supset \Lambda_m )\ .
\ee
A simple example of $N$-modular lattice is $\Lambda_{d,d}[N]\oplus \Lambda_{d,d}$, which is relevant for $N=5$ above. In this case one can parametrize an element in the lattice in $(\IZ^d,N \IZ^d , \IZ^d, \IZ^d)$ and an element of the dual lattice in $(\IZ^d/N, \IZ^d , \IZ^d, \IZ^d)$ and define $\sigma \in O(2d,2d,\IR)$ such that 
\be \frac{\sigma}{\sqrt{N}} = \frac{1}{\sqrt{N}} \left( \begin{array}{cccc} 0&0& \frac{1}{\sqrt{N}}\mathds{1}_{d,d} & 0 \\0&0&0&\sqrt{N}\mathds{1}_{d,d}\\ \sqrt{N}  \mathds{1}_{d,d} & 0&0&0 \\0&\frac{1}{\sqrt{N}}\mathds{1}_{d,d} &0&0 \end{array}\right) = \left( \begin{array}{cccc} 0&0& \frac{1}{N}\mathds{1}_{d,d} & 0 \\0&0&0&\mathds{1}_{d,d}\\  \mathds{1}_{d,d} & 0&0&0 \\0&\frac{1}{N}\mathds{1}_{d,d} &0&0 \end{array}\right)  \ . \ee
The map \eqref{Nmodular2} defines the action $(Q,P)\mapsto (-\sigma\cdot P/\sqrt{N},\sigma^{-1} \cdot Q\sqrt{N})$ of the Fricke S-duality on $\Lambda_{em}$,
which maps $(Q^2, P^2, P\cdot Q)\mapsto (P^2/N, N Q^2, - P\cdot Q)$ and therefore
preserves the quantization conditions \eqref{Q2P2PQ}. 
It also allows to identify $ N \Lambda_m^*$ as a sublattice of $ \Lambda_m$
\be N \Lambda_m^* = \sqrt{N} \sigma \Lambda_m \subset \Lambda_m \ .  \ee  
Electric charge vectors $Q\in \Lambda_m\subset \Lambda_e$ are called untwisted, while vectors
$Q\in \Lambda_e \smallsetminus \Lambda_m$ are called twisted. More generally, we shall call dyonic charge vectors $(Q,P)$ lying in $\Lambda_m\oplus N \Lambda_e \subset \Lambda_e \oplus \Lambda_m$ untwisted, and twisted otherwise.\footnote{Note that this terminology is defined to be consistent with Fricke and $\Gamma_0(N)$ S-duality, but twisted magnetic charges do not correspond to any twisted sector in the conventional sense.}
Untwisted dyons are in particular such that
\be
Q^2 \in 2 \IZ\ ,\quad P^2 \in 2N \IZ\ ,\quad P\cdot Q\in N \IZ\ .
\ee

Half-BPS states exist only when $Q,P$ are collinear. Their mass is then determined in terms
of the charges via 
\be\label{EMBPSmass}
\cM^2(Q,P) = \frac{2}{S_2} (Q_R - SP_R )\, 
\cdot(Q_R - \bar S P_R)
\ee
where, for a vector $Q^I\in \IR^{p,q}$ ($I=1\dots p+q$), we denote by $Q_L^a$ ($a=1\dots p$) and $Q_R^{\hat a}$ ($\hat a=1\dots q$) its projections
on the positive $p$-plane and its orthogonal complement parametrized by the orthogonal
Grassmannian $G_{p,q}$, such that $Q^2=Q_L^2-Q_R^2$.

For primitive purely electric states (such that $Q\in \Lambda_e$ but $Q/d \notin  \Lambda_e$ for all $d>1$), corresponding
to left-moving excitations in the twisted sectors of the perturbative heterotic string, 
it is known that the helicity supertrace $\Omega_4(Q,0)$ is given 
by \cite{Dabholkar:1989jt,Jatkar:2005bh,Dabholkar:2005by,Sen:2005ch,Govindarajan:2009qt}
\be
\label{om4pert_tw}
\Omega_4(Q,0) = c_k\left( -\tfrac{NQ^2}{2} \right)\ ,\quad  
\frac{1}{\Delta_k(\tau)} = \sum_{m\in \IZ\atop m\geq -1} c_k(m) \, q^m
=\frac{1}{q}+ k + \dots ,
\ee
where $q=2^{2\pi\I\tau}$ and $\Delta_k(\tau)$ is the same cusp form \eqref{defg} which enters in the exact $\cR^2$ coupling. In Appendix \ref{sec_F4CHL}, we rederive this result by constructing the one-loop vacuum amplitude for the CHL models under consideration,
and show that primitive purely
electric states corresponding to left-moving excitations in the untwisted sector have an additional contribution (first observed for $N=2$ in \cite{Dabholkar:2005dt})
\be
\label{om4pert_utw}
\Omega_4(Q,0) = c_k\left( -\tfrac{Q^2}{2} \right) + c_k\left( -\tfrac{NQ^2}{2} \right)\ .
\ee
 Invariance under both $\Gamma_0(N)$ and Fricke S-duality implies that the same formulae apply to generic primitive dyons with $Q^2$ being replaced by $\frac{1}{N}{\rm gcd}(N  Q^2, P^2, Q\cdot P)$. It 
 follows that the helicity supertrace for general 1/2 BPS primitive dyons is given by 
\be
\label{Om4twi}
\Omega_4(Q,P) = c_k\left( -\tfrac{{\rm gcd}(NQ^2,P^2,Q\cdot P)}{2} \right)\ .
\ee
for twisted electromagnetic charge $(Q,P) \in (\Lambda_e \oplus \Lambda_m)   \smallsetminus (\Lambda_m \oplus N \Lambda_e)$, and by
\be
\label{Om4utwi}
\Omega_4(Q,P) = c_k\left( -\tfrac{{\rm gcd}(NQ^2,P^2,Q\cdot P)}{2}  \right)
+c_k\left( -\tfrac{{\rm gcd}(NQ^2,P^2,Q\cdot P)}{2N}  \right)\ .
\ee
for untwisted charge $(Q,P) \in \Lambda_m \oplus N \Lambda_e$. In contrast, primitive 1/2-BPS states of the maximal rank theory have a single contribution
\be
\label{Om4max}
\Omega_4(Q,P) = c\left( -\tfrac{{\rm gcd}(Q^2,P^2,Q\cdot P)}{2} \right)\ ,\quad  
\frac{1}{\Delta(\tau)} = \sum_{m\in \IZ\atop m\geq -1} c(m)\, q^m=\frac{1}{q}+24+\dots
\ee

\subsection{Moduli space and 1/2-BPS couplings in $D=3$}
Upon further compactification on a circle, additional moduli arise from the radius $R$ of the circle,
from the holonomies $a^{1I}$ of the $r$ gauge fields, and from the scalars $a^{2I}, \psi$ dual to 
the $r$ Maxwell fields and to
the Kaluza--Klein gauge field in three dimensions, extending \eqref{defM4} to \cite{Marcus:1983hb}
\be
\label{defM3}
\cM_3 = G_{r-4,8} \slash G_3(\IZ) \ .
\ee
The U-duality group $G_3(\IZ)$ includes $G_4(\IZ)$, the Heisenberg group of large gauge transformations acting on $a^{I,i},\psi$, and the 
automorphism group $O(r-5,7,\IZ)$ (or rather a subgroup containing $\widetilde{O}(r-5,7,\IZ)$)
of the Narain lattice $\Lambda_{r-5,7}=\Lambda_{r-6,6}\oplus\sLambda_{1,1}$ corresponding
to T-duality in heterotic string compactified on $T^7$. 
The action of these subgroups is most
easily seen in the vicinity of the cusps $R\to\infty$ and $g_3\to 0$, corresponding to the
decompactification limit to $D=4$ and the weak heterotic coupling limit in $D=3$, 
where \eqref{defM3} reduces to 
\be
\cM_3
\rightarrow
\begin{cases} \IR^+_R \times \cM_4 \times \tilde T^{2r+1} \\
\IR^+_{1/g_3^2} \times \left[\tfrac{O(r-5,7)}{O(r-5)\times O(7)}\slash O(r-5,7,\IZ) \right] \times T^{r+2}
\end{cases}
\ee
Here, $\tilde T^{2r+1}$ is a circle bundle over the torus $T^{2r}$ parametrized by the
holonomies $a^{i,I}$, with fiber parametrized by the NUT potential $\psi$, while $T^{r+2}$
corresponds to the scalars dual to the Maxwell gauge fields after compactifying the heterotic
string on $T^7$. In heterotic perturbation theory, the effective
action in $D=3$ is  singular on codimension-7 loci where $Q_R^2=0$ for a norm 2
vector $Q\in\Lambda_{r-5,7}$, or for a norm $2/N$ vector  $Q\in\Lambda^*_{r-5,7}$.

For $r=28$, it is well-known that
these subgroups generate the automorphism group $O(24,8,\IZ)$ of the `non-perturbative
Narain lattice' $\Lambda_{24,8} = \Lambda_{22,6}\oplus \sLambda_{2,2}$ \cite{Sen:1995wr}. 
To the extent of our knowledge, the U-duality group for CHL models has not  been discussed 
in the literature, but it is natural to expect that it includes the  restricted automorphism group 
$\widetilde{O}(r-4,8,\IZ)$ of an extended 
Narain lattice of the form $\Lambda_{r-4,8}=\Lambda_m \oplus \Lambda_{2,2}$. We find that the following
choice reproduces the correct S and T-dualities in $D=4$:
\be
\label{Narainnp}
\Lambda_{r-4,8} = \Lambda_{m} \oplus \sLambda_{1,1} \oplus \sLambda_{1,1}[N]\ ,
\ee
where $\sLambda_{1,1}[N]$ is the standard hyperbolic lattice with quadratic form rescaled by a factor of $N$, such that $\Lambda_{r-4,8}^*/\Lambda_{r-4,8}\simeq \IZ_N^{k+4}$. In terms of the usual
construction of $\sLambda_{2,2}$ by windings $(n_1,n_2)\in\IZ^2$,  momenta $(m_1,m_2)\in\IZ^2$ and quadratic form  $2m_1n_1+2m_2n_2$,
we define $\sLambda_{1,1} \oplus \sLambda_{1,1}[N]$ as the sublattice of $\sLambda_{2,2}$ where $n_2$ is restricted to be a multiple of $N$. The restricted automorphism group of $\sLambda_{1,1} \oplus \sLambda_{1,1}[N]$ was determined in \cite{Persson:2015jka,Paquette:2016xoo}, and includes 
$\sigma_{T\leftrightarrow S}\ltimes [\Gamma_1(N)\times\Gamma_1(N)]$, acting by fractional linear transformations on the moduli $(T,S)$ parametrizing
$G_{2,2}$, such that $|m_1+S m_2+T n_1+ST n_2|^2/(S_2 T_2)$ is invariant 
(see \cite[\S C]{Gregori:1997hi}, case V for $N=2$, or \cite[\S 3.1.3]{Sen:2007qy} for arbitrary $N$). In the present context, $T$ is interpreted as $\psi+\I R^2$, while $S$ is the
heterotic axiodilaton. Thus, $\widetilde{O}(r-4,8,\IZ)$ contains 
the S-duality group $\Gamma_1(N)$ and T-duality group $\widetilde{O}(r-6,6,\IZ)$ in four dimensions. In addition,  Fricke S-duality in four dimensions
follows from the fact that  the non-perturbative 
lattice \eqref{Narainnp} is itself $N$-modular,
\be
\label{Nmodularnp}
\Lambda_{r-4,8}^*\simeq \Lambda_{r-4,8}[1/N]\ .
\ee
More evidence for the claim \eqref{Narainnp} will come from the analysis of BPS couplings
in $D=3$, to which we now turn.

In this work, we focus on the coupling of the form $F(\Phi) (\nabla\Phi)^4$ in the low energy effective action in $D=3$, where $F(\Phi)$ is a symmetric rank four tensor $F_{abcd}(\Phi)$, and $ (\nabla\Phi)^4$ is a shorthand notation for a particular contraction of the pull-back of the right-invariant one-forms $P_{a\hat a}$ on $G_{r-4,8}$ to $\IR^{2,1}$ (see \eqref{scalarCoupling}). As stated in \cite{Bossard:2016zdx},
and further explained below, supersymmetry requires that the coefficient $F_{abcd}(\Phi)$ satisfies
the tensorial differential equations
\begin{subequations}
\label{wardf3}
\bea
\label{wardf31}
\cD_{(e}\_^{\hat g} \cD_{f)\hat g} \, F_{abcd}  &=&
\tfrac{2-q}{4}\,\delta_{ef}\,F_{abcd}
+\scalebox{0.9}{$(4-q)$} \,\delta_{e)(a}\,F_{bcd)(f} 
+3\,\delta_{(ab}\,F_{cd)ef} +\scalebox{1.1}{$\frac{15k}{(4\pi)^2}$}\delta_{(ab}\delta_{cd}\delta_{ef)}\delta_{q,6}\ ,\nn
 \\ \\
\label{wardf32}
\cD_{[e}{}^{\,[\hat e}  \cD_{f]}{}^{\,\hat f]}  F_{abcd}&=&0\ , \hspace{20mm} 
\cD_{[e}{}^{\hat{a}}  F_{a]bcd} = 0\ ,
\eea
\end{subequations}
where the constant term in the first line occurs from the regularisation in $q=6$ (see \ref{constantTerm}), and where  $\cD_{a\hat b}$ are the covariant derivatives in tangent frame on $G_{p,q}$. In fact, we shall show that all components
of the  tensor $F_{abcd}$ can be recovered from its trace $\trF\hspace{1.5mm}(\Phi)\equiv F_{ab}^{\, ab}(\Phi)$ 
by acting with the differential operators $\cD_{a\hat b}$ (see \eqref{TensorTrace}). 
Supersymmetry requires that $\trF\hspace{1.5mm}(\Phi)$ be an eigenmode of the Laplacian on $G_{r-4,8}$ with a specified
eigenvalue, while U-duality requires that it should be invariant under $\widetilde{O}(r-4,8,\IZ)$. (Note however that the second order differential equations satisfied by $\trF\hspace{1.5mm} (\Phi)$  does not imply \eqref{wardf3}, so
it should not be thought of as a prepotential for $F_{abcd}$.) 

In CHL $\IZ_N$ orbifold of heterotic  string on $T^7$, $F_{abcd}$ 
gets tree-level and one-loop contributions, both of which are solutions of \eqref{wardf3},
invariant under the full T-duality group $O(r-5,7,\IZ)$.
As we show in Appendix A, the one-loop contribution is given by a modular integral\footnote{A
similar computation  for four-graviton couplings in CHL models was performed in \cite{Tourkine:2012vx}.}
\be
\label{f41loop}
F_{abcd}^{\mbox{\tiny (1-loop)}} = \RN
\int_{\Gamma_0(N)\backslash\cH}\! \!\!\frac {\de\tau_1\de\tau_2}{\tau_2^{\, 2}} 
\frac{\Part{r\hspace{-2pt}-\hspace{-2pt}5,7}[P_{abcd}]}{\Delta_k(\tau)}\ ,
\ee 
where $\Delta_k(\tau)$ is the  same cusp form \eqref{defg} which appeared in the $\cR^2$ couplings
in $D=4$, and $\Part{p,q}[P_{abcd}]$ denotes the Siegel--Narain theta series for the lattice $\Lambda_{p,q}$,
\be
\label{defZpq1}
\Part{p,q}[P_{abcd}]=\tau_2^{q/2}\, \sum_{Q\in\Lambda_{p,q}} \, P_{abcd}(Q)\,
e^{\I\pi Q_L^2 \tau - \I \pi Q_R^2 \bar\tau
}\ ,
\ee
with an insertion of the polynomial 
\be
\label{defP4}
P_{abcd}(Q)= Q_{L,a} Q_{L,b} Q_{L,c} Q_{L,d}
-\frac{3}{2\pi\tau_2}\delta_{(ab}Q_{L,c}Q_{L,d)}
+\frac{3}{16\pi^2\tau_2^2}\delta_{(ab}\delta_{cd)},
\ee
$\Gamma_0(N)\backslash\cH$ is any fundamental domain for the action of $\Gamma_0(N)$ on the
Poincar\'e upper half-plane $\cH$, and $\RN$ denotes a suitable regularization prescription (see \eqref{def1looppqFN}).
In view of the form of the one-loop contribution, 
it is therefore natural to conjecture \cite{Obers:2000ta,Bossard:2016zdx} that 
the exact $(\nabla \Phi)^4$ coupling is the obvious generalization of \eqref{f41loop}, where the Narain 
lattice $\Part{r\hspace{-2pt}-\hspace{-2pt}5,7}$ is replaced by its non-perturbative extension \eqref{Narainnp},
\be
\label{f4exact}
F_{abcd}(\Phi) = \RN
\int_{\Gamma_0(N)\backslash\cH}\! \!\!\frac {\de\tau_1\de\tau_2}{\tau_2^{\, 2}} 
\frac{\Part{r\hspace{-2pt}-\hspace{-2pt}4,8}[P_{abcd}]}{\Delta_k(\tau)}\ .
\ee 
A similar formula holds for the trace part $\trF\hspace{1.7mm}(\Phi)  \equiv \delta^{ab} \delta^{cd} F_{abcd}(\Phi)$,
\be
\label{f4exacttr}
\trF\hspace{1.7mm}(\Phi) = \RN
\int_{\Gamma_0(N)\backslash\cH}\! \!\!\frac {\de\tau_1\de\tau_2}{\tau_2^{\, 2}}  
 \, \Part{r\hspace{-2pt}-\hspace{-2pt}4,8} \cdot D_{-k+2} D_{-k} \frac{1}{\Delta_k(\tau)} \ ,
\ee 
where $D_w=\frac{\I}{\pi}(\partial_\tau-\frac{\I w}{2\tau_2})$ is the Maass raising operator,
mapping modular forms of weight $w$ to weight $w+2$. The
proposals  \eqref{f4exact} and \eqref{f4exacttr}   are manifestly
invariant (or covariant) under the full automorphism group $O(r-4,8,\IZ)$ of 
the non-perturbative lattice \eqref{Narainnp}, which contains the true
U-duality group in $D=3$. Moreover, since the latter is
$N$-modular, $\Part{r\hspace{-2pt}-\hspace{-2pt}4,8}$ is invariant under the combined
action of the Fricke involution on $\cH$ and the
rotation $\sigma\in O(r-4,8,\IR)$  realizing the isomorphism
\eqref{Nmodularnp},
\be
\label{FrickeGamma}
\Part{r\hspace{-2pt}-\hspace{-2pt}4,8}(\Phi,\tau)[P_{abcd}] = 
\left(-\I \tau\sqrt{N}\right)^{-k} \Part{r\hspace{-2pt}-\hspace{-2pt}4,8}[P_{abcd}]
\left(\sigma\cdot\Phi,-\frac{1}{N\tau}\right)\ .
\ee
Since $\Delta_k$ is also an eigenmode of the Fricke involution on $\cH$, and since the
fundamental domain $\Gamma_0(N)\backslash\cH$ can be chosen to be invariant under this involution,
it follows that $F_{abcd}(\Phi)$ (and therefore $\trF\hspace{1.7mm}(\Phi)$) is covariant (invariant) under
the action of $\sigma$ on $G_{r-4,8}$. As already anticipated, this action descends to Fricke S-duality in $D=4$. 

It is also important to note that the couplings \eqref{f4exact} and \eqref{f4exacttr}
are singular on codimension-$8$ loci where $Q_R^2=0$ for some norm 2 vector $Q\in 
\Lambda_{r-4,8}$, or norm $2/N$  vector $Q\in \Lambda^*_{r-4,8}$. When the vector $Q$ 
is of the form $Q=(0,\CQ,0) \in \Lambda_{r-4,8}$ with $\CQ\in \Lambda_{r-5,7}$, this singularity
is visible at the level of the one-loop correction to the $(\nabla\Phi)^4$ coupling, and is due to 
additional states becoming massless. However, the one-loop correction is singular in real 
codimension 7, while the full non-perturbative coupling (assuming that \eqref{f4exact} is 
correct) is singular in real codimension 8. Indeed,  the invariant norm $Q_R^2=\CQ_R^2 + \tfrac12 g_3^2 (\CQ\cdot a)^2$ 
vanishes only when both $\CQ_R^2=0$ and $\CQ\cdot a=0$. This partial resolution may be seen as an analogue of the resolution of the conifold singularity on the vector multiplet branch in type II strings compactified on a CY threefold times a circle, or equivalently on the hypermultiplet branch
in the mirror description \cite{Ooguri:1996me}. Singularities associated to 
generic vectors $Q\in \Lambda_{r-4,8}$ are not visible at any order in perturbation theory, 
and are associated to `exotic' particles in $D=3$ becoming 
massless \cite{Elitzur:1997zn,Obers:1998fb}.

\section{Establishing and solving supersymmetric Ward identities\label{sec_susy}}

In this section, we establish the supersymmetric Ward identities \eqref{wardf3}, from
linearized superspace considerations, relate the components of the tensor $F_{abcd}$
to its trace $\trF\hspace{1.7mm} \equiv F_{ab}^{\,ab}$, and  show that the genus-one modular integral
\eqref{f4exact} obeys this identity. For completeness, we solve the first equation of \eqref{wardf3} in appendix \ref{eqDiffDecomp}, and show that it is satisfied by each Fourier mode of $F_{abcd}$.

\subsection{$(\nabla \Phi)^4$ type invariants in three dimensions}
In three dimensional supergravity with half-maximal supersymmetry, 
the linearised superfield $W_{\hat{a}a}$ satisfies the constraints \cite{Marcus:1983hb,deWit:1992up,Greitz:2012vp}
\be 
D_\alpha^i W_{\hat{a}a} = (\Gamma_{\hat{a}})^{i\hat{\jmath}} \chi_{\alpha\hat{\jmath}a}\ ,  \qquad 
D_\alpha^i  \chi_{\beta\hat{\jmath}a} =-\I (\sigma^\mu)_{\alpha\beta}  (\Gamma^{\hat{a}})_{\hat{\jmath}}{}^i \partial_\mu  W_{\hat{a}a} \ , 
\ee
with $\hat{a}=1$ to $8$ for the vector of $O(8)$, $i=1$ to $8$ for the positive chirality Weyl spinor of $Spin(8)$ and  $\hat{\imath}=1$ to $8$ for the negative chirality Weyl spinor. The 1/2 BPS linearised invariants are defined using harmonics  of $Spin(8)/U(4)$ parametrizing a $Spin(8)$ group element $u^r{}_i, u_{r i}$ in the Weyl spinor representation of positive chirality  \cite{Howe:1994ms}, 
\be 2u_{r(i} u^r{}_{j)}= \delta_{ij} \ , \qquad \delta^{ij} u_{ri} u^s{}_j = \delta_r^s \ , \qquad  \delta^{ij} u_{ri} u_{sj} =0 \ ,\qquad  \delta^{ij} u^{r}{}_{i} u^s{}_j = 0  \ ,  \ee
$u_{r\hat{\imath}}, u^r{}_{\hat{\imath}}$ in the Weyl spinor representation of negative chirality, 
\be 2u_{r(\hat{\imath}} u^r{}_{\hat{\jmath})}= \delta_{\hat{\imath}\hat{\jmath}} \ , \qquad \delta^{\hat{\imath}\hat{\jmath}} u_{r\hat{\imath}} u^s{}_{\hat{\jmath}} = \delta_r^s \ , \qquad  \delta^{\hat{\imath}\hat{\jmath}} u_{r\hat{\imath}} u_{s\hat{\jmath}} =0 \ ,\qquad  \delta^{\hat{\imath}\hat{\jmath}} u^{r}{}_{\hat{\imath}} u^s{}_{\hat{\jmath}} = 0  \ ,  \ee
and $u^+{}_{\hat{a}}, u^{rs}{}_{\hat{a}}, u^-{}_{\hat{a}}$ in the vector representation, 
\bea &&2u^+{}_{(\hat{a}} u^-{}_{\hat{b})} +\frac{1}{2}\varepsilon_{rstu} u^{rs}{}_{\hat{a}} u^{tu}{}_{\hat{b}} = \delta_{\hat{a}\hat{b}}  \ , \quad \delta^{\hat{a}\hat{b}}  u^+{}_{\hat{a}} u^-{}_{\hat{b}} = 1 \ , \quad  \delta^{\hat{a}\hat{b}}  u^{rs}{}_{\hat{a}} u^{tu}{}_{\hat{b}} = \frac{1}{2}\varepsilon^{rstu} \ , \nn \\
&& \delta^{\hat{a}\hat{b}}  u^+{}_{\hat{a}} u^+{}_{\hat{b}} = 0\ , \quad  \delta^{\hat{a}\hat{b}}  u^{+}{}_{\hat{a}} u^{rs}{}_{\hat{b}}=0\ , \quad  \delta^{\hat{a}\hat{b}}  u^{-}{}_{\hat{a}} u^{rs}{}_{\hat{b}}=0\ , \quad  \delta^{\hat{a}\hat{b}}  u^-{}_{\hat{a}} u^-{}_{\hat{b}} = 0\ , 
\eea
with $r=1$ to $4$ of $U(4)$. They are related through the relations
\bea u^+{}_{\hat{a}} u_{ri} (\Gamma^{\hat{a}})^{i\hat{\jmath}} &=& \sqrt{2} u_{r\hat{\imath}} \delta^{\hat{\imath}\hat{\jmath}} \ , \quad u_{ri} u_{s\hat{\jmath}} (\Gamma_{\hat{a}})^{i\hat{\jmath}}  =  \varepsilon_{rstu}u^{tu}{}_{\hat{a}} \ , \quad u^{rs}{}_{\hat{a}} u_{ti} (\Gamma^{\hat{a}})^{i\hat{\jmath}} = 2 \delta_t^{[r} u^{s]}{}_{\hat{\imath}} \delta^{\hat{\imath}\hat{\jmath}} \ , \nn \\
 u_{ri} u^s{}_{\hat{\jmath}} (\Gamma_{\hat{a}})^{i\hat{\jmath}} &=& \sqrt{2} u^{-}{}_{\hat{a}} \delta_r^s \ , \quad  u^r{}_i u_{s \hat{\jmath}} (\Gamma_{\hat{a}})^{i\hat{\jmath}} = \sqrt{2}Êu^{+}{}_{\hat{a}} \delta^r_s \ , \quad u^r{}_i u^s{}_{\hat{\jmath}} (\Gamma_{\hat{a}})^{i\hat{\jmath}} = 2 u^{rs}{}_{\hat{a}} \, . 
\eea
The superfield $W^+_a \equiv  u^{+\hat{a}} W_{\hat{a}a} $ then  satisfies the G-analyticity property 
\be 
u^r{}{}_i D_\alpha^i u^{+\hat{a}} W_{\hat{a}a} \equiv D_\alpha^r W^+_a = 0 \ . 
\ee
One can obtain a linearised invariant from the action of the eight derivatives $D_{\alpha r} \equiv u_{ri} D_\alpha^i$'s on any homogeneous function of the $W^+_a$'s. After integrating over the harmonic variables  with the normalisation $ \int \de u = 1$ and using
\be  \int 
\de u \, u^-{}_{\hat{a}_1}\dots u^-{}_{\hat{a}_n} W^{+a_1} \cdots W^{+a_{n}} = \frac{6!\,n!}{(6+2n)(5+n)!} W^{a_1}{}_{(a_1} \cdots W^{a_n}{}_{a_n)^\prime}  \ ,  
\ee
with the projection $(\hat{a}_1\dots \hat{a}_n)^\prime$ on the traceless symmetric component (recall that $u^-{}_{\hat{a}} u^{-\hat{a}} = 0$), one gets \footnote{In particular for a single vector multiplet \begin{multline}  [D^8]\frac{1}{(n+4)!}  (W^+)^{n+4} = \frac{1}{n!}(W^+)^{n} \bigl(  2 \partial_\mu W_{rs} \partial_\nu W^{rs} \partial^\mu W_{tu} \partial^\nu W^{tu} -   \partial_\mu W_{rs} \partial^\mu W^{rs} \partial_\nu W_{tu} \partial^\nu W^{tu} \bigr) \\- \frac{8}{(n+1)!} (W^+)^{n+1} \partial_\mu W_{rs}Ê\partial_\nu W^{rs} \partial^\mu \partial^\nu W^- +  \frac{8}{(n+2)!} (W^+)^{n+2} \partial_\mu \partial_\nu W^- Ê \partial^\mu \partial^\nu W^- +\dots \ . \end{multline}}
\bea 
&& \frac{(6+2n)(5+n)!}{6!\, n!} \int \de u \,u^-{}_{\hat{a}_1}\dots u^-{}_{\hat{a}_n}[ D^8]   \tfrac{1}{(n+4)!} c_{a_1\dots a_{n+4}} W^{+a_1} \dots W^{+a_{n+4}} 
\nn\\
&=& \frac{1}{n!}c_{a_1\dots a_{n}abcd} W^{a_1}{}_{(\hat{a}_1} W^{a_2}{}_{\hat{a}_2}\dots  W^{a_n}{}_{\hat{a}_n)^\prime}  \cL^{(0)abcd} \nn\\
&& +\frac{1}{(n-1)!}c_{a_1\dots a_{n}abcd} W^{a_2}{}_{(\hat{a}_2} W^{a_3}{}_{\hat{a}_3}\dots  W^{a_{n}}{}_{\hat{a}_{n}}  \cL^{(0)a_{1}abcd}_{\hat{a}_1)^\prime}+ \dots  \nn\\
& &\quad +  \frac{1}{(n-4)!} c_{a_1\dots a_{n}abcd} W^{a_5}{}_{(\hat{a}_5} W^{a_6}{}_{\hat{a}_6}\dots  W^{a_{n}}{}_{\hat{a}_{n} } \cL^{(0)  a_1a_2a_3a_4abcd }_{\hat{a}_1\hat{a_2}\hat{a_3}\hat{a_4})^\prime} Ê+\partial ( \dots ) \label{LinearisedGene}
\;  , \qquad \eea
where the $\cL^{n+4}_n$ are symmetric tensors consisting of a 
homogeneous polynomial of order $4+n$ in $\partial_\mu W^{a\hat{a}}$, $\chi_{\alpha \hat{\imath} a}$ and $\partial_\mu \chi_{\alpha \hat{\imath} a}$, {\it {\it i.e.}}
\bea 
\cL^{(0)abcd} &=&2 \partial_\mu W^{(a}{}_{\hat{a}}  \partial^\mu W^{b}{}_{\hat{b}} \partial_\nu W^{c|\hat{a}} \partial^\nu W^{d)\hat{b}}-\partial_\mu W^{(a}{}_{\hat{a}}  \partial^\mu W^{b|\hat{a}} \partial_\nu W^{c}{}_{\hat{b}} \partial^\nu W^{d)\hat{b}}+ \dots  \nn\\
 \cL^{(0)abcde}_{\hat{a}} &\sim& 4\times  \chi^2 (\partial W)^3+ 5\times \chi^3 \partial \chi \partial W  \nn \\
  \cL^{(0)a_1a_2abcd}_{\hat{a}_1\hat{a}_2} &\sim& 6\times  \chi^4 (\partial W)^2+ 2\times \chi^5 \partial \chi  \nn \\
    \cL^{(0)a_1a_2a_3abcd}_{\hat{a}_1\hat{a}_2\hat{a}_3} &\sim&  \chi^6 \partial W \nn \\
\cL^{(0)  a_1a_2a_3a_4abcd }_{\hat{a}_1\hat{a_2}\hat{a_3}\hat{a_4}} Ê&\sim&  \chi^{8}
\eea
where we only wrote the bosonic part of the first polynomial, and only indicated the number of independent structures for the others, such that $\chi^8$ is for example the unique Lorentz singlet in the irreducible representation of $O(8)$ with four symmetrised indices without trace and eight symmetrised $O(r-8)$ indices. A total derivative has been extracted in \eqref{LinearisedGene} 
in order to remove all second derivative terms $\partial_\mu \partial_\nu W^{a\hat{a}}$.

At the non-linear level, derivatives of the scalar fields only appear through the pull-back of the right-invariant form $  P_{a\hat{b}}$ defined from the Maurer--Cartan form
\be \label{MCform} 
\de g\,  g^{-1} = \left( \begin{array}{cc} \, \de p_{L a}{}^I \eta_{IJ} p_{L b}{}^J \, &
\,   -\de p_{L a}{}^I \eta_{IJ} p_{R \hat{b}}{}^J \, \\ \, \de p_{R \hat{a}}{}^I \eta_{IJ} p_{L b}{}^J \, &\, 
  -\de p_{R \hat{a}}{}^I \eta_{IJ} p_{R \hat{b}}{}^J \, \end{array}\right) 
  \equiv  \left( \begin{array}{cc} \, - \omega_{ab} \, &\,  P_{a\hat{b}}Ê \, \\ \,  P_{b\hat{a}} \, &\,  - \omega_{\hat{a}\hat{b}} \, \end{array}\right) \; , 
  \ee
where $\eta_{IJ}$ is the  $O(r-8,8)$ metric and  $p_{L,a}{}^I$, $p_{R,\hat b}{}^I$ are the left and right projections parametrised by the  Grassmaniann $G_{r-8,8}$. The right-invariant metric on $G_{r-8,8}$ is defined as $G_{\upmu\upnu} = 2 P_{\upmu a \hat{b}} P_{\upnu}^{a \hat{b}} $ and the covariant derivative in tangent frame acts on a symmetric tensor 
as 
\be \label{derivativeAction}\cD_{a\hat{b}} A_{a_1\dots a_m,\hat{b}_1\dots \hat{b}_n} \equiv P_{\upmu a\hat{b}} G^{\upmu\upnu} ( \scalebox{0.9}{$\partial_\upnu  A_{a_1\dots a_m,\hat{b}_1\dots \hat{b}_n} + m \omega_{\upnu (a_1}{}^c A_{a_2\dots a_m)c,\hat{b}_1\dots \hat{b}_n} + n \omega_{\upnu (\hat{b}_1}{}^{\hat{c}} A_{a_1\dots a_m,|\hat{b}_2\dots \hat{b}_n)\hat{c} } $}) \ . \ee 
The supersymmetry invariant associated to a tensor $F_{abcd}$ on the Grassmanian defines a Lagrange density  $\cL$ that decomposes naturally as 
\begin{multline} 
\cL = F_{a_1a_2a_3a_4} \cL^{a_1a_2a_3a_4} + \cD_{(a_1}{}^{\hat{a}} F_{a_2a_3a_4a_5)} \cL^{a_1\dots a_5}{}_{\hat{a}} +\cD_{(a_1}{}^{\hat{a}_1} \cD_{a_2}{}^{\hat{a}_2}   F_{a_{3}a_4a_5 a_6)} \cL^{a_1\dots a_{6}}{}_{\hat{a}_1\hat{a}_2} \\ +\cD_{(a_1}{}^{\hat{a}_1} \cD_{a_2}{}^{\hat{a}_2} \cD_{a_3}{}^{\hat{a}_3}   F_{a_{4}a_5a_6 a_7)} \cL^{a_1\dots a_{7}}{}_{\hat{a}_1\hat{a}_2 \hat{a}_3}  \\ + \cD_{(a_1}{}^{\hat{a}_1} \cD_{a_2}{}^{\hat{a}_2} \cD_{a_3}{}^{\hat{a}_3} \cD_{a_4}{}^{\hat{a}_4}   F_{a_{5}\dots a_{8})} \cL^{a_1\dots a_{8}}{}_{\hat{a}_1\dots \hat{a}_4}  \ , \label{InvAns} \end{multline}
where the $\cL^{n+4}{}_{n}$ are $O(r-8,8)$ invariant polynomial functions of the 
following covariant fields:
\be P_{\mu\, a \hat{b}} = \partial_\mu \phi^\upmu P_{\upmu\, a \hat{b}}  \, , \quad  \chi_{\alpha {\hat{i}} a} \,  , \quad \cD_\mu  \chi_{\alpha {\hat{\imath}} a} = \nabla_\mu    \chi_{\alpha {\hat{\imath}} a}  + \partial_\mu \phi^\upmu \Bigl( \omega_{\upmu\, a}{}^b \chi_{\alpha {\hat{\imath}} a} + \frac{1}{4} \omega_{\upmu\, \hat{a}\hat{b}}( \Gamma^{\hat{a}\hat{b}})_{\hat{\imath}}{}^{\hat{\jmath}}  \chi_{\alpha {\hat{\jmath}} a} \Bigr)    \ , \label{CovFields} \ee
and the dreibeins and the gravitini fields. Because  
non-linear invariants define a linear invariant by truncation to lowest order in the
 fields \eqref{CovFields}, the 
covariant densities $\cL^{4+n}_n$ reduce at lowest order to homogeneous polynomials of 
degree $n+4$ in the covariant fields \eqref{CovFields} that coincide with the linearised 
polynomials $\cL^{(0)n+4}_n$, in particular 
\be \label{scalarCoupling}\cL^{abcd}=  \sqrt{-g} \bigl(   2 P_\mu^{(a}{}_{\hat{a}}  P^{\mu \, b}{}_{\hat{b}}  P_\nu^{c|\hat{a}}  P^{\nu \, d)\hat{b}} -    P_\mu^{(a}{}_{\hat{a}}  P^{\mu \, b|\hat{a}}  P_\nu^{c}{}_{\hat{b}}  P^{\nu \, d)\hat{b}}   + \dots \bigr) \ . 
\ee
The important conclusion to draw from the linearised analysis is that the $O(r-8,8)$ right-invariants  tensors $\cL^{n+4}_{n}$ appearing in the ansatz \eqref{InvAns} are symmetric in both sets of indices and traceless in the $O(8)$ indices. Checking the supersymmetry invariance (modulo a total derivative) of $\cL$ in this basis, one finds that there is no term to cancel  the supersymmetry variation 
\be \delta F_{abcd} = \bigl(  \overline{\epsilon}_i (\Gamma^{\hat{f}})^{i\hat{\jmath}} \chi_{\hat{\jmath}}^{e} \bigr)  \cD^{e\hat{f}} F_{abcd} \ee 
of the tensor $F_{abcd}$  and of its derivative when open $O(r-8)$ indices are antisymmetrized, hence the tensor $F_{abcd}$ must satisfy the constraints 
\be 
\label{LinearF4}  \cD_{[a}{}^{[\hat{a}} \cD_{b]}{}^{\hat{b}]} F_{cdef} = 0 \ , \qquad \cD_{[e}{}^{\hat{a}}  F_{a]bcd} = 0 \; . 
\ee
Similarly, because the $\cL^{n+4}_n$ are traceless in the $O(8)$ indices, the $O(8)$ singlet component of $\delta (\cD F )\cL^5_1$ can only be cancelled by terms coming from $F \delta \cL^4$, {\it {\it i.e.}}
\be F_{abcd} \delta \cL^{abcd}Ê+ \frac{1}{8}\cD_e{}^{\hat{a}} \cD_{f\hat{a}}  F_{abcd}  ( \overline{\epsilon}\, Ê\Gamma^{\hat{c}} \chi^e)  \cL_{\hat{c}}^{abcdf} \sim 0 
\ee
modulo terms arising from the supercovariantisation,\footnote{The same construction in superspace implies that the lift of $\cL$ in superspace is $\de$-closed \cite{Gates:1997kr}, such that $\de_\omega \cL^{abcd} = \frac{15}{16} P^{\hat{c}}{}_e \wedge \cL_{\hat{c}}^{abcde} - \frac{5}{8} P^{\hat{c}(a} \wedge \cL_{\hat{c}}^{bcd)e}{}_e $, in agreement with equation \eqref{VaryL}. Therefore, the terms associated to the variation of the gravitini that we disregard here do not spoil the argument \cite{Bossard:2014aea}.} so that the covariant components must satisfy
\be \delta \cL^{abcd}Ê+ \frac{5b_1}{8}   ( \overline{\epsilon}\, Ê\Gamma^{\hat{c}} \chi_e)  \cL_{\hat{c}}^{abcde}+ \frac{5b_2}{8}   ( \overline{\epsilon}\, Ê\Gamma^{\hat{c}} \chi^{(a})  \cL_{\hat{c}}^{bcd)e}{}_e = \nabla_\mu ( \dots ) \label{VaryL} \ee
and the tensor $F_{abcd}$ an equation of the form 
\be
\label{D2F4susy} 
 \cD_{e}{}^{\hat{a}} \cD_{f\hat{a}}  F_{abcd} =  5b_1 \delta_{e(f} F_{abcd)} + 5 b_2 \, \delta_{(fa} F_{bcd)e} \ , \ee
for some numerical constants $b_1, b_2$ which are fixed by consistency. In particular the integrability condition on the component antisymmetric in $e$ and $f$ implies $b_2 = 2b_1 + 4$. 

Before determining the constants $b_i$, it is convenient to generalize $F_{abcd}$ to a 
completely symmetric tensor $F_{abcd}^{(p,q)}$ on a general Grassmanian $G_{p,q}$, 
which would arise by considering a superfield in  $D=10-q$ dimensions with $3\le q \le 6$, with harmonics parametrizing similarly the Grassmannian $G_{q-2,2}$ \cite{Bossard:2009sy}. The corresponding 
invariant takes the form $\cL = F_{abcd}^{(p,q)}\, \cL^{abcd}+\dots$
with 
\begin{multline} \label{F4allD} \cL^{abcd} = \sqrt{-g} \Bigl( F^{(a}_{\mu\nu} F^{b|\nu\sigma} F^b_{\sigma\rho} F^{d)\rho\mu} - \frac{1}{4} F^{(a}_{\mu\nu} F^{b|\mu\nu} F^b_{\sigma\rho} F^{d)\sigma\rho} \\+ ( 4 F^{(a}_{\mu\sigma} F^b_{\nu}{}^{|\sigma} - \eta_{\mu\nu} F^{(a}_{\sigma\rho} F^{b|\sigma\rho}) P^{\mu |c}{}_{\hat{a}} P^{\nu| d)\hat{a}}Ê\\+ 2 P_\mu^{(a}{}_{\hat{a}}  P^{\mu \, b}{}_{\hat{b}}  P_\nu^{c|\hat{a}}  P^{\nu \, d)\hat{b}} -    P_\mu^{(a}{}_{\hat{a}}  P^{\mu \, b|\hat{a}}  P_\nu^{c}{}_{\hat{b}}  P^{\nu \, d)\hat{b}}  +\dots  \Bigr) \end{multline}
where $F_{abcd}$ is subject to 
the constraints \eqref{LinearF4} 
and 
\be 
\label{EqD2F}  
\cD_{e}{}^{\hat{a}} \cD_{f\hat{a}}  F^{(p,q)}_{abcd} = b_1\, \delta_{ef} F^{(p,q)}_{abcd}+ 2 b_2 \delta_{f(a} F^{(p,q)}_{bcd)e} +(2b_2 -q)  \delta_{e(a} F^{(p,q)}_{bcd)f} + 3 b_3 \,\delta_{(ab} F^{(p,q)}_{cd)ef}   \; . 
\ee
with coefficients $b_1,b_2,b_3$ a priori depending on $p,q$.

A first integrability condition  for \eqref{EqD2F} is obtained through
\bea 
0 =  \cD_e{}^{\hat{a}} ( \cD_{f\hat{a}} F^{(p,q)}_{abcd} - \cD_{(a|\hat{a}} F^{(p,q)}_{bcd)f} ) &=&\Bigl( b_1- \frac{2b_2 -q}4 \Bigr) ( \delta_{ef} F^{(p,q)}_{abcd} - \delta_{e(a} F^{(p,q)}_{bcd)f} )\nn\\
&& + \frac{3}{2} ( b_2 - b_3) ( \delta_{f(a} F^{(p,q)}_{bcd)e} - \delta_{(ab} F^{(p,q)}_{cd)ef}) \; ,  
\eea
which implies $b_1 = \frac{b_2-q}{4}$ and $b_3=b_2$, consistently with \eqref{D2F4susy}. Similarly, considering 
\bea &&\cD_g{}^{\hat{a}} \bigl(\cD_e{}^{\hat{b}} \cD_{f\hat{b}} F^{(p,q)}_{abcd}\bigr)  - \cD_f{}^{\hat{a}} \bigl( \cD_e{}^{\hat{b}} \cD_{g\hat{b}} F^{(p,q)}_{abcd}\bigr)= 
2b_1 \delta_{e[f} \cD_{g]}{}^{\hat{a}} F^{(p,q)}_{abcd} 
+ 2 b_2 \delta_{a)[f} \cD_{g]}{}^{\hat{a}} F^{(p,q)}_{e(bcd} \nn\\
&=& [ \cD_g{}^{\hat{a}}  ,  \cD_e{}^{\hat{b}} ]   \cD_{f\hat{b}} F^{(p,q)}_{abcd} -   [ \cD_f{}^{\hat{a}}  ,  \cD_e{}^{\hat{b}} ]   \cD_{g\hat{b}} F^{(p,q)}_{abcd}+\cD_e{}^{\hat{b}} [ \cD_{[g}{}^{\hat{a}} , \cD_{f]\hat{b}}] F^{(p,q)}_{abcd} \nn\\
&=&  \frac{2-q}{2} \delta_{e[f} \cD_{g]}{}^{\hat{a}} F^{(p,q)}_{abcd} + 2 \delta_{a)[f} \cD_{g]}{}^{\hat{a}} F^{(p,q)}_{e(bcd}  \ ,   \eea
and therefore $b_1=\frac{2-q}{4}$ and $b_2=1$ and so $b_3=1$ so that 
\be
\label{D2FFinal} 
 \cD_{e}{}^{\hat{a}} \cD_{f\hat{a}}  F^{(p,q)}_{abcd} =  5\frac{2-q}{4} \delta_{e(f} F^{(p,q)}_{abcd)} + 5 \delta_{(fa} F^{(p,q)}_{bcd)e} \ . 
 \ee
Taking traces of this equation one can show that the entire tensor is determined by its trace component $\trF^{(p,q)} \equiv F_{ab}^{(p,q)ab}$ through 
\be 
\label{TensorTrace} 
{F}_{abcd}^{(p,q)}  = \tfrac{1}{(8+p-q)(6+p-q)} \Bigl( 2 \cD_{(a}{}^{\hat{e}}  \cD_{b|\hat{e}}  \cD_c{}^{\hat{f}}  \cD_{d)\hat{f}}  +(2q-7)  \delta_{(ab} \cD_c{}^{\hat{e}}  \cD_{d)\hat{e}} +  \tfrac{3(q-2)(q-4)}{8} \delta_{(ab} \delta_{cd)} \Bigr) \trF^{(p,q)}  \ . 
 \ee
The function $\trF^{(p,q)}$ is an eigenmode of the Laplacian $\Delta_{G_{p,q}}  \equiv 2\cD_{a\hat{b}} \cD^{a\hat{b}}$ on $G_{p,q}$, and satisfies 
\be  \label{D2scalar} 
\Delta_{G_{p,q}} \trF^{(p,q)} =  - \frac12 (p+4)(q-6) \trF^{(p,q)} \  , \qquad  \cD_{[a}{}^{[\hat{a}} \cD_{b]}{}^{\hat{b}]} \trF^{(p,q)} = 0\  . \ee
It is worth noting, however, that Eq. \eqref{D2FFinal} for the tensor defined by \eqref{TensorTrace} 
is an additional constraint on the function $\trF\hspace{1.5mm}$, which does not follow by integrability from the two equations \eqref{D2scalar}.

Finally, let us note that the discussion so far only applies to the local Wilsonian effective action. As we shall see in the next subsection, the Ward identity satisfied by the renormalized coupling $\hat{F}_{abcd}$ is corrected in four dimensions (for $q=6$) because  of the 1-loop divergence of the supergravity amplitude \cite{Fischler:1979yk}, leading
to the source term in \eqref{wardf3}.

\subsection{The modular integral solves the Ward identities\label{modularIntegral}}

In this subsection we shall prove that the  modular integral \eqref{f4exact} is a solution of
the supersymmetric Ward identities \eqref{wardf3}. More generally, we shall show that 
the modular integral
\be
\label{def1looppq}
F^{(p,q)}_{abcd}(\Phi) = \RN
\int_{\Gamma_0(N)\backslash\cH}\! \!\!\frac {\de\tau_1\de\tau_2}{\tau_2^{\, 2}} 
\frac{\Part{p,q}[P_{abcd}]}{\Delta_k(\tau)}\ ,
\ee
where $\Delta_k(\tau)$ is the cusp form \eqref{defg} 
of weight $k$ under $\Gamma_0(N)$, 
$\Lambda_{p,q}$ is a level $N$ even lattice of signature $(p,q)$ with $\tfrac{p-q}{2}+4=k$,
and $P$ is the quartic polynomial \eqref{defP4}, satisfies the constraints \eqref{LinearF4} 
and \eqref{EqD2F}. Moreover, its trace $\delta^{ab}\delta^{cd}F^{(p,q)}_{abcd}(\Phi)$ is given by
\be
\label{f4exactpqtr}
\trF^{(p,q)}(\Phi) = \RN
\int_{\Gamma_0(N)\backslash\cH}\! \!\!\frac {\de\tau_1\de\tau_2}{\tau_2^{\, 2}} 
\Part{p,q}\,\cdot  D_{-k+2} D_{-k}
\frac{1}{\Delta_k(\tau)}\ .
\ee

Before going into the proof however, it will be useful to spell out the regularization prescription which
we use to define these otherwise divergent modular integrals. 
We follow the procedure developed in \cite{MR656029,Angelantonj:2011br,Angelantonj:2013eja}, whereby the integral is first carried out on the truncated  fundamental domain
$\cF_{N,\Lambda}=\cF_{N} \cap \{ \tau_2 < \Lambda\} \cap \{ \frac{\tau_2}{N|\tau|^2}>\Lambda\}$, where $\cF_{N}$ is the standard fundamental domain for $ \Gamma_0(N)\backslash\cH$, invariant under the Fricke involution $\tau\mapsto -1/(N\tau)$, and then the limit $\Lambda\to\infty$ is taken
after subtracting any divergent term in $\Lambda$. In the case of the integral \eqref{def1looppq},
the divergent term originates from the contribution of the vector $Q=0$ in $\Part{p,q}[P_{abcd}]$, so the regularized integral is defined for $q\neq 6$ by 
\be
\label{def1looppqFN}
F^{(p,q)}_{abcd}(\Phi) = \lim_{\Lambda\to\infty} \left[ 
\int_{\cF_{N,\Lambda}}\! \!\!\frac {\de\tau_1\de\tau_2}{\tau_2^{\, 2}} 
\frac{\Part{p,q}[P_{abcd}]}{\Delta_k(\tau)}\, 
- \frac{3 (2k)}{16\pi^2} \frac{\Lambda^{\frac{q-6}{2}}}{\frac{q-6}{2}}\, \delta_{(ab} \delta_{cd)} \right]\ ,
\ee
where $2k=c(0)$ for the heterotic string toroidal compactification, and $2k=c_k(0)$ for prime CHL models.
For $q<6$, no subtraction is necessary, as long as the integral is carried out first along $\tau_1\in[-\tfrac12,\tfrac12]$ in the region $\tau\to\infty$. 
For $q=6$, the integral is logarithmically divergent, and the regularized integral is defined 
instead by 
\be
\label{def1looppqFN6}
\widehat{F}^{(p,6)}_{abcd}(\Phi) = \lim_{\Lambda\to\infty} \left[ 
\int_{\cF_{N,\Lambda}}\! \!\!\frac {\de\tau_1\de\tau_2}{\tau_2^{\, 2}} 
\frac{\Part{p,6}[P_{abcd}]}{\Delta_k(\tau)}\, 
- \frac{3 (2k)}{16\pi^2} \log\Lambda\, \delta_{(ab} \delta_{cd)} \right]\ .
\ee
The logarithmic divergence at $q=6$ is consistent with the expected divergence in the one-loop
scattering amplitude of four gauge bosons in $D=4$ supergravity \cite{Fischler:1979yk}.
Equivalently, following \cite{0919.11036} one may consider the modular integral
\be
\label{def1looppqFNs}
F^{(p,q)}_{abcd}(\Phi,\epsilon) = \int_{SL(2,\IZ)\backslash \cH}\! \!\!\frac {\de\tau_1\de\tau_2}{\tau_2^{\, 2-\epsilon}} 
\sum_{\gamma\in\Gamma_{\scriptscriptstyle 0}(N)\backslash SL(2,\IZ)}\left.\frac{\Part{p,q}[P_{abcd}]}{\Delta_k(\tau)}\right|_\gamma\ , 
\ee
which converges for $\Re(\epsilon)<\tfrac{6-q}{2}$, and defines the renormalized integral
as the constant term in the Laurent expansion at $\epsilon=0$ of the analytical continuation of $F^{(p,q)}_{abcd}(\Phi,\epsilon)$. The result will then
differ from \eqref{def1looppqFN6} by an irrelevant additive constant. 
In what follows, we shall often abuse notation and omit the hat in $\hat F^{(p,q)}_{abcd}$ when
stating properties valid for arbitrary $q$. It is also important to note that while  the regularized integral \eqref{def1looppqFN} or \eqref{def1looppqFN6} 
is finite at generic points on $G_{p,q}$, it diverges on a real codimension-$q$ loci of $G_{p,q}$, where $Q_{R,\hat a}=0$ for a  vector $Q\in\Lambda_{p,q}$ with $Q^2=2$, or for a  vector $Q\in\Lambda^*_{p,q}$ with $Q^2=2/N$ (see \eqref{Fpqsing}).

In order to establish that $F_{abcd}^{(p,q)}$ satisfies the constraints \eqref{EqD2F}, we shall first establish differential equations for a general class of
lattice partition functions 
\be
\label{generalPoly}
\Part{p,q}[P]=\tau_2^{\frac{q}{2}}\, \sum_{Q\in\Lambda_{p,q}} \, P(Q)\,
e^{\I\pi Q_L^2 \tau - \I \pi Q_R^2 \bar\tau}\ ,
\ee
where the polynomial $P(Q)$ is obtained by acting with the operator $\tau_2^n e^{-\frac{\Delta}{8\pi\tau_2}}$, with 
\be \Delta\equiv \sum_a \left( \frac{\partial\; }{\partial {Q_{L}^{a}}}\right)^2+\sum_{\hat a} \Bigl( \frac{\partial\; }{\partial {Q_{R}^{\hat a}}}\Bigr)^2 \ , \ee on a homogeneous polynomial of bidegree $(m,n)$ in 
 $(Q_L,Q_R)$, respectively. As shown in  \cite{0919.11036}, $\Part{p,q}[P]$ satisfies
\be
\label{LatStrans}
\Part{p,q}[P](-1/\tau)=
\frac{(-\I)^{\frac{p-q}{2}} \, \tau^{\frac{p-q}{2}+m-n}}{\sqrt{|\Lambda_{p,q}^*/\Lambda_{p,q}|}} 
\PartD{p,q}[P](\tau)\ ,
\ee
which implies that it transforms as a modular form of weight $\tfrac{p-q}{2}+m-n$ under $\Gamma_0(N)$. 
More specifically, we shall consider $\Part{p,q}\big[P_{a_1\ldots a_m,\hb_1\ldots \hb_n}\big]$
with
\be \label{DefBordPol} 
P_{a_1\ldots a_m,\hat b_1\ldots\hat b_n}=\tau_2^n e^{-\frac{\Delta}{8\pi\tau_2}}\big(Q_{L,a_1}\ldots Q_{L,a_m}\,Q_{R,\hb_1}\ldots Q_{R,\hb_n}\big)\ .
\ee
The quartic polynomial $P_{abcd}$ defined in \eqref{defP4} arises in the  case $(m,n)=(4,0)$,
so that   $\Part{p,q}[P_{abcd}]$ is a modular form of weight $\tfrac{p-q}{2}+4=k$, ensuring
the modular invariance of  the integrands in \eqref{def1looppq} and \eqref{f4exactpqtr}. Upon contracting the indices, it is easy to check that $ \delta^{ab} \delta^{cd}
\Part{p,q}[P_{abcd}]=D_{k-2} D_{k-4}\Part{p,q}[1]$, so the claim that \eqref{f4exactpqtr}
gives the trace of \eqref{def1looppq} follows by integration by parts. 
 
To obtain the differential equations satisfied by \eqref{def1looppq}, 
we shall act with the covariant derivative 
$\cD_{a\hat b}$, defined in \eqref{MCform} and \eqref{derivativeAction}. 
As mentioned below \eqref{EMBPSmass}, $p_{L,a}{}^I$, $p_{R,\hat b}{}^I$ are the left and right orthogonal projectors on the Grassmaniann $G_{p,q}=O(p,q)/\left[O(p)\times O(q)\right]$. Using the derivative rules
\be\label{projRules}
\cD_{a\hat b}\,p_{L,c}{}^I=\frac{1}{2}\delta_{ac}\, p_{R,\hat b}{}^I\ ,\hspace{15mm}  \cD_{a\hat b}\,p_{R,\hat c}\_^I=\frac{1}{2}\delta_{\hat b\hat c}\, p_{L,a}{}^I\ ,
\ee
one can effectively define the action of the covariant derivative on a function that only depends on $Q_L$ and $Q_R$ as
\be \label{effectiveDab} \cD_{a\hat b} = \frac{1}{2}Ê\bigl( Q_{L, a} \partial_{\hat{b}} + Q_{R,\hat{b}} \partial_a \bigr) \ , \ee
where  $\partial_a=\frac{\partial}{\partial Q_L^a}$, $\partial_{\hat b}=\frac{\partial}{\partial Q_R^{\hat b}}$. Acting with $\cD_{e\hat g}$ on \eqref{generalPoly} we get  
\be 
\cD_{e\hat g} \Part{p,q}\big[  P_{a_1\ldots a_m,\hb_1\ldots \hb_n}\big]=\Part{p,q}\left[\left(\cD_{e\hat g}-2\pi\tau_2 \,Q_{L,e}Q_{R,\hat g}\right)\,P_{a_1\ldots a_m\hb_1\ldots \hb_n}\right]\ .
\ee
Using \eqref{effectiveDab}, one computes the commutation relations 
\begin{align}
\label{commutationRelations}
&[\Delta,\cD_{e\hat g}]=2\partial_e\partial_{\hat g}\,,&\quad&[\Delta, Q_{L,e}Q_{R,\hat g}]=4\cD_{e\hat g}\,,\\
&[\Delta,Q_{L,e}Q_{L,f}]=2\delta_{ef}+4Q_{L,(e}\partial_{f)}\,,&\quad&[\Delta, Q_{L,(e}\partial_{f)}]=2\partial_e\partial_f\, . 
\end{align}
Using them along with the Baker-Campbell-Hausdorff formula
\be
\label{BCH}
e^{\frac{\Delta}{8\pi\tau_2}}\cO\,e^{-\frac{\Delta}{8\pi\tau_2}}=\cO+\frac{1}{8\pi\tau_2}[\Delta,\,\cO]+\frac{1}{2!}\frac{1}{(8\pi\tau_2)^2}[\Delta,\,[\Delta,\,\cO]]+\ldots\,,
\ee
one easily obtains
\be \label{Dehf}
\cD_{e\hat g} \Part{p,q}\big[  P_{a_1\ldots a_m,\hb_1\ldots \hb_n}\big]=-2\pi\tau_2\,\Part{p,q}\Big[e^{-\frac{\Delta}{8\pi\tau_2}}\Big(Q_{L,e}Q_{R,\hat g}-\frac{1}{\scalebox{0.8}{$(4 \pi \tau_2)^2$}}\partial_e\partial_{\hat g}\Big)e^{\frac{\Delta}{8\pi\tau_2}} P_{a_1\ldots a_m,\hb_1\ldots \hb_n}\Big]\ .
\ee
Note that the similarity transformation is such that the operator acts on the simple monomial in $Q_{a_1}\dots Q_{a_m} Q_{\hat{b}_1} \dots Q_{\hat{b}_n}$ according to \eqref{DefBordPol}, such that it directly follows from \eqref{Dehf} that 
\be \cD_{e\hat g} \Part{p,q}\big[  P_{a_1\ldots a_m,\hb_1\ldots \hb_n}\big]=\Part{p,q}\Big[-2\pi\ P_{ea_1\dots a_m,\hat{g} \hat{b}_1\dots \hat{b}_n} + \tfrac{mn}{8\pi}  \delta_{e(a_1} P_{a_2\ldots a_m),(\hb_2\ldots \hb_n} \delta_{\hb_1)\hat{g}} \Big]\ .
\ee
Upon antisymmetrizing in $(e,a_1)$, we get 
\be 
\cD_{[e}\_^{\hat g} \Part{p,q}\big[  P_{a_1]\ldots a_m,\hb_1\ldots \hb_n}\big]=\frac{1}{8\pi^2\tau_2^2}\,\Part{p,q}\Big[e^{-\frac{\Delta}{8\pi\tau_2}}\partial_{[e}\partial^{\hat g}\,e^{\frac{\Delta}{8\pi\tau_2}} P_{a_1]\ldots a_m,\hb_1\ldots \hb_n}\Big] . 
\ee
which vanishes when $n=0$ since $e^{\frac{\Delta}{8\pi\tau_2}} P_{a_1\ldots a_m}$ does not depend on $Q_R$. Acting a second time with $\cD_{a\hat{b}}$ and antisymmetrizing, we get
\be 
\cD_{[e}\_^{[\hat{e}}\cD_{f]}\_^{\hat{f}]} \Part{p,q}\big[  P_{a_1\ldots a_m,\hb_1\ldots \hb_n}\big]=-2\Part{p,q}\Big[e^{-\frac{\Delta}{8\pi\tau_2}}\,Q_{L,[e}Q_{R}\_^{[\hat e}\partial_{f]}\partial^{\hat f]}\,e^{\frac{\Delta}{8\pi\tau_2}} P_{a_1\ldots a_m,\hb_1\ldots \hb_n}\Big],
\ee
which similarly vanishes when $n=0$. 
Setting $m=4$, we conclude that the modular integral \eqref{def1looppq}
satisfies
\be
\cD_{[e}{}^{\hat{a}}  F_{a]bcd} =0\ ,\qquad \cD_{[e}{}^{\,[\hat e}  \cD_{f]}{}^{\,\hat f]}  F_{abcd}= 0 \ , 
\ee
which therefore establishes the last two equations in \eqref{wardf3}. Note that these two equations do not rely on any particular property of the function $1/\Delta_k$. 

Now, the first equation of \eqref{wardf3} arises from applying the quadratic operator $\cD^2_{ef}\equiv \cD_{(e}{}^{\hat g}\cD_{f)\hat g}$ on the partition function with polynomial insertion,
\be \begin{split}
4\cD^2_{ef} \Part{p,q}\big[  P_{a_1\ldots a_m,\hb_1\ldots \hb_n}\big]&=\Part{p,q}\Big[ \Big(4 \cD^2_{ef}-8\pi\tau_2 Q_{L,(e}Q_R\_^{\hat g}\cD_{f)\hat g}
 \\
&\left.+16\pi^2\tau_2^2\left(Q_{L,e}Q_{L,f}-\tfrac{\delta_{ef}}{4\pi\tau_2}\right)
\left(Q_{R}^2-\tfrac{q}{4\pi\tau_2}\right)-q\delta_{ef}\Big)P_{a_1\ldots a_m\hb_1\ldots \hb_n}\right],
\end{split}
\ee
which gives, using \eqref{commutationRelations} and \eqref{BCH}
\be 
\label{D2efG}
\begin{split}
&4\cD^2_{ef} \Part{p,q}\big[ P_{a_1\ldots a_m,\hb_1\ldots \hb_n}\big]=
\Part{p,q} \Big[ 
e^{-\frac{\Delta}{8\pi\tau_2}}
\Big(16\pi^2\tau_2^2 \,Q_R^2\,Q_{L,e}Q_{L,f}+\frac{\partial_e\partial_f\partial_R^2}{16\pi^2\tau_2^2}
\\
&\qquad\qquad\qquad
-Q_{L,(e}\partial_{f)}(2 Q_R\_^{\hat g} \partial_{\hat g}+q)-\delta_{ef}(Q_R\_^{\hat g} \partial_{\hat g}+q)\Big)e^{\frac{\Delta}{8\pi\tau_2}}P_{a_1\ldots a_m\hat b_1\ldots\hat b_n}
\Big] 
\end{split}
\ee
The first term on the r.h.s. can be rewritten as the action of 
the Maass lowering operator $\bar D_w=-\I\pi\tau_2^2\partial_{\bar\tau}$ mapping modular forms of weight $w$ to weight $w-2$. Indeed,
\be
\begin{split}
\bar D_w \Part{p,q}\left[ P_{efa_1\ldots a_m,\hb_1\ldots \hb_n} \right]=&
-\pi^2\tau_2^2 \, \Part{p,q}\left[  \left(Q_R^2-\tfrac{q+2n}{4\pi\tau_2}\right)
P_{efa_1\ldots a_m, \hat b_1\ldots \hat b_n}\right]
\\
&\quad+\frac{1}{16}\Part{p,q}\left[\Delta\, P_{efa_1\ldots a_m,\hb_1\ldots \hb_n} \right]
\\=& 
 \Part{p,q}\left[
e^{-\frac{\Delta}{8\pi\tau_2}}\left(\tfrac{1}{16} \partial_L^{\; 2}-(\pi \tau_2Q_R)^2\right)e^{\frac{\Delta}{8\pi\tau_2}}P_{efa_1\ldots a_m,\hb_1\ldots \hb_n}\right].
\end{split}
\ee
where in the second line, we used the fact that  $\Delta$ commutes with $e^{-\frac{\Delta}{8\pi\tau_2}}$.  The r.h.s. of \eqref{D2efG} can thus be written as 
\be \label{polyEq}
\begin{split}
&4\cD^2_{ef}\Part{p,q}\left[ P_{a_1\ldots a_m,\hb_1\ldots \hb_n}\right]=
(2-q(1+n))\delta_{ef}
\Part{p,q}\left[ P_{a_1\ldots a_m,\hb_1\ldots \hb_n}\right]\\
&+m(4-q(1+2n))\delta_{|e)(a_1}
\Part{p,q}\left[ P_{a_2\ldots a_m)(f|,\hb_1\ldots \hb_n}\right]
+m(m-1)\delta_{(a_1a_2}
\Part{p,q}\left[ P_{a_3\ldots a_m)ef,\hb_1\ldots \hb_n}\right]\\
&+\tfrac{m(m-1)n(n-1)}{16\pi^2}\delta_{e(a_1}\delta_{|f|a_2}
\Part{p,q}\left[ P_{a_3\ldots a_m),(\hb_1\ldots \hb_{n-2}}\right]
\delta_{\hat b_{n-1}\hat b_{n})}-16 \bar D_w
\Part{p,q}\left[P_{efa_1\ldots a_m,\hb_1\ldots \hb_n}\right],
\end{split}
\ee
where only the last term remains to be computed explicitely. 
Specializing to the case of main interest, we obtain 
\be
\label{polyEqF4}
\Box_{ef} \cdot \Part{p,q}\left[P_{abcd}\right]= -4\bar D_w \, \Part{p,q}\left[P_{abcdef}\right]
\ee
where, for any tensor $F_{abcd}$, we denote
\be
\label{defBoxef}
\Box_{ef} \cdot F_{abcd} \equiv 
\cD^2_{ef} F_{abcd} + \frac{(q-2)}{4}\delta_{ef}\, F_{abcd}
+ (q-4)\delta_{(e|(a} F_{bcd)|f)}-3\delta_{(ab}
F_{cd)ef}
\ee
We can now integrate  both sides of \eqref{polyEqF4} times $1/\Delta_k$ on the truncated fundamental domain $\cF_{N,\Lambda}$, leading to 
\be
\Box_{ef} \, \int_{\cF_{N,\Lambda}}\frac{\de \tau_1\de\tau_2}{\tau_2^2} 
\frac{\Part{p,q}\left[ P_{abcd}\right]}{\Delta_k}= 
-4\int_{\cF_{N,\Lambda}}\frac{\de \tau_1\de\tau_2}{\tau_2^2}
\frac{1}{\Delta_k}\bar D_{k+2}\Part{p,q}\left[ P_{abcdef}\right]
\ee
The r.h.s. is a boundary term, because $\bar D_{-k}(1/\Delta_k)=0$ by holomorphicity. To compute the boundary term we use Stokes' theorem in the form
\be\label{StokesTheorem}
\int_{\partial\cF_{N,\Lambda}}f\,g\,\de\tau=\int_{\cF_{\Lambda}}\de(f\,g\,\de\tau)=\frac{2}{\pi}\int_{\cF_{N,\Lambda}}\frac{\de \tau_1\de\tau_2}{\tau_2^2}(\bar D_w f\,g+f\,\bar D_{w'}\,g),
\ee
where $f$ and $g$ are any modular forms of weight $w$ and $w'=-w+2$ and $2\de\tau_1\de\tau_2=\I \de\tau\wedge \de\bar\tau$. By modular invariance, the boundary term reduces to an integral along the segment $\{1/2\leq\tau_1<1/2, \,\tau_2=\Lambda\}$ and its image under the Fricke involution (for $N>1$). The
latter can be mapped to the former upon using \eqref{LatStrans}. At generic points on the Grassmannian $G_{p,q}$, the contributions of non-zero vectors in $\Lambda_{p,q}$ and 
$\Lambda^*_{p,q}$ are exponentially suppressed, leaving only the contribution of $Q=0$:
\be
\Box_{ef} \, \int_{\cF_{N,\Lambda}}\frac{\de \tau_1\de\tau_2}{\tau_2^2} 
\frac{\Part{p,q}\left[ P_{abcd}\right]}{\Delta_k}
=\Lambda^{\frac{q-6}{2}}\,\frac{15\,(2k)}{2(4\pi)^2}\delta_{(ab}\delta_{cd}\delta_{ef)},
\ee
where we recall that $2k=c(0)$ for the heterotic string compactifications and $2k=2c_k(0)=\tfrac{24}{N+1}$ for CHL models, see table \ref{TableauCHL}. 
Physically, $2k-2$ is the number of vector multiplets. Acting with the same operator $\cD^2_{ef}$ on the subtraction in \eqref{def1looppqFN},
we see that the term proportional to $\Lambda^{(q-6)/2}$ cancels, except for $q=6$ where
the substraction in \eqref{def1looppqFN6} leaves a finite remainder. Thus, we find, as claimed earlier, that the modular integral
\eqref{def1looppq} is annihilated by the second-order differential 
operator $\Box_{ef}$ defined in \eqref{defBoxef},
up to a constant source term present when $q=6$,
\be\label{constantTerm}
\Box_{ef} \,  F_{abcd}^{(p,q)} = \frac{15(2k)}{2(4\pi)^2}\delta_{(ab}\delta_{cd}\delta_{ef)}\, \delta_{q,6}\ .
\ee
In \ref{eqDiffDecomp}, as a consistency check we show 
that this equation is verified by each Fourier mode
in the degeneration limit $O(p,q)\rightarrow O(p-1,q-1)$.

\section{Weak coupling expansion of exact $(\nabla\Phi)^4$ couplings\label{pertLimit}}

In this section, we study the  expansion of the proposal \eqref{f4exact} in the limit where the heterotic string coupling $g_3$ goes to zero, and show that it reproduces the known tree-level and one-loop amplitudes, along with an infinite series of  NS5-brane, Kaluza--Klein monopole and H-monopole instanton corrections. We start by analyzing the expansion of the tensorial modular integral defining the coupling and its trace
\begin{subequations}
\label{f4pq}
\bea
\label{f4pqtens}
F^{(p,q)}_{abcd}(\Phi)&=&\RN
\int_{\Gamma_0(N)\backslash\cH}\! \!\!\frac {\de\tau_1\de\tau_2}{\tau_2^{\, 2}}\,
\frac{\Part{p,q}[P_{abcd}]}{\Delta_k(\tau)}\ ,
\\
\label{f4pqscal}
\trF^{(p,q)}(\Phi) &=& \RN
\int_{\Gamma_0(N)\backslash\cH}\! \!\!\frac {\de\tau_1\de\tau_2}{\tau_2^{\, 2}} \,
\Part{p,q}\,  D_{-k+2} D_{-k} 
\frac{1}{\Delta_k(\tau)}\ ,
\eea
\end{subequations}
for a level $N$ even lattice $\Lambda_{p,q}$ of arbitrary signature $(p,q)$, 
 in the limit near the cusp where $O(p,q)$ is broken to $O(1,1)\times O(p-1,q-1)$, so that the moduli space decomposes into 
\be
\label{decomp1}
G_{p,q} \to \IR^+ \times G_{p-1,q-1} \ltimes \IR^{p+q-2}\ .
\ee
For simplicity, we first discuss the maximal rank case $N=1$, $p-q=16$,  where the integrand is invariant under the full modular group, before dealing with the case of $N$ prime, where the integrand is invariant under the Hecke congruence subgroup $\Gamma_0(N)$. The reader uninterested by the details of the derivation may skip to \S\ref{sec_pertlim}, where we specialize to the values $(p,q)=(r-4,8)$ relevant for the $(\nabla\Phi)^4$ couplings in $D=3$ and interpret the various contributions as 
perturbative and non-perturbative effects in heterotic string theory compactified on $T^7$. 
In \S\ref{sec_iibpert} we 
discuss the case $(p,q)=(r-7,5)$ relevant for $H^4$ couplings in type IIB string theory compactified on $K3$.

\subsection{$O(p,q)\to O(p-1,q-1)$ for even self-dual lattices\label{sec_pertmax}}
We first consider the case where the lattice $\Lambda_{p,q}$  is even self-dual and
 factorizes in the limit \eqref{decomp1} as 
\be
\Lambda_{p,q} \to {\Lambda}_{p-1,q-1} \oplus \sLambda_{1,1}\ .
\ee
We shall denote by $R$ the coordinate on $\IR^+$ and by $a^I$, $I=2\ldots p+q-1$ the coordinates on $\IR^{p+q-2}$. $R$ parametrizes a one-parameter subgroup $e^{R H_0}$ in $O(p,q)$, such that the action of the non-compact Cartan generator $H_0$ on the Lie algebra $\mf{so}_{p,q}$ decomposes into 
\be
\label{sopqDec1}
\mf{so}_{p,q}\simeq {(\bf p+q-2)}^{(-2)}\oplus( \mf{gl}_1
\oplus\mf{so}_{p-1,q-1})^{(0)}\oplus({\bf p+q-2})^{(2)}\ .
\ee
while the coordinates $a^I$ parametrize the unipotent subgroup obtained by exponentiating 
the grade $2$ component in this decomposition. A  generic charge vector $Q_{\cI} \in\Lambda_{p,q}\simeq {\bf 1}^{(-2)} \oplus ({\bf p+q-2})^{(0)}\oplus {\bf 1}^{(2)}$ decomposes into $Q_\cI=(m,\CQ_I,n)$ where $(m,n)\in \sLambda_{1,1} = \IZ^{2}$  and $\CQ_I\in\Lambda_{p-1,q-1}$, such that $Q^2=-2mn+\CQ^2$. The orthogonal projectors defined by $Q_L\equiv p^\cI_{L}Q_\cI$ and $Q_R\equiv p^\cI_{R}Q_\cI$ decompose according to 
\be
\label{pLRdec1}
\begin{split}
p^\cI_{L,1}Q_\cI = & \frac{1}{R\sqrt 2}\left( m + a\cdot \CQ + \frac12 a\cdot a \,n\right) - \frac{R}{\sqrt{2}} n,\\
p^\cI_{L,\alpha}Q_\cI =&  \widetilde p^I_{L,\alpha}(\CQ_I+ n a_I),
\\
p^\cI_{R,1}Q_\cI = & \frac{1}{R\sqrt 2}\left( m + a\cdot \CQ + \frac12 a\cdot a \,n\right) + \frac{R}{\sqrt{2}} n, \\
p^\cI_{R,\hat\alpha}Q_\cI =& \widetilde p^I_{R,\hat\alpha}(\CQ_I+ n a_I),
\end{split}
\ee
where  $\widetilde p^I_{L,\alpha}, \widetilde p^I_{R,\hat\alpha}$ ($\alpha=2\dots d+16$, $\hat\alpha=2\dots d$) are orthogonal projectors in $G_{p-1,q-1}$ satisfying $\CQ^2=
\CQ_L^2-\CQ_R^2$. In the following we shall 
denote $| Q_R|\equiv \sqrt{\CQ_{R}^2}$.

To study the behavior of \eqref{f4pq} in the limit $R\gg 1$,\footnote{Since $1/\Delta$ grows as $e^{\frac{2\pi}{\tau_2}}$ at $\tau_2\to 0$,  the following treatment which relies on exchanging the sum and the integral for unfolding is justified for $R^2> 2$.} it is useful to perform a Poisson resummation on $m$. For a lattice 
partition function $\Part{p,q}$ with no insertion, as in the scalar integral 
\eqref{f4pqscal}, this gives
\be
\label{Poisson1}
\Part{p,q} = R\, \tau_2^{\frac{q-1}{2}} \sum_{(m,n)\in \IZ^2}\, \sum_{\CQ\in \Lambda_{p-1,q-1}}e^{-\frac{\pi R^2 |n\tau+m|^2}{\tau_2}}
e^{2\pi\I m(a\cdot \CQ+\frac12 a\cdot a\, n)}\, q^{\frac12 \CQ_L^2}\, q^{\frac12 \CQ_R^2}
\ee
In the case of a lattice sum with momentum insertion, as in the tensor integral $F^{(p,q)}_{abcd}$ \eqref{f4pqtens}, we must distinguish whether the indices $abcd$ lie along the direction $1$ or along the directions $\alpha$. Denoting by $h$ the number of indices along direction $1$, the previous result generalizes to 
\begin{multline}\label{poissonRessummedP}
\Part{p,q} \left[ e^{-\frac{\Delta}{8\pi\tau_2}} \left[ (Q_{L,1})^h Q_{L,\alpha_1} \ldots Q_{L,\alpha_{4-h}} \right] \right] 
= R\, \sum_{(m,n)\in \IZ^2}
\left( \frac{ R(n\bar \tau+ m)}{\I\tau_2\sqrt{2}}\right)^h\, 
e^{-\frac{\pi R^2 |n\tau+m|^2}{\tau_2}}\\
\times 
\ShiftPart{p\hspace{-1.5pt}-\hspace{-2pt}1,p\hspace{-1.5pt}-\hspace{-2pt}1}{na}\Big[e^{-\frac{\Delta}{8\pi\tau_2}} 
\left[ \CQ_{L,\alpha_1}\ldots \CQ_{L,\alpha_{4-h}}\right] 
e^{2\pi\I m(\CQ-\frac12 a\, n)\cdot a} \Big] .
\end{multline}
In this representation, modular invariance is manifest, since a transformation $\tau\mapsto 
\frac{a\tau+b}{c\tau+d}$ can be compensated by a linear transformation $(n,m)\mapsto 
(n,m)\big(\colvec[0.7]{a&b\\c&d}\big)$, under which the  second line of \eqref{poissonRessummedP} transforms with weight $12-h$. As a relevant example for what follows, consider the case $(n,m)=k(c,d)$, $k={\rm gcd}(m,n)$, then using an transformation $\big(\colvec[0.7]{a&b\\c&d}\big)\in SL(2,\IZ)$
\begin{multline}
\sum_{\CQ\in\Lambda_{p-1,q-1}+kc\,a} e^{-\frac{\Delta}{8\pi\tau_2}} 
\left[ \CQ_{L,\alpha_1}\ldots \CQ_{L,\alpha_{4-h}}\right] 
e^{2\pi\I \,kd(\CQ-\frac12 a\, kc)\cdot a}\,q^{\frac12 \CQ_L^2}\,\bar q^{\frac12\CQ_R^2} =
\\
(c\tau+d)^{12-h} \, \sum_{\CQ\in\Lambda_{p-1,q-1}}
e^{-\frac{\Delta}{8\pi\tau_2}} \left[ \CQ_{L,\alpha_1}\ldots \CQ_{L,\alpha_{4-h}}\right] 
e^{2\pi\I k\,\CQ\cdot a}\,q^{\frac12 \CQ_L^2}\,\bar q^{\frac12\CQ_R^2} \ .
\end{multline}
We can therefore compute the integral using the orbit method \cite{MR993311,McClain:1986id,Dixon:1990pc}, namely decompose the sum over $(m,n)$ into various orbits under $SL(2,\IZ)$, and for each orbit $\cO$, retain the contribution of a particular element $\varsigma\in \cO$ at the expense of extending the integration domain $\cF_1= SL(2,\IZ)\backslash\cH$ to $\Gamma_\varsigma\backslash \cH$, where $\Gamma_\varsigma$ is the stabilizer of $\varsigma$ in $SL(2,\IZ)$,\footnote{This unfolding procedure requires particular care since
the integrand is not of rapid decay near the cusp. We suppress these details here, and refer to 
\cite{MR656029,0919.11036,1004.11021,Angelantonj:2011br,Angelantonj:2012gw,Angelantonj:2013eja} for rigorous treatments.} by using the identity 
\be\label{unfoldingIdentity}
\bigcup_{\gamma\in\Gamma_\varsigma\backslash SL(2,\IZ)}\gamma\cdot \cF_1=\Gamma_\varsigma\backslash \cH.
\ee
The coset representative $\varsigma\in\cO$, albeit  arbitrary, is usually chosen so as to make the unfolded domain $\Gamma_\varsigma\backslash \cH$ as simple as possible.
In the present case, there are two types of orbits:

\paragraph{The trivial orbit}  $(n,m)=(0,0)$ produces, up to a factor of $R$, the integrals \eqref{f4pq} for the lattice $\Lambda_{p-1,q-1}$, provided none of the indices $abcd$ lie along the direction 1,
\be
\label{Fdec0}
F_{\alpha\beta\gamma\delta}^{(p,q),0} = R\, F^{(p-1,q-1)}_{\alpha\beta\gamma\delta}\ ,\quad
\trF^{(p,q),0} = R \, \trF^{(p-1,q-1)}\ ,
\ee
while it vanishes otherwise ({\it i.e.} when $h>0$). 

\paragraph{The rank-one orbit} corresponds to terms with $(n,m)\neq (0,0)$. Setting $(n,m)=k(c,d)$,  with $\gcd(c,d)=1$ and
$k\neq 0$, the doublet $(c,d)$ can always be rotated by an element of $SL(2,\IZ)$ into $(0,1)$, whose stabilizer inside $SL(2,\IZ)$ is $\Gamma_\infty=\{\big(\colvec[0.7]{1&n\\0&1}\big), n\in\IZ\}$.
Thus, doublets $(c,d)$ with $\gcd(c,d)=1$
are in one-to-one correspondence with elements of $\Gamma_\infty\backslash SL(2,\IZ)$. For each $k$, one can therefore unfold the integration domain $SL(2,\IZ)\backslash\cH$ to $\cS=\Gamma_{\infty}\backslash\cH  = \IR^+_{\tau_2}\times (\IR/\IZ)_{\tau_1}$, the unit width strip, provided one keeps only the term $(c,d)=(0,1)$ in the sum. The resulting contribution to the tensor integral \eqref{f4pqtens}
are
\be
\label{pertRank1orbit}	
\begin{split}
&F_{\alpha\beta\gamma\delta}^{(p,q),1} = R  \int_{\IR^+}\frac{\de \tau_2}{\tau_2^2}\int_{\IR/\IZ}\hspace{-2mm}\de \tau_1 \, 
 \sum_{k\neq 0} e^{-\pi R^2 k^2/\tau_2}
\frac{\Part{p\hspace{-1.5pt}-\hspace{-2pt}1,q\hspace{-1.5pt}-\hspace{-2pt}1}\left[ \tilde P_{\alpha\beta\gamma\delta}\, e^{2\pi\I k a^I \CQ_I}\right]}{\Delta}\ ,
\\
&F_{11\gamma\delta}^{(p,q),1} = R\int_{\IR^+}\frac{\de \tau_2}{\tau_2^2}\int_{\IR/\IZ}\hspace{-2mm}\de \tau_1 \,  
\sum_{k\neq 0}
\left( \frac{Rk}{\I\tau_2\sqrt2}\right)^2 e^{-\pi R^2 k^2/\tau_2}
\frac{\Part{p\hspace{-1.5pt}-\hspace{-2pt}1,q\hspace{-1.5pt}-\hspace{-2pt}1}\left[\tilde P_{\alpha\beta}\, e^{2\pi\I k a^I \CQ_I}\right]}{\Delta}\ ,
\\
&F_{1111}^{(p,q),1} = R\int_{\IR^+}\frac{\de \tau_2}{\tau_2^2}\int_{\IR/\IZ}\hspace{-2mm}\de \tau_1 \,  \sum_{k\neq 0}
\left( \frac{Rk}{\I\tau_2\sqrt2}\right)^4 e^{-\pi R^2 k^2/\tau_2}
\frac{\Part{p\hspace{-1.5pt}-\hspace{-2pt}1,q\hspace{-1.5pt}-\hspace{-2pt}1}\left[ e^{2\pi\I k a^I \CQ_I}\right]}{\Delta}\ ,
\end{split}
\ee
where
\be
\tilde P_{\alpha_1\dots \alpha_{4-h}} = e^{-\frac{\Delta}{8\pi\tau_2}} \left[ \CQ_{L,\alpha_1} \dots \CQ_{L,\alpha_{4-h}} \right]\ ,
\ee 
while the contribution to its trace is
\be
\trF^{(p,q),1} = R  \int_{\IR^+}\frac{\de \tau_2}{\tau_2^2}\int_{\IR/\IZ}\hspace{-2mm}\de \tau_1 \,  
\sum_{k\neq 0} e^{-\pi R^2 k^2/\tau_2}
\,
\Part{p\hspace{-1.5pt}-\hspace{-2pt}1,q\hspace{-1.5pt}-\hspace{-2pt}1}\left[  e^{2\pi\I k a^I \CQ_I}\right]\, D^2\left(\frac{1}{\Delta}\right)\ .
\ee
The integral over $\cS$ can be computed by inserting the Fourier expansion 
\be
\label{FourierDelta}
\frac{1}{\Delta}=\sum_{\substack{m\in\IZ\\m\geq -1}} c(m) \, q^m\ ,\quad 
D^2 \frac{1}{\Delta} = a_2\, c(0) + \sum_{\substack{m\in\IZ-\{0\}\\m\geq -1}}
\sum_{\ell=0}^2 a_\ell\, m^{2-\ell} c(m)\, q^m \tau_2^{-\ell}
\ee
where 
\be
a_0=4\ ,\quad a_1=\frac{p-q+6}{\pi}\ ,\quad a_2=\frac{(p-q+6)(p-q+8)}{16\pi^2} \ .
\ee
The integral over $\tau_1$ picks up the Fourier coefficient $c(m)$ with $m=-\tfrac12 \CQ^2$. The remaining integral over $\tau_2$ can be computed after expanding 
$\tilde P_{\alpha_1\dots \alpha_{4-h}} =\sum_{\ell= 0}^{\scalebox{0.60}{$\left\lfloor\frac{4-h}{2}\right\rfloor$}}\tilde P^{(\ell)}_{\alpha_1\dots \alpha_{4-h}}\tau_2^{-\ell}$, where  $\tilde P^{(\ell)}_{\alpha_1\dots \alpha_{4-h}}$ is a polynomial in $\CQ$ of degree $4-h-2\ell$, or zero when $2\ell>4-h$. Contributions with $\CQ=0$ lead to power-like terms,
\be
\label{Fdec1} 
\begin{split}
&F_{\alpha\beta\gamma\delta}^{(p,q),1,0}=  R^{q-6} \, \xi(q-6) \,\frac{3c(0)}{8\pi^2}\delta_{(\alpha\beta}\delta_{\gamma\delta)},\quad \\
& F_{11\alpha\beta}^{(p,q),1,0} = R^{q-6} \, \xi(q-6)\,(7-q) \frac{c(0)}{8\pi^2}\delta_{\alpha\beta}\ ,
\\
&F_{1111}^{(p,q),1,0} = R^{q-6} \, \xi(q-6)\,(7-q)(9-q) \frac{c(0)}{8\pi^2}\ ,
\end{split}
\ee
while the result vanishes for an odd number of indices along the direction 1, and for its trace
\be
\label{Fdec1s}
\trF^{(p,q),1,0} =R^{q-6} \, \xi(q-6) \, (p-q+6)(p-q+8)\, \frac{c(0)}{8\pi^2}\ .
\ee
Here we used $\tilde P^{(2)}_{abcd}(0)=\frac{3}{16\pi^2}\delta_{(ab}\delta_{cd)}$, $\tilde P^{(1)}_{ab}(0)=-\frac{1}{4\pi}\delta_{ab}$, and $\tilde P^{(0)}=1$. Note that \eqref{Fdec1s} and \eqref{Fdec1} have
a simple pole at $q=6$, which is subtracted by the regularization prescription mentioned 
below \eqref{def1looppqFNs}.  For $q=7$, the pole in \eqref{Fdec1s}, \eqref{Fdec1} cancels against the pole from the
zero orbit contribution \eqref{Fdec0}.

In contrast, non-zero vectors $\CQ$ lead to exponentially suppressed contributions, which depend on the axions through a phase factor $e^{2\pi \I k a\cdot \CQ}$. After rescaling $\CQ\mapsto Q/k$, we find that the Fourier
coefficient with charge $Q\in\Lambda_{p-1,q-1}\smallsetminus\{0\}$ 
is given by  
\be\label{Fdec1NP}
\begin{split}
&F_{\alpha\beta\gamma\delta}^{(p,q),1,Q} = 4 
\,\bar c(Q)\,R^{\tfrac{q-1}{2}}\sum_{\ell=0}^2 \, 
 \frac{\tilde P^{(\ell)}_{\alpha\beta\gamma\delta}(Q)}{R^\ell}\, \frac{K_{\frac{q-3}{2}-\ell}\left( 2\pi\, R \sqrt{2|Q_R|^2} \right)}{\sqrt{2|Q_R|^2}^{\tfrac{q-3}{2}-\ell}}
 \\
&F_{1\alpha\beta\gamma}^{(p,q),1,Q} = 4 
\,\bar c(Q)\,R^{\tfrac{q-1}{2}}\sum_{\ell=0}^1 \, 
 \frac{\tilde P^{(\ell)}_{\alpha\beta\gamma}(Q)}{\I\sqrt{2} R^\ell}\, \frac{K_{\frac{q-5}{2}-\ell}\left( 2\pi \, R  \sqrt{2|Q_R|^2} \right)}{\sqrt{2|Q_R|^2}^{\frac{q-5}{2}-\ell}}
 \\
 &\vdots
 \\
 &F_{1111}^{(p,q),1,Q} = 4 
 \bar c(Q)\,R^{\tfrac{q-1}{2}}\,\frac{\tilde P^{(0)}}{4} \,\frac{K_{\frac{q-11}{2}}\left( 2\pi \, R \sqrt{2|Q_R|^2} \right)}{\sqrt{2|Q_R|^2}^{\frac{q-11}{2}}}	
\end{split}
\ee
for the tensor integral, and 
\be
\trF^{(p,q),1,Q} = 4 
\,\bar c(Q)\, R^{\frac{q-1}{2}}\,  
\sum_{\ell=0}^2\, \frac{a_\ell}{R^{\ell}}
 \left(-\tfrac{Q^2}{2}\right)^{2-\ell}
\frac{K_{\frac{q-3}{2}-\ell}\left( 2\pi\, R \sqrt{2|Q_R|^2} \right)}{{\sqrt{2|Q_R|^2}^{\frac{q-3}{2}-\ell}}}
\ee
for its trace. In either case, 
\be\label{Fdec1Measure}
\bar c(Q) = \sum_{d|Q} c\left(-\tfrac{Q^2}{2d^2}\right)\, d^{q-7}\ .
\ee
The physical interpretation of these results will be discussed in \S\ref{sec_pertlim}, after generalizing
them to $\IZ_N$ orbifolds.

\subsection{Extension to $\IZ_N$ CHL orbifolds \label{sec_ZNp1}}
The degeneration limit  \eqref{decomp1} of the modular integrals  \eqref{f4pq}
for $\IZ_N$ CHL models with $N=2,3,5,7$ can be treated similarly by adapting 
the orbit method to the case where the integrand is invariant under the Hecke congruence subgroup $\Gamma_0(N)$ \cite{Mayr:1993mq,Trapletti:2002uk,Angelantonj:2013eja}. 
In  \eqref{f4pq},  
$\Delta_k$ is the cusp form of weight $k=\tfrac{24}{N+1}$ defined in \eqref{defg},
and $\Part{p,q}$ is the partition function 
for a lattice 
\be\label{ZNpertLimit}
\Lambda_{p,q}=\tilde \Lambda_{p-1,q-1}\oplus\sLambda_{1,1}[N]\ ,
\ee 
where $\tilde \Lambda_{p-1,q-1}$ is a level $N$ even lattice of signature $(p-1,q-1)$.  The lattice $\sLambda_{1,1}[N]$ is obtained from
the usual unimodular lattice $\sLambda_{1,1}$ by restricting the winding and momentum 
to $(n,m)\in N\IZ\oplus \IZ$. After Poisson resummation on $m$, Eq. \eqref{Poisson1} and
\eqref{poissonRessummedP} continue to 
hold, except for the fact that $n$ is restricted to run over $N\IZ$. The 
sum over $(n,m)$ can then be decomposed into orbits of $\Gamma_0(N)$:\footnote{Since $1/\Delta_k$ grows as $e^{\frac{2\pi}{N\tau_2}}$ at $\tau_2\to 0$,  the following treatment which relies on exchanging the sum and the integral for unfolding is justified for $NR^2> 2$.}

\paragraph{Trivial orbit} The term $(n,m)=(0,0)$ produces the same modular integral, up to a factor of $R$,
\be
\label{trivialOrbZN}
F_{\alpha\beta\gamma\delta}^{(p,q),0} = R\, \tilde F^{(p-1,q-1)}_{\alpha\beta\gamma\delta}\ ,\quad
\trF^{(p,q),0} = R \, \ttrF^{(p-1,q-1)}\ ,
\ee
where $\tilde F^{(p-1,q-1)}_{\alpha\beta\gamma\delta}$, $\ttrF^{(p-1,q-1)}$ are the integrals
\eqref{f4pq} for the lattice $\tilde \Lambda_{p-1,q-1}$ defined by \eqref{ZNpertLimit}.
\paragraph{Rank-one orbits} Terms with $(n,m)=k(c,d)$ with $k\neq 0$ and ${\rm gcd}(c,d)=1$ fall into two different classes of orbits under $\Gamma_0(N)$:
\begin{itemize} 
\item Doublets $k(c,d)$ such that $c=0\mod N$ and $k\in \IZ$ can be rotated by an element of $\Gamma_0(N)$
into $(0,1)$, whose stabilizer in $\Gamma_0(N)$ is $\Gamma_\infty=\{\big(\colvec[0.7]{1&n\\0&1}\big), n\in\IZ\}$. For these elements, one can unfold the integration domain $\Gamma_0(N)\backslash\cH$ into the unit width strip 
$\cS=\Gamma_{\infty}\backslash\cH=\IR^+_{\tau_2}\times (\IR/\IZ)_{\tau_1}$; 

\item Doublets $k(c,d)$ such that $c\neq 0\mod N$ and $k=0\mod N$ can be rotated by an element of $\Gamma_0(N)$ into $(1,0)$, whose stabilizer in $\Gamma_0(N)$ is $S\,\Gamma_{\infty,N}\,S^{-1}$, where $\Gamma_{\infty,N}=\{\big(\colvec[0.7]{1&n\\0&1}\big), n\in N\IZ\}$ and $S=\big(\colvec[0.7]{0&-1\\1&0}\big)$. One can unfold the integration domain $\Gamma_0(N)\backslash\cH$ into $S\,\Gamma_{\infty,N}\,S^{-1}\backslash\cH$, and change  variable
$\tau\to-1/\tau$ so as to reach $\cS_N= \Gamma_{\infty,N}\backslash\cH = \IR^+_{\tau_2}\times (\IR/N\IZ)_{\tau_1}$, the width-$N$ strip. Under this change of variable, the level-$N$ weight-$k$ cusp form transforms as $\Delta_k(-1/\tau)=N^{-\frac{k}{2}}(-\I\tau)^k\Delta_k(\tau/N)$, while the partition function  for the sublattice $\tilde \Lambda_{p-1,q-1}$ transforms as 
\be \tPart{p\hspace{-1.5pt}-\hspace{-2pt}1,q\hspace{-1.5pt}-\hspace{-2pt}1}[P_{\alpha\beta\gamma\delta}](-1/\tau)=\tilde\upsilon N^{-\frac{k}{2}-1} (-\I)^{\frac{p-q}{2}}\tau^k\,\tPartD{p\hspace{-1.5pt}-\hspace{-2pt}1,q\hspace{-1.5pt}-\hspace{-2pt}1}[P_{\alpha\beta\gamma\delta}](\tau)\ ,
\ee
where $\tPartD{p\hspace{-1.5pt}-\hspace{-2pt}1,q\hspace{-1.5pt}-\hspace{-2pt}1}(\tau)$ denotes the sum over the dual lattice $\tilde \Lambda_{p-1,q-1}^*$, and $\tilde\upsilon N^{-\frac{k}{2}-1}=\big|\tilde \Lambda^*_{p-1,q-1}/\tilde \Lambda_{p-1,q-1}\big|^{-1/2}$ (Note that $\tilde\upsilon=N^{1-\delta_{q,8}}$ for $q\leq8$ in the cases of interest).
\end{itemize}
For the simplest component $F^{(p,q),1}_{\alpha\beta\gamma\delta}$, the sum of the two classes of orbits then reads 
\begin{multline}
\label{ZNunfoldingStep}
F^{(p,q),1}_{\alpha\beta\gamma\delta} =  R  \int_{\IR^+}\frac{\de \tau_2}{\tau_2^2}\int_{\IR/\IZ}\de \tau_1\,  
\frac{1}{\Delta_k(\tau)}\sum_{k\neq 0} e^{-\pi R^2 k^2/\tau_2}
\tPart{p\hspace{-1.5pt}-\hspace{-2pt}1,q\hspace{-1.5pt}-\hspace{-2pt}1}\left[  e^{2\pi\I k a^I \CQ_I}\,P_{\alpha\beta\gamma\delta}\right] 
\\
+  R\int_{\IR^+}\frac{\de \tau_2}{\tau_2^2}\int_{\IR/(N\IZ)}\de \tau_1\,  
\frac{\tilde\upsilon}{N}\, \frac{1}{\Delta_k(\tau/N)}\sum_{\substack{k\neq 0 \\ k=0 \mod N}} e^{-\pi R^2 k^2/\tau_2}
\tPartD{p\hspace{-1.5pt}-\hspace{-2pt}1,q\hspace{-1.5pt}-\hspace{-2pt}1}\left[  e^{2\pi\I k a^I  \CQ_I}\,P_{\alpha\beta\gamma\delta}\right] \ .
\end{multline}
The contributions from $\CQ=0$ lead to power-like terms, 
\be
\label{Fdec1N} 
\begin{split}
&F_{\alpha\beta\gamma\delta}^{(p,q)(1,0)} =\,R^{q-6}\xi(q-6)\left(1+\tilde\upsilon N^{q-7}\right)\frac{3c_k(0)}{8\pi^2}\delta_{(\alpha\beta}\delta_{\gamma\delta)}\ ,
\\
&F_{11\alpha\beta}^{(1,0)} = R^{q-6} \, \xi(q-6)\,(7-q) \left(1+\tilde\upsilon N^{q-7}\right)\frac{c_k(0)}{8\pi^2}\delta_{\alpha\beta}\ ,
\\
&F_{1111}^{(1,0)} = R^{q-6} \, \xi(q-6)\,(7-q)(9-q)\left(1+\tilde\upsilon N^{q-7}\right) \frac{c_k(0)}{8\pi^2}\ ,
\end{split}
\ee
for the tensor integral and 
\be
\label{Fdec1Ns} 
\trF^{(p,q)(1,0)} =\,R^{q-6}\xi(q-6)(p-q+6)(p-q+8)\left(1+\tilde\upsilon N^{q-7}\right)\frac{c_k(0)}{8\pi^2}
\ee
for its trace, where $c_k(0)=k$ is the constant term in $1/\Delta_k$. As in \eqref{Fdec1s}
and \eqref{Fdec1}, the pole at $q=6$ is subtracted by the regularization prescription \eqref{def1looppqFN}, while the pole at $q=7$ cancels against the pole from the
zero orbit contribution \eqref{trivialOrbZN}.

The terms with non-zero vector $\CQ$ produce exponentially suppressed corrections of the same form  as in the maximal rank case \eqref{Fdec1NP}, but with a different summation measure,
namely
\be\label{measureZNWeak}
\bar c_k(Q)=\sum_{\substack{d\geq 1,\\ Q/d\in\tilde\Lambda_{p-1,q-1 }}}
c_k\Big(-\frac{Q^2}{2d^2}\Big)\, d^{q-7}+\hspace{2mm} \, \tilde\upsilon\, \hspace{-6mm} 
\sum_{\substack{d\geq 1,\\ Q/d\in N\tilde\Lambda^*_{p-1,q-1 } }} \hspace{-4mm} c_k\Big(-\frac{Q^2}{2Nd^2}\Big) \left(Nd\right)^{q-7},
\ee
where the first term, arising from the first class of orbits, has support on $\tilde\Lambda_{p-1,q-1}$, and the second term, arising from the second class of orbits, has support on the sublattice  $N
\tilde\Lambda_{p-1,q-1}^*\subset\tilde\Lambda_{p-1,q-1}$. In the latter contribution, notice that one factor of $N$ in the numerator of the Fourier coefficient comes from the matching condition with $1/\Delta_k(\tau/N)$, and two factors of $N$ in its denominator come from all the divisors being originally multiples of $N$.

It will also be useful to consider a different degeneration limit of the type \eqref{decomp1} 
where the lattice decomposes as
\be\label{ZNpertLimitNonOrb}
\Lambda_{p,q}=\Lambda_{p-1,q-1}\oplus\sLambda_{1,1}\ ,
\ee
where $\sLambda_{1,1}$ is the usual unimodular even lattice, with no restriction on the windings and momenta $(n,m)$, and $\Lambda_{p-1,q-1}$ is a  level $N$ even lattice of signature $(p-1,q-1)$, not to be confused with the lattice $\tilde\Lambda_{p-1,q-1}$ above. The sum over $(n,m)\in\IZ\oplus\IZ$ can then be decomposed into orbits of $\Gamma_0(N)$. The trivial orbit is similar to \eqref{trivialOrbZN}, but now $F^{(p-1,q-1)}_{\alpha\beta\gamma\delta}$ and $\trF^{(p-1,q-1)}$ are the modular integrals for the lattice $\Lambda_{p-1,q-1}$. For the rank-one orbit, the discussion goes as before, except that the second class of orbits $(m,n)=k(c,d)$ with $k=\gcd(m,n)$ and $c\neq0\mod N$ has no restriction on $k$. For the simplest component $F^{(p,q),1}_{\alpha\beta\gamma\delta}$, the sum of the two classes of orbits then reads
\begin{multline}
\label{ZNunfoldingStep2}
F^{(p,q),1}_{\alpha\beta\gamma\delta} =  R  \int_{\IR^+}\frac{\de \tau_2}{\tau_2^2}\int_{\IR/\IZ}\de \tau_1\,  
\frac{1}{\Delta_k(\tau)}\sum_{k\neq 0} e^{-\pi R^2 k^2/\tau_2}
\Part{p\hspace{-1.5pt}-\hspace{-2pt}1,q\hspace{-1.5pt}-\hspace{-2pt}1}\left[  e^{2\pi\I k a^I \CQ_I}\,P_{\alpha\beta\gamma\delta}\right] 
\\
+  R\int_{\IR^+}\frac{\de \tau_2}{\tau_2^2}\int_{\IR/(N\IZ)}\de \tau_1\,  
\frac{1}{\Delta_k(\tau/N)}\frac{\upsilon}{N}\sum_{\substack{k\neq 0}} e^{-\pi R^2 k^2/\tau_2}
\PartD{p\hspace{-1.5pt}-\hspace{-2pt}1,q\hspace{-1.5pt}-\hspace{-2pt}1}\left[  e^{2\pi\I k a^I  \CQ_I}\,P_{\alpha\beta\gamma\delta}\right] \ ,
\end{multline}
where $\upsilon N^{-\frac{k}{2}-1}=\big|\Lambda^*_{p-1,q-1}/\Lambda_{p-1,q-1}\big|^{-1/2}$ (which now simplifies to $\upsilon=N^{-\delta_{q,8}}$ for $q\leq8$ in the cases of interest).
The contributions from $\CQ=0$ lead to power-like terms, 
\be
\label{Fdec1NNonOrb} 
\begin{split}
&F_{\alpha\beta\gamma\delta}^{(p,q)(1,0)} =\,R^{q-6}\xi(q-6)\big(1+\upsilon\big)\frac{3c_k(0)}{8\pi^2}\delta_{(\alpha\beta}\delta_{\gamma\delta)},
\\
&F_{11\alpha\beta}^{(1,0)} = R^{q-6} \, \xi(q-6)\,(7-q)\big(1+\upsilon\big)\frac{c_k(0)}{8\pi^2}\delta_{\alpha\beta},
\\
&F_{1111}^{(1,0)} = R^{q-6} \, \xi(q-6)\,(7-q)(9-q)\big(1+\upsilon\big)\frac{c_k(0)}{8\pi^2}
\end{split}
\ee
for the tensor integral and
\be
\trF^{(p,q)(1,0)} =\,R^{q-6}\xi(q-6)(p-q+6)(p-q+8)\big(1+\upsilon\big)\frac{c_k(0)}{8\pi^2}
\ee
for its trace, where $c_k(0)=k$ is the constant term in $1/\Delta_k$.

The terms with non-zero vector $\CQ$ produce exponentially suppressed corrections of the same form  as in the maximal rank case \eqref{Fdec1NP}, but with a different summation measure,
namely
\be\label{measureZNWeakNonOrb}
\bar c_k(Q)=\sum_{\substack{d\geq 1,\\ Q/d\in\Lambda_{p-1,q-1 }}}
c_k\Big(-\frac{Q^2}{2d^2}\Big)\, d^{q-7}+\hspace{2mm} \,\upsilon \hspace{-3mm} 
\sum_{\substack{d\geq 1,\\ Q/d\in \Lambda^*_{p-1,q-1 } }} \hspace{-3mm} c_k\Big(-\frac{NQ^2}{2d^2}\Big)\, d^{q-7},
\ee
where the first term, arising from the first class of orbits, has support on $\Lambda_{p-1,q-1}$, and the second term, arising from the second class of orbits, has support on the dual lattice $\Lambda_{p-1,q-1}^*$. In the latter contribution, notice that one factor of $N$ in the numerator of the Fourier coefficient comes from the matching condition with $1/\Delta_k(\tau/N)$.

\subsection{Perturbative limit of exact $(\nabla\Phi)^4$ couplings in $D=3$ \label{sec_pertlim}}

Specializing to $(p,q)=(2k,8)=(r-4,8)$, and decomposing as $\Lambda_{2k,8}=\Lambda_{2k-1,7} \oplus \sLambda_{1,1}[N]$,
the limit \eqref{decomp1} studied in this section corresponds to the 
expansion of the exact $(\nabla\Phi)^4$ couplings in $D=3$ in the limit where the heterotic
string coupling $g_3=1/\sqrt{R}$ becomes weak. To interpret the resulting contributions in 
the language of heterotic perturbation theory, one should remember that the U-duality function
$F^{(2k,8)}_{abcd}(\Phi)$ is the coefficient of the $(\nabla\Phi)^4$ coupling in the low-energy
action written in Einstein frame, such that the metric $\gamma_E$ is inert under U-duality,
\be
S_3 = \int \de^3 x \, \sqrt{-\gamma_E}\, \left[ \cR[\gamma_E] - ( 2 \delta_{\hat{a}\hat{b}} \delta_{\hat{c}\hat{d}} -\delta_{\hat{a}\hat{c}} \delta_{\hat{b}\hat{d}}) F^{(2k,8)}_{abcd}(\Phi) \, 
\gamma_E^{\mu\rho} \gamma_E^{\nu\sigma}
P_\mu^{a\hat a} P_\nu^{b\hat b}  P_\rho^{c\hat c}P_\sigma^{ d\hat d} \right] +\dots\ .
\ee
In terms of the string frame metric $\gamma=\gamma_E g_3^4$, 
one finds
\be
S_3=\int \de^3 x \, \sqrt{-\gamma}\, \left[ \frac{1}{g_3^2} \cR[\gamma] -g_3^2 \, ( 2 \delta_{\hat{a}\hat{b}} \delta_{\hat{c}\hat{d}} -\delta_{\hat{a}\hat{c}} \delta_{\hat{b}\hat{d}}) F^{(2k,8)}_{abcd}(\Phi) \, 
\gamma^{\mu\rho} \gamma^{\nu\sigma}
P_\mu^{a\hat a} P_\nu^{b\hat b}  P_\rho^{c\hat c}P_\sigma^{ d\hat d} \right] +\dots \ .
\ee
Using $c_k(0)=k$ for CHL orbifolds with $N>1$  or $c(0)=2k$ in the maximal rank case, 
and $\xi(2)=\tfrac{\pi}{6}$, the results from \S\ref{sec_pertmax} and \S\ref{sec_ZNp1} read
\be
\label{Fexppert}
g_3^2\, F^{(2k,8)}_{abcd} = \frac{3}{2\pi g_3^2} \delta_{(ab}\delta_{cd)}
  + F^{(2k-1,7)}_{abcd}+
 \sum'_{Q\in\Lambda_{2k-1,7}} \, \bar c_k(Q) e^{-\frac{2\pi \sqrt2\,  
|Q_R|}{g_3^2}+2\pi \I a\cdot Q}P^{(*)}_{abcd}\ ,
\ee
where we omit the detailed form of exponentially suppressed corrections, and 
the summation measure is read off from \eqref{measureZNWeak}
\be
\label{measureNS5KK}
\bar c_k(Q) = \sum_{\substack{d\geq 1,\\ Q/d\in\Lambda_{2k-1,7 }}}
d\, c_k\Big(-\frac{Q^2}{2d^2}\Big)+
\sum_{\substack{d\geq 1,\\ Q/d\in N \Lambda^*_{2k-1,7 } }} N\,d\, c_k\Big(-\frac{Q^2}{2Nd^2}\Big)\ ,
\ee
The first two terms in \eqref{Fexppert}, originating from the zero orbit and 
rank-one orbit, respectively, should match the tree-level and one-loop contributions,
respectively. Indeed, the  dimensional reduction of the tree-level $\mathcal{R}^2+(\Tr F^2 )^2$ coupling in ten-dimensional heterotic string theory \cite{Gross:1986mw,Bergshoeff:1989de} leads to a tree-level $(\nabla\Phi)^4$ coupling in $D=3$, 
with a coefficient which is by construction independent of $N$. A more detailed analysis of the ten-dimensional
origin of this term will be given in \S\ref{sec_eff4}.
The second term in \eqref{Fexppert} of course matches
the one-loop contribution  \eqref{f41loop} by construction. 
The remaining non-perturbative terms can be interpreted as heterotic NS5-brane, KK5-brane and
H-monopoles wrapped on any possible $T^6$ inside $T^7$ \cite{Obers:2000ta}. More precisely,
NS5-brane and KK5-brane charges correspond to momentum and winding charges in the hyperbolic part  $\sLambda_{1,1}[N]\oplus\sLambda_{k-2,k-2}$ of $\Lambda_m\oplus\sLambda_{1,1}$, while H-monopoles
correspond to charges in the gauge lattice $\Lambda_{k,8-k}$ (for the heterotic string compactification on $T^7$, these sublattices must be replaced by  $\sLambda_{7,7}$ and $E_8\oplus E_8$ or $D_{16}$, respectively).
Note that  \cite{Obers:2000ta} studied these corrections on a special locus in moduli space, corresponding to $T^4/\IZ_2$
realization of K3 surfaces on the type II side, and did not keep track of all gauge charges,
which resulted in a different summation 
measure. 

\subsection{Decompactification limit of  one-loop $F^4$ couplings\label{sec_pertdeclim}}

For general $(p,q)=(d+2k-8,d)=(d+r-12,d)$ with $q\leq 7$, the modular integral \eqref{f4pqtens} is interpreted as the one-loop $F^4$ amplitude in a heterotic CHL orbifold compactified down to dimension $D=10-d$. The 
decomposition \eqref{ZNpertLimit} corresponds to the case (a) where the radius $R$ of a circle in $T^d$ orthogonal to the $\IZ_N$ orbifold action becomes large, while the limit \eqref{ZNpertLimitNonOrb} corresponds to the case (b) where the radius $R$ of the circle in $T^d$  singled out by the $\IZ_N$ orbifold 
action becomes large in string units.

The power-like terms contributions in $R$ come 
in part from the trivial orbit, and from the zero-charge contribution to the rank-one orbit:
\be
\label{decompZNqleq7}
\begin{split}
a):\quad & F_{\alpha\beta\gamma\delta}^{(p,q)} =\,R F^{(p-1,q-1)}_{\alpha\beta\gamma\delta}+R^{q-6}\xi(q-6)\frac{3(2k)}{8\pi^2}\delta_{(\alpha\beta}\delta_{\gamma\delta)}+\ldots\,
\\
b):\quad &
 F_{\alpha\beta\gamma\delta}^{(p,q)} =\,R \tilde F^{(p-1,q-1)}_{\alpha\beta\gamma\delta} 
+R^{q-6}\xi(q-6)\frac{3k(1+N^{q-6})}{8\pi^2}\delta_{(\alpha\beta}\delta_{\gamma\delta)}+\ldots\,
\end{split}
\ee
The first term reproduces, up to a volume factor of $R$,  the one-loop $F^4$ amplitude in $D+1$ dimensions \eqref{Fdec0}, either in the same CHL model (case a), or in the full heterotic string compactification (case b). Indeed, in the
latter case, the  partition function $\Part{p\hspace{-1.5pt}-\hspace{-2pt}1,q\hspace{-1.5pt}-\hspace{-2pt}1}$ factorizes into $\sPart{d\hspace{-2pt}+\hspace{-2pt}k\hspace{-2pt}-\hspace{-2pt}9,d\hspace{-2pt}+\hspace{-2pt}k\hspace{-2pt}-\hspace{-2pt}9}\times \Part{k,8\hspace{-2pt}-\hspace{-2pt}k}$. The fundamental domain $\Gamma_0(N)\backslash\cH$ can be extended to $SL(2,\IZ)\backslash\cH$, at the expense of replacing $\Gamma_{\Lambda_{k,8-k}}/\Delta_k$ by the sum over its images under $\Gamma_0(N)\backslash SL(2,\IZ)=\{1,S,TS,\dots, T^{N-1}S\}$. As explained in \S\ref{sec_F4CHL}, this sum reproduces $\Part{d\hspace{-2pt}+\hspace{-2pt}15,d\hspace{-2pt}-\hspace{-2pt}1}/\Delta$, the partition function for the maximal rank theory in dimension $D+1$.

The second term, originating from  the zero-charge contribution to the rank-one orbit,
can instead be understood  as the limit $s\to 0$ of an infinite tower of terms 
of the schematic form $\sum_{m\neq 0} (\tfrac{m^2}{R^2} - s)^{3-\tfrac{d}{2}} F^4$ in the low-energy effective action, where $s$ is a Mandelstam variable, arising from threshold contributions of Kaluza--Klein excitations of the massless supergravity states in dimension $D+1$. In the limit $R\to \infty$, this infinite series along with the term $m=0$ from the non-local part of the action in dimension $D$ sums up to the contribution of massless supergravity states to the non-local part of the action in dimension $D+1$. The pole at $q=6$ in the second term of \eqref{decompZNqleq7} originates from the logarithmic infrared divergence in the local part of the string effective action in dimension $D=4$, and matches the expected ultraviolet divergence in 4-dimensional supergravity. The apparent pole at $q=7$ cancels against a pole in the first term, due to the same logarithmic divergence. Indeed, the $1/\epsilon$ pole of the full amplitude $F_{abcd}^{(p,6)}(\Phi,\epsilon)$ can be extracted from its Laurent expansion at $\epsilon=0$, namely
\be\label{4Dpole}
F^{(p,6)}_{abcd}(\Phi,\epsilon) = - \frac{3 (2k)}{16\pi^2\epsilon}\, \delta_{(ab} \delta_{cd)} +\cO(1)
\ee
In addition, massive perturbative BPS states with non-vanishing charge $Q\in \Lambda_{d+2k-9,d-1}$ in dimension $D+1$ and mass $\cM(Q)$ lead to exponentially suppressed terms of order $e^{-2\pi R \cM(Q)}$, weighted by the helicity supertrace $\Omega_4(Q)$, as expected on general grounds.

\subsection{Perturbative limit of exact $H^4$ couplings in type IIB on $K3$ \label{sec_iibpert}} 
Here we briefly consider the case $q=5$, $N=1$, corresponding to type IIB string theory compactified on $K3$. In Einstein frame, the low energy effective action takes the form
\be
S_6 = \int \de^6 x \, \sqrt{-\gamma_E}\, \left[ \cR[\gamma_E] -F^{21,5}_{abcd}(\Phi) \, 
 H^a_{\mu\nu\kappa} H^b_{\rho\sigma}{}^\kappa H^{c \, \mu\nu\lambda} H^{d \rho\sigma}{}_\lambda  \right] + \dots 
\ee
where the three-form $H^\alpha$  with $\alpha\neq 1$ are the self-dual field-strengths of  the reduction
of the RR two-form, four-form and six-form
on the self-dual part of the homology lattice $H^{\rm even}(K3)=
E_8\oplus E_8\oplus \sLambda_{4,4}$, while $H^1$ is the self-dual component of the NS-NS
two-form field-strength. 
We shall restrict for simplicity to the components $\alpha,\beta,\gamma,\delta \neq 1$. In terms
of the string frame metric $\gamma=g_s\gamma_E$ and setting $\cH^a=g_s H^a$ (since
Ramond-Ramond field are normalized as $H\sim 1/g_s$ in type II perturbation theory),
we get 
\be
S_6 = \int d^6 x \, \sqrt{-\gamma}\, \left[ \frac{1}{g_s^2} \cR[\gamma] - \frac{1}{g_s}\, F^{21,5}_{\alpha\beta\gamma\delta}\, (\Phi)  \,  H^\alpha_{\mu\nu\kappa} H^\beta_{\rho\sigma}{}^\kappa H^{\gamma \, \mu\nu\lambda} H^{\delta \rho\sigma}{}_\lambda  \right] + \dots
\ee
Identifying $R=1/g_s$, the large radius expansion of $F^{21,5}_{\alpha\beta\gamma\delta}$ becomes, schematically,
\be
\frac{1}{g_s}\, F^{21,5}_{\alpha\beta\gamma\delta} = 
\frac{1}{g_s^2} F^{20,4}_{\alpha\beta\gamma\delta}(\Phi) +  \frac{3}{2\pi} \delta_{(\alpha\beta}\delta_{\gamma\delta)}
+ \sum'_{Q\in\Lambda_{20,4}} \, 
\bar c(Q) e^{-\frac{2\pi \sqrt2\,  |Q_R|^2}{g_s}-2\pi \I a\cdot Q}P^*_{\alpha\beta\gamma\delta}  \ .
\ee
The first term proportional to $F^{20,4}_{abcd}$ is now recognized as a tree-level correction in type II on $K3$, 
the second term  is a one-loop correction which to our knowledge has not been computed independently yet, 
and the remaining terms originate from D3, D1, D(-1) branes wrapped on $K3$ \cite{Kiritsis:2000zi}.
It is worth noting that decompactification limits of the form $O(2k,8)\to O(2k-3,5)$ exist in principle 
for all CHL models listed in Table  \ref{TableauCHL}, however,  they cannot be interpreted
in terms of six-dimensional chiral string vacua, due to anomaly cancellation constraints.

\section{Large radius expansion of exact $(\nabla\Phi)^4$ couplings\label{decompLimit}}

In this section, we study the  expansion of the proposal \eqref{f4exact} in the limit where
the radius $R$ of one circle in the internal space goes to infinity. We show that it reproduces the known $F^4$ and $\cR^2$ couplings in $D=4$, along with an infinite series of  $\cO(e^{-R})$
corrections from 1/2-BPS dyons whose wordline winds around the circle, as well as an 
infinite series of  $\cO(e^{-R^2})$
corrections from Taub-NUT instantons.
We start by analyzing the  expansion of genus-one
modular integrals \eqref{f4pqscal} and \eqref{f4pqtens} for arbitrary values of $(p,q)$, in the limit 
near the cusp where $O(p,q)$ is broken to $O(2,1)\times O(p-2,q-2)$, so 
that the moduli space decomposes into 
\be
\label{Gpqdec2}
G_{p,q} \to \IR^+ \times \left[\frac{SL(2)}{SO(2)} \times G_{p-2,q-2}\right] \ltimes \IR^{2(p+q-4)} \times \IR\ 
\ee
As in the previous section, we first discuss the maximal rank case $N=1$, $p-q=16$, where
the integrand is invariant under the full modular group, before dealing with the case of $N$ prime. 
The reader uninterested by the details of the derivation may skip to \S\ref{sec_decomp}, where we  specialize to the values $(p,q)=(r-4,8)$ relevant for the $(\nabla\Phi)^4$ couplings in 
$D=3$, and interpret the various contributions arising in the decompactification limit to $D=4$. 

\subsection{$O(p,q)\to O(p-2,q-2)$ for even self-dual lattices\label{sec_decompmax}}
We first consider the case where the lattice $\Lambda_{p,q}$  is even self-dual and
 factorizes in the limit \eqref{Gpqdec2} as 
 \be
\Lambda_{p,q} \to \Lambda_{p-2,q-2} \oplus \sLambda_{2,2}\ .
\ee
In order to study the behavior of the modular integral \eqref{f4pqtens} in the limit \eqref{Gpqdec2}, 
we denote by $\RV,S,\phi,a^{I,i},\psi$ the coordinates for each factors in  \eqref{Gpqdec2}, where $i=1,2$ and $I=3,\ldots, p+q-2$. 
The coordinate $\RV$ (not to be confused with the one used in \S\ref{pertLimit}) parametrizes a one-parameter subgroup 
$e^{\RV H_1}$ in $O(p,q)$, such that
the action of the non-compact Cartan generator $H_1$ on the Lie algebra $\mf{so}_{p,q}$ decomposes into \be\label{sopqDec}
\mf{so}_{p,q}\simeq \,\ldots\,\oplus(\mf{gl}_1\oplus\mf{so}_{p-2,q-2})^{(0)}\oplus({\bf 2}\otimes({\bf p+q-4}))^{(1)}\oplus {\bf 1}^{(2)},
\ee
while $(a^{iI},\psi)$ parametrize the unipotent subgroup obtained by exponentiating the
grade $1$ and $2$ components in this decomposition. We parametrize  the $SO(2)\backslash SL(2,\IR)$ coset representative $v_\mu{}^i$ and the symmetric $SL(2,\IR)$ element $M \equiv v^T v$ by the complex upper half-plane coordinate $S = S_1 + \I S_2 $
\be
\label{torusVielbein}
v_\mu{}^i=\frac{1}{\sqrt{S_2}} \begin{pmatrix} 1 & S_1 \\ 0 & S_2 \end{pmatrix}\ ,\quad 
M^{ij}= \delta^{\mu\nu}v_\mu{}^iv_{\nu}{}^j=\frac{1}{S_2} \begin{pmatrix} 1 & S_1 \\ S_1 & |S|^2 \end{pmatrix}\ .
\ee
A  generic charge vector $Q_{\cI} \in\Lambda_{p,q}
\simeq {\bf p+q}\simeq {\bf 2}^{(-1)} \oplus ({\bf p+q-4})^{(0)}\oplus {\bf 2}^{(1)}$ decomposes 
into $Q=(m^i,\CQ_I,n_j)$, where $(m^i,n_i)\in\sLambda_{2,2}$ and $\CQ_I\in
\Lambda_{p-2,q-2}$ such that $Q^2=-2m^i n_i+\CQ^2$. The 
projectors defined by $Q_L\equiv p^\cI_{L}Q_\cI$ and $Q_R\equiv p^\cI_{R}Q_\cI$ decompose according to 
\be
\label{pLRdec2}
\begin{split}
p^\cI_{L,\mu}Q_\cI = & \frac{v_{i\mu}^{-1}}{\RV\sqrt{2}}\left( m^i + a^i \cdot \CQ + (\psi\epsilon^{ij}+\frac12 a^i \cdot a^j)  n_j \right)- \frac{\RV}{\sqrt{2}}v_\mu{}^i n_i \\ 
p^\cI_{L,\alpha}Q_\cI = & \tilde p^{\,I}_{L,\alpha}(\CQ_I+ n_i a^i_I)\\
p^\cI_{R,\mu}Q_\cI = & \frac{v_{i\mu}^{-1}}{\RV\sqrt{2}}\left( m^i + a^i \cdot \CQ + (\psi\epsilon^{ij}+\frac12 a^i \cdot a^j)  n_j \right)+ \frac{\RV}{\sqrt{2}}v_\mu{}^i n_i  \\ 
p^\cI_{R,\hat\alpha} Q_\cI = & \tilde p^{\,I}_{R,\hat\alpha}(\CQ_I + n_i a^i_I)
\end{split}
\ee
where $\tilde p^I_{L,\alpha}, \tilde p^I_{R,\hat\alpha}$ ($\alpha=3\dots p$, $\hat\alpha=3\dots q$)
are orthogonal projectors in $G_{p-2,q-2}$ satisfying $\CQ^2=
\CQ_L^2-\CQ_R^2$. 

In order to study the region $\RV\gg 1$ it is useful to perform a Poisson resummation on the momenta $m^i$ along $\sLambda_{2,2}$. Note that this analysis is in principle valid for a region containing $R> \sqrt2$. 
In the case of the scalar integral \eqref{f4pqscal}, one  obtains 
\be
\label{Poisson2}
\Part{p,q} = \RV^2\, \tau_2^{\frac{q-1}{2}} \sum_{A\in \IZ^{2\times 2}}\, 
\Part{p\hspace{-1.5pt}-\hspace{-2pt}2,q\hspace{-1.5pt}-\hspace{-2pt}2}\Big[
e^{-\frac{\pi}{\tau_2}\frac{\RV^2}{S_2}\left|(1,S)
	A \big(\colvec[0.75]{ \tau\\ 1 }\big) \right|^2
	 -2\pi \I\,(\psi+\I R^2)\, \det\, 
	A+2\pi \I\,m_i(\CQ\cdot a^i+\frac{a^i\cdot a^j}{2}n_j)}\Big]\ ,
\ee
where $A=\big(\colvec[0.75]{ n_1\,m_1\\n_2\,m_2 }\big)$.
In the case of \eqref{f4pqtens}, we must distinguish whether the indices $abcd$ lie along
the direction $1,2$ or along the directions $\alpha$. Denoting by $h$ the number of indices of the first kind, we get
\begin{multline}
\label{poissonRessummedP2} 
\Part{p,q}\left[ e^{-\frac{\Delta}{8\pi\tau_2}} \left( \prod_{i=1}^h (Q_{L,\mu_i} )Q_{L,\alpha_1} \dots  
Q_{L,\alpha_{4-h}} \right)  \right]
= {\RV^2}\hspace{-2mm} \sum_{A\in \IZ^{2\times 2}}  \left(\frac{\RV}{\I\sqrt{2}}\right)^{h}
\\
\times
{\rm exp}\left(-\frac{\pi}{\tau_2}\frac{\RV^2}{S_2}
     \left|(1,S)	 A
	 \big(\colvec[0.75]{ \tau\\ 1 }\big) \right|^2
	 -2\pi \I\,T\, \det\, A 
	 \right)\prod_{k=1}^h  \left[\tfrac{1}{\sqrt{S_2}}
\big(\colvec[0.75]{1&S_1\\0&S_2}\big) A
\big(\colvec[0.75]{\bar\tau\\1}\big) \right]_{\mu_k}
\\
\times
\ShiftPart{p\hspace{-1.5pt}-\hspace{-2pt}2,q\hspace{-1.5pt}-\hspace{-2pt}2}{n_ia^i}\Big[ e^{-\frac{\Delta}{8\pi\tau_2}}  \left[  \CQ_{L,\alpha_1} \dots \CQ_{L,\alpha_{4-h}}  \right] e^{2\pi \I\,m_i(\CQ \cdot a^i-\frac{a^i\cdot a^j}{2}n_j)}\Big]
\end{multline}
In this representation, modular invariance is manifest, since a transformation $\tau\to\frac{a\tau+b}{c\tau+d}$ can be compensated by a linear action   $A\to A\big(\colvec[0.7]{d&-b\\-c&a}\big)$,
under which the last line of \eqref{poissonRessummedP2} transforms with weight $12-h$.
We can therefore decompose the sum over $A$ into various orbits under $SL(2,\IZ)$ and 
apply the unfolding trick to each orbit:

\paragraph{The trivial orbit}  $A=0$ produces, up to a factor of $\RV^2$, the integrals \eqref{f4pq} or  for the lattice $\Lambda_{p-2,q-2}$, provided none of the indices $abcd$ lie along the direction 1 or 2,
\be
\label{F34zero}
F^{(p,q),0}_{\alpha\beta\gamma\delta}= {\RV^2}\, F^{(p-2,q-2)}_{\alpha\beta\gamma\delta}\ ,\quad
\trF^{(p,q),0} = {\RV^2}\, \trF^{(p-2,q-2)}\ ,
\ee
while it vanishes otherwise ({\it i.e.} when $h>0$).

\paragraph{Rank-one orbit:} Matrices with $\det A=0$ but $A\neq 0$ can be decomposed into $A = \big(\colvec[0.8]{0&j\\0&p}\big)\big(\colvec[0.8]{a&b\\c&d}\big)$, where $(j,p)\in\IZ^2\smallsetminus(0,0)$ and  $\big(\colvec[0.8]{a&b\\c&d}\big)\in \Gamma_\infty\backslash SL(2,\IZ)$. As before the fundamental domain $SL(2,\IZ)\backslash\cH $ can be unfolded to the strip $\cS=\Gamma_\infty\backslash \cH=\IR^+_{\tau_2}\times (\IR/\IZ)_{\tau_1}$ using \eqref{unfoldingIdentity}, leading to
\be
\begin{split}
&F^{(p,q),1}_{\mu_1\ldots\mu_h\alpha_1\ldots\alpha_{4-h}}={\RV^2}\sum'_{\substack{(j,p)}}
\,\prod_{i=1}^h \left(\frac{\RV}{\I\sqrt{2}}\right)^{h}\left[\tfrac{1}{\sqrt{S_2}}
\Big(\colvec[0.9]{1&S_1\\0&S_2 }\Big)
\Big(\colvec[0.9]{j\\p}\Big)
\right]_{\mu_i}
	 \\
	 &\qquad\qquad\qquad\qquad\times \int_{\IR^+}\frac{\de\tau_2}{\tau_2^{2+h}}\int_{\IR/\IZ}\hspace{-3mm}\de\tau_1\,
	 \frac{e^{-\frac{\pi}{\tau_2}\frac{\RV^2}{S_2}\left|j+pS\right|^2}}{\Delta}\,\Part{p\hspace{-1.5pt}-\hspace{-2pt}2,q\hspace{-1.5pt}-\hspace{-2pt}2}\left[	 
	 \tilde P_{\alpha_{1}\ldots \alpha_{4-h}}
	 e^{2\pi \I\,(j\CQ\cdot a^1+p\CQ\cdot a^2) }\right],
\\
&\trF^{(p,q),1}={\RV^2}\sum'_{\substack{(j,p)}} 
 \int_{\IR^+}\frac{\de\tau_2}{\tau_2^{2+h}}\int_{\IR/\IZ}\hspace{-3mm}\de\tau_1
e^{-\frac{\pi}{\tau_2}\frac{\RV^2}{S_2}\left|j+pS\right|^2}
\,
\Part{p\hspace{-1.5pt}-\hspace{-2pt}2,q\hspace{-1.5pt}-\hspace{-2pt}2}\left[ e^{2\pi \I\,(j\CQ\cdot a^1+p\CQ\cdot a^2)} \right]\, D^2\left(\frac{1}{\Delta}\right)\ ,
\end{split}
\ee
for the tensor integral with $0\leq h\leq 4$ indices along the large torus and its trace respectively. Inserting the Fourier expansion \eqref{FourierDelta}, the integral over $\tau_1$ picks up the Fourier coefficient $c(m)$ with $m=-\tfrac12 \CQ^2$. The remaining integral over $\tau_2$ can be computed after expanding $\tilde P_{\alpha_1\dots \alpha_{4-h}} =\sum_{\ell=0}^{\scalebox{0.60}{$\left\lfloor\frac{4-h}{2}\right\rfloor$}} \tilde P^{(\ell)}_{\alpha_1\dots \alpha_{4-h}} \tau_2^{-\ell}$, where  $\tilde P^{(\ell)}_{\alpha_1\dots \alpha_{4-h}}$ is a polynomial in $\CQ$ of degree $4-h-2\ell\geq0$, or vanishing otherwise. The contribution of $\CQ=0$ produces power-like terms in ${\RV^2}$,
\be 
\label{F34deg0}
\begin{split}
&F_{\alpha\beta\gamma\delta}^{(p,q),1,0} =  {\RV}^{q-6}
\frac{3c(0)}{8\pi^2}\,\cE^\star(\tfrac{8-q}{2},S)\,\delta_{(\alpha\beta}\delta_{\gamma\delta)},
\\
&F_{\mu\nu\gamma\delta}^{(p,q),1,0} =  {\RV}^{q-6}
\frac{c(0)}{4\pi^2}\left[\tfrac{8-q}{4}\delta_{\alpha\beta}\delta_{\mu\nu}-\delta_{\alpha\beta}\cD_{\mu\nu}\right]\cE^\star(\tfrac{8-q}{2},S),
\\
&F_{\mu\nu\rho\sigma}^{(p,q),1,0} =  {\RV}^{q-6}
\frac{c(0)}{2\pi^2}\left[\cD^2_{\mu\nu\rho\sigma}-\tfrac{10-q}{2}\delta_{(\mu\nu}\cD_{\rho\sigma)}+\left(\tfrac{8-q}{2}\right)\left(\tfrac{10-q}{2}\right)\tfrac{3}{8}\delta_{(\mu\nu}\delta_{\rho\sigma)}\right]\cE^\star(\tfrac{8-q}{2},S)
\end{split}
\ee
for the tensor integral, and
\be
\label{F34deg0s}
\trF^{(p,q),1,0} =  \RV^{q-6} 
\, \frac{c(0)}{8\pi^2}\, (p-q+6)(p-q+8)  \, \cE^\star(\tfrac{8-q}{2},S)\ ,
\ee
for its trace. Here, $\cE^\star(s,S)$ is the completed weight 0 non-holomorphic Eisenstein series,
\be
\cE^\star(s,S)=\frac12
\pi^{-s}\, \Gamma(s)\, 
\sum_{(m,n)\in\IZ^2}'\frac{S_2^s}{|nS+m|^{2s}} \equiv
\xi(2s)\, \cE(s,S)\ ,
\ee
$\cD_{\mu\nu}$ is the  traceless differential operator on $\frac{SL(2,\IR)}{SO(2)}$ defined in appendix \ref{tensEinsensteinSeries}, and $\cD^2_{\mu\nu\rho\sigma}=\cD_{(\mu\nu}\cD_{\rho\sigma)}-\frac{1}{4}\delta_{(\mu\nu}\delta_{\rho\sigma)}\cD_{\tau\kappa}\cD^{\tau\kappa}$ is the traceless operator of degree 2 in the symmetric representation. The equalities used to write \eqref{F34deg0} are detailed in \eqref{EseriesRelationsBis}, and similar expressions using non-holomorphic series of non-zero weight are given in \eqref{EseriesRelations2}. Recall
that $\cE^\star(s,S)$ is invariant under
 $s\mapsto 1-s$, and has simple poles at $s=0$ and $s=1$.
As in the previous
section, the pole at $q=6$ is subtracted by the regularization prescription mentioned 
below \eqref{def1looppqFNs}, while the pole at $q=8$ cancels against the pole from the
zero orbit contribution \eqref{F34zero}.

Contributions of non-zero vectors $\CQ\in\Lambda_{p-2,q-2}$, on the other hand, lead to exponentially
suppressed contributions, {\it e.g.} for the trace of the tensor integral
\begin{multline} 
2   \RV^{\frac{q}{2}}
\sum'_{\CQ\in\Lambda_{p-2,q-2}}\sum'_{(j,p)}
e^{2\pi \I (j\CQ\cdot a_1+p\CQ\cdot a_2)}\sum_{\ell=0}^2\, \frac{a_\ell}{{\RV^2}^\ell} \, 
\left(-\tfrac{\CQ^2}{2}\right)^{2-\ell}\,
c\left(-\tfrac{\CQ^2}{2}\right)\,
\left(\frac{2\CQ_R^2 S_2}{|j+pS|^2}\right)^{\frac{q-4-2\ell}{4}}\, 
\\
 \times K_{\frac{q-4}{2}-\ell}\left( 2\pi \sqrt{\tfrac{2{\RV^2}}{U_2}} |j+pU|  |\CQ_R| \right)
\end{multline}
Defining $(Q,P)=(j,p)\CQ$, we see that the Fourier expansion
with respect to $(a_1,a_2)$ has support on collinear vectors $(Q,P)$ with $Q,P\in \Lambda_{p-2,q-2}$. Extracting
the greatest common divisor of $(j,p)$, we find that the
Fourier coefficients with charge $Q'^{i}=(Q,P)$ and mass $\cM(Q,P)=\sqrt{2Q'^{i}_RQ'^{j}_RM_{ij}}$ 
defined in \eqref{EMBPSmass} are given by
\bea \label{F34deg}
F_{\alpha\beta\gamma\delta}^{(p,q),1,Q'} &=&  4
\RV^{\frac{q}{2}} \,\bar c(Q'^i)\, 
\sum_{\ell=0}^2\frac{\cP^{(\ell)}_{\alpha\beta\delta\gamma}(Q'^i,S)}{{\RV^2}^\ell}\frac{K_{\tfrac{q-4}{2}-\ell}\left( 2\pi \RV\cM(Q,P) \right)}{\cM(Q,P)^{\frac{q-4}{2}-\ell} }
\nn \\
 F_{\mu\alpha\beta\gamma}^{(p,q),1,Q'} &=&  4
\RV^{\frac{q}{2}} \,\bar c(Q'^i)\, 
\sum_{\ell=0}^1\frac{\cP^{(\ell)}_{\mu\alpha\beta\delta}(Q'^i,S)}{\I\sqrt{2} {\RV^2}^\ell}\frac{K_{\tfrac{q-6}{2}-\ell}\left( 2\pi \RV\cM(Q,P) \right)}{\cM(Q,P)^{\frac{q-6}{2}-\ell} } 
\nn \\
& \vdots&\nn 
\\
 F_{\mu\nu\rho\sigma}^{(p,q),1,Q'} &=&  4
\RV^{\frac{q}{2}} \,\bar c(Q'^i)\, 
\frac{\cP^{(0)}_{\mu\nu\sigma\rho}(Q'^i,S)}{4}\frac{K_{\tfrac{q-12}{2}}\left( 2\pi \RV\cM(Q,P) \right)}{\cM(Q,P)^{\frac{q-12}{2}} } 
\eea
for the tensor integral, and
\be
 \trF^{(p,q),1,Q'} =  4  \, 
\RV^{\frac{q}{2}} \,\bar c(Q'^i)\,\sum_{\ell=0}^2\, \frac{a_\ell}{{\RV^2}^\ell} \left[ -\frac{{\rm gcd}(Q'^{i}\cdot Q'^{j})}{2}\right]^{2-\ell}\,
\frac{K_{\tfrac{q-4}{2}-\ell}\left( 2\pi \RV\cM(Q,P) \right)}{\cM(Q,P)^{\frac{q-4}{2}-\ell} }
\ee
for its trace.
The covariantized versions of $P_{abcd}(Q)$ with respect to the torus' metric,  $\cP^{(\ell)}_{\alpha\beta\gamma\delta},\ldots, \cP^{(\ell)}_{\mu\nu\sigma\rho}$ are given in appendix \ref{DecompPolynomials}. Finally the degeneracy is given by
\be
\label{muP34}
\bar c(Q,P) = \sum_{\substack{(Q,P)/d\in\Lambda_{p-2,q-2}^{\oplus2}}}\, 
\left( \frac{d^2}{{\rm gcd}(Q^2,Q\cdot P,P^2)}\right)^{\tfrac{q-8}{2}}\, 
c\left(-\tfrac{{\rm gcd}(Q^2,Q\cdot P,P^2)}{2d^2}\right),
\ee
with support $(Q,P)\in\Lambda_{p-2,q-2}\oplus \Lambda_{p-2,q-2}$.

\paragraph{Rank-two orbit} Finally, rank-two matrices can  be uniquely decomposed as $A=\big(\colvec[0.7]{k &j\\0 &p }\big)\big(\colvec[0.7]{a &b\\c &d }\big)$
where $k>j\geq 0$ and $p\neq 0$ and $\big(\colvec[0.7]{a &b\\c &d }\big)\in SL(2,\IZ)$.
The matrices $A$ can therefore be restricted to $A=\big(\colvec[0.7]{k &j\\0 &p }\big)$,
provided the integral is extended to the double cover of the upper half-plane $\cH$. 
This leads to
\begin{multline}
F^{(p,q),1}_{\mu_1\ldots\mu_h\alpha_1\ldots\alpha_{4-h}} =2{\RV^2} 
\sum_{\substack{k>j\geq0\\p\neq0}}
\left(\frac{\RV}{\I\sqrt{2}}\right)^{h}\,
e^{-2\pi \I kp(\psi+\I\RV^2)}\int_{\IR^+}\frac{\de\tau_2}{\tau_2^{2+h}}\int_{\IR}\de \tau_1 \,
\frac{e^{-\frac{\pi}{\tau_2}\frac{\RV^2}{S_2}\left|k\tau+j+pS\right|^2}}{\Delta}
\\
\times 
 \prod_{l=1}^h \left[\tfrac{1}{\sqrt{S_2}}
\Big(\colvec[0.9]{1&S_1\\0&S_2}\Big)
\Big(\colvec[0.9]{k \bar\tau +j \\p}\Big)
\right]_{\mu_l} \ShiftPart{p\hspace{-1.5pt}-\hspace{-2pt}2,q\hspace{-1.5pt}-\hspace{-2pt}2}{n_ia^i}\Big[P_{\alpha_1\ldots\alpha_{4-h}}\,
	 e^{2\pi \I\,\left(j(\CQ-\frac12 ka_1)\cdot a_1+p(\CQ-\frac12 ka_1)\cdot a_2\right)}\Big]
\end{multline}
for the tensor integral, and to
\begin{multline}
\trF^{(p,q),1}=2{\RV^2} 
\sum_{\substack{k>j\geq0\\p\neq0}} e^{-2\pi \I kp(\psi+\I\RV^2)}
 \int_{\IR^+}\frac{\de\tau_2}{\tau_2^{2}}\int_{\IR}\de \tau_1\,
e^{-\frac{\pi}{\tau_2}\frac{\RV^2}{S_2}\left|k\tau+j+pS\right|^2}
\\\times 
\ShiftPart{p\hspace{-1.5pt}-\hspace{-2pt}2,q\hspace{-1.5pt}-\hspace{-2pt}2}{n_ia^i}\Big[ e^{2\pi \I\,\left(j(\CQ-\frac12 ka_1)\cdot a_1+p(\CQ-\frac12 ka_1)\cdot a_2\right)}\Big]\, 
D^2\left(\frac{1}{\Delta}\right)
\end{multline}
for its trace.

Inserting the Fourier expansion \eqref{FourierDelta}, the integral over $\tau_1$ is Gaussian
while the integral over $\tau_2$ is of Bessel type. The sum over $0\leq j<k$ enforces a Kronecker delta function modulo $k$,  
\be 
\label{jphaseTerm}
\sum_{j=0}^{k-1}\,{\rm exp}\left[2\pi \I\,\frac{j}{k}\left(\tfrac{\CQ^2}{2}+m\right)\right]=
\left\{
\begin{split}
&k\quad {\rm if}\quad\tfrac{\CQ^2}{2}+m=l k,\quad l \in\IZ,
\\
&0\quad{\rm otherwise}
\end{split}\,\right.
\ee 
Relabelling the charges as $p\CQ\to P$, $kp\to -M_1$ and $lp\to -M_2$, and defining 
$D=-\tfrac{P^2}{2}+M_1M_2$ one obtains, for the trace of the tensor integral,
\begin{multline}
\label{FQNA}
\trF^{(p,q),2} = \sum_{\substack{M_1\ne 0 ,M_2 \\
P\in\Lambda_{p-2,q-2}}}
\trF^{(p,q),2,M_1}\left(P-M_1 a_1,M_2-a_1\cdot P+\tfrac12 (a_1\cdot a_1)M_1\right)\, 
\\
\times
e^{2\pi\I (P\cdot a_2+M_1(\psi-\frac12 a_1 \cdot a_2)+
(M_2-a_1\cdot P+\tfrac12 (a_1\cdot a_1)M_1) S_1 )}
\end{multline}
where $\trF^{(p,q),2,M_1}$ is the non-Abelian Fourier coefficient,
\be
\label{FQM1M2scal}
\trF^{(p,q),2,M_1}(P,M_2) = 4 
({\RV^2} S_2)^{\frac{q-2}{2}}\, \bar c(M_1,M_2,P)
\sum_{\ell=0}^2 \frac{a_\ell\, D^{2-\ell}}{ 
{({\RV^2} S_2)^{\ell}}}
\left(\frac{2\pi}{S_{\rm cl}}\right)^{\frac{q-5}{2}-\ell}
K_{\frac{q-5}{2}-\ell}( S_{\rm cl})\ ,
\ee
$S_{\rm cl}$ is the classical action
\be\label{Scl}
S_{\rm cl}(M_1,M_2,P)=2\pi \sqrt{\left( {\RV^2}M_1+S_2 M_2\right)^2+2{\RV^2} {S_2} P_R^2}\quad  , 
\ee
and $\bar c(M_1,M_2,P)$ the summation measure
\be\label{nonAbDegeneracy}
\bar c(M_1,M_2,P)=\sum_{\substack{d|(M_1,M_2)\\P/d\in\Lambda_{p-2,q-2}}}c\left(\tfrac{D}{d^2}\right)\,d^{q-7} 
\ .
\ee
It is worth noting that \eqref{FQNA} is the general expansion of a function of $(S_1,a_1,a_2,\psi)$
invariant under discrete shifts  $T_{b, \epsilon_1,\epsilon_2, \kappa}$ acting as 
\be
(S_1,a_1, a_2,\psi) \mapsto \left(S_1 + b, a_1 + \epsilon_1, a_2+\epsilon_2+ b a_1,
\psi+ \kappa + \tfrac12[ \epsilon_2 (a_1+\epsilon_1) - \epsilon_1 (a_2+b a_1)] \right)
\ee
with $b, \kappa \in \IZ$ and $\epsilon_1,\epsilon_2\in \IZ^{p-2,q-2}$. Invariance
under $T_{b,0,\epsilon_2,\kappa}$  is manifest, while invariance under $T_{0,\epsilon_1,0,0}$ 
is realized by shifting $P\mapsto P+M_1 \epsilon_1, M_2\mapsto M_2+\epsilon_1 P + \tfrac12 M_1 \epsilon_1^2$, which leaves $D$ and $\tilde M_2=M_2-a_1\cdot P+\tfrac12 (a_1\cdot a_1)M_1$ invariant.  It is worth noting that in the special case  $p=2$, $P_R^2$ 
vanishes identically  so \eqref{Scl} simplifies to $S_{\rm cl}=2\pi | R^2 M_1 + S_2 M_2|$.

Similarly, for the tensor integral, we get
\be
\label{FQM1M2}
\begin{split}
F_{\alpha\beta\gamma\delta}^{(p,q),2,M_1}(P,M_2)&=4 
\,({\RV^2} S_2)^\frac{q-2}{2}\,\bar c(\scalebox{0.8}{$M_1,M_2,P$})
\, 
\sum_{\ell=0}^2 \frac{\tilde P^{(\ell)}_{\alpha\beta\gamma\delta}(P)}{({\RV^2} S_2)^{\ell}}\left(\frac{2\pi}{S_{\rm cl}}\right)^{\frac{q-5}{2}-\ell}K_{\frac{q-5}{2}-\ell}( S_{\rm cl})
\\
F_{2\alpha\beta\gamma}^{(p,q),2,M_1}(P,M_2)&=4 
\,({\RV^2} S_2)^\frac{q-2}{2}\,\bar c(\scalebox{0.8}{$M_1,M_2,P$})
\, 
\sum_{\ell=0}^1 \frac{\tilde P^{(\ell)}_{\alpha\beta\gamma}(P)}{\I\sqrt2({\RV^2} S_2)^{\ell-\frac{1}{2}}}\left(\frac{2\pi}{ S_{\rm cl}}\right)^{\frac{q-7}{2}-\ell}\hspace{-1.5mm} K_{\frac{q-7}{2}-\ell}( S_{\rm cl})
\\
&\vdots
\\
F_{2222}^{(p,q),2,M_1}(P,M_2)&=4 
\,({\RV^2} S_2)^\frac{q-2}{2}\,\bar c(\scalebox{0.8}{$M_1,M_2,P$})
\frac{\tilde P^{(0)}}{4({\RV^2} S_2)^{-2}}\left(\frac{2\pi}{S_{\rm cl}}\right)^{\frac{q-13}{2}}K_{\frac{q-13}{2}}
(S_{\rm cl}),
\end{split}
\ee
where we restricted to the cases $\mu,\nu,\ldots =2$ for simplicity.

\subsection{Extension to $\IZ_N$ CHL orbifolds  \label{sec_ZNp2}}
The degeneration limit  \eqref{Gpqdec2} of the modular integrals \eqref{f4pq} for $\IZ_N$ CHL models with $N=2,3,5,7$ can be treated similarly by 
applying the orbit method. In \eqref{f4pq},  $\Delta_k$ is the cusp form of weight $k=\tfrac{24}{N+1}$ defined in \eqref{defg}, 
and $\Part{p,q}[P_{abcd}]$ is the partition function with insertion of $P_{abcd}$ for a lattice 
\be\label{CHLlatticeDecomp}
\Lambda_{p,q}=\Lambda_{p-2,q-2}\oplus\sLambda_{1,1}\oplus\sLambda_{1,1}[N]\ ,
\ee
where $\Lambda_{p-2,q-2}$ is a lattice of level $N$. The lattice $\sLambda_{1,1}\oplus\sLambda_{1,1}[N]$ is obtained from
the usual unimodular lattice $\sLambda_{2,2}$ by restricting the windings and momenta to $(n_1,n_2,m_1,m_2)\in \IZ \oplus N\IZ\oplus \IZ \oplus \IZ$, hence breaking the automorphism group $O(2,2,\IZ)$ to $\sigma_{S\leftrightarrow T}\ltimes[\Gamma_0(N) \times \Gamma_0(N)]$. After Poisson resummation on $m_2$, Eq. \eqref{Poisson2} and \eqref{poissonRessummedP2} continue to 
hold, except for the fact that $n_2$ is restricted to run over $N\IZ$. The sum over $A=\big(\colvec[0.7]{n_1&m_1\\n_2&m_2}\big)$ can then be decomposed into orbits of $\Gamma_0(N)$:\,\footnote{Note that the subsequent analysis is valid in the region of the moduli space where $N\RV^2>2S_2$}

\paragraph{Trivial orbit} The contribution of $A=0$ reduces, up to a factor of $\RV^2$, to the integrals \eqref{f4pq} for the lattice $\Lambda_{p-2,q-2}$, 
\be\label{trivialOrbitDecompZN}
F^{(p,q),0}_{\alpha\beta\gamma\delta}= {\RV^2}\, F^{(p-2,q-2)}_{\alpha\beta\gamma\delta}\ ,\quad 
\trF^{(p,q),0} = {\RV^2}\, \trF^{(p-2,q-2)}\ ,
\ee

\paragraph{Rank-one orbits} Matrices $A$ of rank-one fall into two different classes of orbits under $\Gamma_0(N)$. For simplicity, let us first consider the case where $(n_2,m_2)\neq(0,0)$, and denote $(m_2,n_2)=p(n_2',m_2')$, with $p=\gcd(n_2,m_2)$:
\begin{itemize} 
\item Matrices with $n_2'=0\mod N$, as they are required to be rank-one, can be decomposed as $\big(\colvec[0.8]{n_1&m_1\\n_2&m_2}\big)= \big(\colvec[0.8]{0&j\\0&p}\big)\big(\colvec[0.8]{a&b\\c&d}\big)$ with $(j,p)\in\IZ^2\smallsetminus \{(0,0)\}$, $p\neq0$ and  $\big(\colvec[0.8]{a&b\\c&d}\big)\in\Gamma_\infty\backslash\Gamma_0(N)$. For this class of orbit, one can thus unfold directly the domain $\Gamma_0(N)\backslash\cH$ into the unit strip $\cS=\Gamma_{\infty} \backslash \cH=\IR_{\tau_2}^+\times (\IR/\IZ)_{\tau_1}$.
\item Matrices with $n_2'\neq 0\mod N$ can be decomposed as   $\big(\colvec[0.8]{n_1&m_1\\n_2&m_2}\big)= \big(\colvec[0.8]{j&0\\p&0}\big)\big(\colvec[0.8]{a&b\\ c&d}\big)$ with $(j,p)\in\IZ\oplus N\IZ\smallsetminus\{(0,0)\}$, $p\neq0$ and  $\big(\colvec[0.8]{a&b\\ c&d}\big)\in S\,\Gamma_{\infty,N}\,S^{-1}\backslash\Gamma_0(N)$, where $\Gamma_{\infty,N}=\{\big(\colvec[0.8]{1&n\\0&1}\big), n\in N\IZ\}$. One can then unfold the fundamental domain $\Gamma_0(N)\backslash \cH$ into $S\,\Gamma_{\infty,N}\,S^{-1} \backslash \cH$, and change variable $\tau\to-1/\tau$ as in the weak coupling case \eqref{ZNunfoldingStep} to recover the integration domain $\cS_N=\Gamma_{\infty,N}\backslash\cH=\IR_{\tau_2}^+\times (\IR/N\IZ)_{\tau_1}$, the width-$N$ strip.
\end{itemize}
The remaining contributions $A$ with $(n_2,m_2)=(0,0)$ belong to the two classes of orbits above. Let $(n_1,m_1)=j(n_1',m_1')$, where $j=\gcd(n_1,m_1)$ and $j\in\IZ$, then contributions with $n_1'=0\mod N$ correspond to the cases $(j,p)=(j,0)$ in the first class above; contributions with $n_1'\neq 0\mod N$ correspond to $(j,p)=(j,0)$ in the second class above.

After unfolding and changing variable, the result for the simplest component $F^{(p,q),1}_{\alpha\beta\gamma\delta}$ reads (similarly to \eqref{ZNunfoldingStep})
\begin{multline}
\label{ZNunfoldingStepDecomp}
F^{(p,q),1}_{\alpha\beta\gamma\delta} =  \RV^2  \int_{\IR^+}\hspace{-1mm}\frac{\de\tau_2}{\tau_2^2}\int_{\IR/\IZ}\hspace{-2mm}\de\tau_1\,  
\frac{1}{\Delta_k(\tau)}\sum'_{\substack{(j,p)\in\IZ^2} }
e^{-\frac{\pi\RV^2}{\tau_2\,S_2} |j+pS|^2}
\Part{p\hspace{-1.5pt}-\hspace{-2pt}2,q\hspace{-1.5pt}-\hspace{-2pt}2}\left[  e^{2\pi\I( j \CQ\cdot a_1+p\CQ) \cdot a_2}\,P_{\alpha\beta\gamma\delta}\right] 
\\
+   \RV^2 \int_{\IR^+}\hspace{-1mm}\frac{\de\tau_2}{\tau_2^2}\int_{\IR/\IZ}\hspace{-2mm}\de\tau_1\,  
\frac{1}{\Delta_k(\tau/N)}\frac{\upsilon}{N}\sum'_{\substack{(j,p)\in\IZ^2 \\ p=0 \mod N}} e^{-\frac{\pi \RV^2}{\tau_2\,S_2} |j+pS|^2}
\PartD{p\hspace{-1.5pt}-\hspace{-2pt}2,q\hspace{-1.5pt}-\hspace{-2pt}2}\left[e^{2\pi\I( j \CQ\cdot a_1+p\CQ) \cdot a_2}\,P_{\alpha\beta\gamma\delta}\right] \ ,
\end{multline}
where $\PartD{p\hspace{-1.5pt}-\hspace{-2pt}2,q\hspace{-1.5pt}-\hspace{-2pt}2}$ is the partition function of the dual lattice $\Lambda_{p-2,q-2}^*$ and where \linebreak $\upsilon =N^{k/2+1} |\Lambda_{p-2,q-2}^*/\Lambda_{p-2,q-2}|^{-1/2}$ (which reduces to $\upsilon=N^{1-\delta_{q,8}}$ for $q\leq 8$ in the cases of interest). The contributions from $\CQ=0$ thus give 
\be 
\label{Z2F34deg0}
\begin{split}
&F_{\alpha\beta\gamma\delta}^{(p,q),1,0} =  \RV^{q-6}
\frac{3(2c_k(0))}{8\pi^2}\,\frac{1}{2}
\left(\cE^\star_{\frac{8-q}{2}}(S)+vN^{\frac{q-8}{2}}\cE^\star_{\frac{8-q}{2}}(NS)\right)\,\delta_{(\alpha\beta}\delta_{\gamma\delta)},
\\
&F_{\mu\nu\gamma\delta}^{(p,q),1,0} =  \RV^{q-6}
\frac{2c_k(0)}{4\pi^2}\left[\tfrac{8-q}{4}\delta_{\alpha\beta}\delta_{\mu\nu}-\delta_{\alpha\beta}\cD_{\mu\nu}\right]\frac{1}{2}\left(\cE^\star_{\frac{8-q}{2}}(S)+vN^{\frac{q-8}{2}}\cE^\star_{\frac{8-q}{2}}(NS)\right),
\\
&F_{\mu\nu\rho\sigma}^{(p,q),1,0} =  \RV^{q-6}
\frac{2c_k(0)}{2\pi^2}
\\
&\quad\times\left[\cD^2_{\mu\nu\rho\sigma}-\tfrac{10-q}{2}\delta_{(\mu\nu}\cD_{\rho\sigma)}+\left(\tfrac{8-q}{2}\right)\left(\tfrac{10-q}{2}\right)\tfrac{3}{8}\delta_{(\mu\nu}\delta_{\rho\sigma)}\right]\frac{1}{2}
\left(\cE^\star_{\frac{8-q}{2}}(S)+vN^{\frac{q-8}{2}}\cE^\star_{\frac{8-q}{2}}(NS)\right),
\end{split}
\ee
for the tensor integral, and 
\be
\trF^{(p,q),1,0} =  \RV^{q-6}
(p-q+6)(p-q+8)\frac{2c_k(0)}{8\pi^2}\,
\frac{1}{2}\left(\cE^\star_{\frac{8-q}{2}}(S)+vN^{\frac{q-8}{2}}\cE^\star_{\frac{8-q}{2}}(NS)\right),
\ee
for its trace. Recall $c_k(0)=\tfrac{24}{N+1}=k$ is the zero mode of $1/\Delta_k=\sum_m c_k(m) q^m$. As in \eqref{F34deg0s} and \eqref{F34deg0},  the pole at $q=6$ is minimally subtracted by the regularization prescription mentioned below \eqref{def1looppqFNs}, while the pole at $q=8$ cancels against the pole from the zero orbit contribution \eqref{trivialOrbitDecompZN}.

The contributions with $\CQ\neq0$ are exponentially suppressed at large $\RV$, and have similar Fourier coefficients as in the full rank case \eqref{F34deg}, except for a different summation measure. Let us label the electromagnetic charges by $(Q,P)=(j,p)\CQ=(j',p')\hat Q$ where 
$(j',p')$ are coprime integers. It will be useful to  classify all possible rank-one charges $(Q,P)$ in orbits of the S-duality group $\Gamma_0(N)$ 
acting as $\big(\colvec[0.8]{Q\\P}\big)\to\big(\colvec[0.8]{a&b\\c&d}\big)\big(\colvec[0.8]{Q\\P}\big)$, where $\big(\colvec[0.8]{a&b\\c&d}\big)\in\Gamma_0(N)$. 
\begin{itemize}
\item Charges $(Q,P)$ such that $p'=0\mod N$ are in the same orbit as purely electric
charges $(\hat Q,0)$. Their Fourier coefficient gets contributions from both   terms in \eqref{ZNunfoldingStepDecomp}  with $d = \mbox{gcd}(j,p)$ and $\frac{\hat{Q}}{d} = \tilde{Q} \in \Lambda_{p-2,q-2}$ in the first case and $\frac{\hat{Q}}{d} = \tilde{Q} \in \Lambda^*_{p-2,q-2}$ in the second, such that they are weighted by the measure
\be\label{electricZNDeg}
\bar c_k(Q,P)=\hspace{-2mm}\sum_{\substack{d\geq 1\\\hat Q/d\in\Lambda_{p-2,q-2}}}\hspace{-4mm}c_k\Big(-\frac{\hat Q^2}{2d^2}\Big)\left(\frac{d^2}{\hat Q^2}\right)^{\frac{q-8}{2}}+\hspace{2mm} \upsilon\,\hspace{-6mm} \sum_{\substack{d\geq 1\\\hat Q/d\in\Lambda^*_{p-2,q-2}}}\hspace{-6mm}c_k\Big(-\frac{N\hat Q^2}{2d^2}\Big)\,\, \left(\frac{d^2}{N\hat Q^2}\right)^{\frac{q-8}{2}}\ ,
\ee
where the first contribution has support $Q\in\Lambda_{p-2,q-2}\subset\Lambda^*_{p-2,q-2}$, while the second has support on $Q\in\Lambda_{p-2,q-2}^*$. Notice that the latter is matched against $1/\Delta_k(\tau/N)$, which explains the $N$ factor in the argument of $c_k$.

\item Charges $(Q,P)$ such that $p'\neq 0\mod N$ are in the same orbit as purely magnetic
charges $(0,\hat P)$, where we relabelled $\hat{Q}$ as $\hat{P}$ for convenience. Their Fourier coefficient gets contributions from both terms in \eqref{ZNunfoldingStepDecomp}  with $d = \mbox{gcd}(j,p)$ and $\frac{\hat{P}}{d} = \tilde{Q} \in \Lambda_{p-2,q-2}$ in the first case and $N d = \mbox{gcd}(j,p)$ (because $j=0$ mod $N$) and $\frac{\hat{P}}{Nd} = \tilde{Q} \in \Lambda^*_{p-2,q-2}$ in the second, such that they are weighted by the measure
\be
\label{magneticZNDeg}
\bar c_k(Q,P)=\hspace{-2mm}\sum_{\substack{d\geq 1\\\hat P/d\,\in\Lambda_{p-2,q-2}}}\hspace{-4mm}c_k\Big(-\frac{\hat P^2}{2d^2}\Big)\left(\frac{d^2}{\hat P^2}\right)^{\frac{q-8}{2}}+ \hspace{2mm} \upsilon\,\hspace{-8mm}\sum_{\substack{d\geq 1\\\hat P/d\,\in N\Lambda^*_{p-2,q-2}}}\hspace{-6mm}c_k\Big(-\frac{\hat P^2}{2Nd^2}\Big)\left(\frac{Nd^2}{\hat P^2}\right)^{\frac{q-8}{2}},
\ee
where the first contribution has support $P\in\Lambda_{p-2,q-2}$, while the second has $P\in N\Lambda_{p-2,q-2}^*\subset\Lambda_{p-2,q-2}$. In the latter contribution, one $N$ factor in the argument of $c_k$ comes from the matching condition, and two $N$ factors in its denominator come from all divisors $d$ being originally multiples of $N$.
\end{itemize}

\paragraph{Rank-two orbit} For the rank-two matrices $A$, the two classes of orbits are similarly given by studying $(n_2,m_2)=p(n_2',m_2')$, where $p=\gcd(n_2,m_2)$.
\begin{itemize}
\item Contributions for which $(n_2',m_2')=(0,1)\mod N$ can be decomposed as $A=\big(\colvec[0.8]{k&j\\0&p}\big)\big(\colvec[0.8]{a&b\\c&d}\big)$, $0\leq j < k$, $p\in\IZ\smallsetminus\{0\}$ and $\big(\colvec[0.8]{a&b\\c&d}\big)\in SL(2,\IZ)$, where its representative has trivial stabilizer. For this first class of orbits, the fundamental domain can be unfolded to the full upper half-plane $\cH=\IR^+_{\tau_2}\times \IR_{\tau_1}$.

\item Contributions for which $(n_2',m_2')=(1,0)\mod N$ can have $A=\big(\colvec[0.8]{j&k\\p&0}\big)\big(\colvec[0.8]{a&b\\c&d}\big)$, $0\leq j < Nk$, $p\in N\IZ\smallsetminus\{0\}$ and $\big(\colvec[0.8]{a&b\\c&d}\big)\in SL(2,\IZ)$, where representative has trivial stabilizer. For this second class of orbits, the fundamental domain can be unfolded to $\cH=\IR^+_{\tau_2}\times \IR_{\tau_1}$ as well and the integrand can be brought back to the standard lattice sum representation using a transformation $\tau\to-1/\tau$, in the spirit of \eqref{ZNunfoldingStepDecomp}. 
\end{itemize}

Both classes of contributions lead to the same type of non-Abelian Fourier coefficient as in the unorbifolded case \eqref{FQM1M2scal} and \eqref{FQM1M2}, except for a different summation
measure $\bar c(M_1,M_2,P)$. The first class have support $(M_1,M_2,P)\in\IZ \oplus \IZ\oplus \Lambda_{p-2,q-2}$, whereas the second class have support $(M_1,M_2,P)\in N\IZ \oplus N \IZ \oplus N\Lambda^*_{p-2,q-2}$. In fine the summation measure reads
\be\label{Z2Rank2untwist}
\bar c_k(M_1,M_2,P)=\sum_{\substack{d|(M_1,M_2)\\P/d\in\Lambda_{p-2,q-2}}}\hspace{-4mm}c_k\Big(\frac{D}{d^2}\Big)\,d^{q-7}+\hspace{2mm}\upsilon\hspace{-4mm}\sum_{\substack{Nd|(M_1,M_2)\\P/d\in N\Lambda_{p-2,q-2}^*}}\hspace{-4mm}c_k\Big(\frac{D}{Nd^2}\Big)\,(Nd)^{q-7},
\ee
where we recall that $D=-\tfrac12P^2+M_1M_2$. For the second class of orbits, one factor of $N$  in the argument of $c_k$ comes from the matching condition, and two factors of $1/N$  come from the fact that all divisors  were originally multiples of $N$.

\subsection{Large radius  limit and BPS dyons \label{sec_decomp}}
Specializing to $(p,q)=(2k,8)=(r-4,8)$, and  choosing $\Lambda_{p-2,q-2}=\Lambda_m$,
the degeneration studied in this section corresponds to the limit  of the exact $(\nabla \Phi)^4$ amplitude in heterotic string on $T^7$ in the limit where a circle inside $T^7$,
orthogonal to the $\IZ_N$ action, decompactifies. The coordinate $\RV$ is identified as 
the radius  of the large circle  in
units of the four-dimensional Planck length $l_P=g_4 l_H$.
The contributions from the various orbits discussed in \S\ref{sec_decompmax} and \S\ref{sec_ZNp2}
are then interpreted as follows:

\subsubsection{Effective action in $D=4$ \label{sec_eff4}}
In the large $R$ limit, $F_{\alpha\beta\gamma\delta}^{(2k,8)}$ should reproduce the exact four-dimensional $F^4$ coupling, up to exponentially suppressed corrections. As already mentioned below 
\eqref{F34deg0} and \eqref{Z2F34deg0}, the contribution of the vector $\widetilde Q=0$ to the
rank-one orbit has a pole at $q=8$. Using the regularisation \eqref{def1looppqFNs}, that formally sets $q=8+2\epsilon$, one obtains 
\bea 
F_{\alpha\beta\gamma\delta}^{(2k,8),1,0}(\epsilon) &=&  \RV^{2+2\epsilon} \frac{3(2k)}{(4\pi)^2}\,
\left(\cE^\star_{-\epsilon}(S)+N^{\epsilon} \cE^\star_{-\epsilon }(NS)\right)\,\delta_{(\alpha\beta}\delta_{\gamma\delta)}  \\
&=& R^2  Ê\frac{3}{2(2\pi)^2} \Bigl( \frac{k}{\epsilon}  - \log( S_2^{\; k} |\Delta_k(S)|^2)+ k \Bigl( \log\Bigl(  \frac{R^2}{4\pi} \Bigr)   - \gamma\Bigr) \Bigr)\,\delta_{(\alpha\beta}\delta_{\gamma\delta)} + \mathcal{O}(\epsilon) \ ,  \nn
 \eea
However, this pole cancels against the pole  \eqref{4Dpole} in the trivial zero-orbit
contribution  \eqref{F34zero}, \eqref{trivialOrbitDecompZN}, leaving
the finite result 
\be F^{(2k,8)}_{\alpha\beta\gamma\delta} = R^2 \biggl(- \frac{3}{2(2\pi)^2}Ê\Bigl( \log( S_2^{\; k} |\Delta_k(S)|^4) - 2 k \log R \Bigr) \delta_{(\alpha\beta} \delta_{\gamma\delta)} + \hat{F}^{(2k-2,6)}_{\alpha\beta\gamma\delta}(\Phi) \biggr) +\dots 
 \qquad 
 \ee
where $\hat F^{(2k-2,6)}_{\alpha\beta\gamma\delta}$ is the renormalized 1-loop
coupling, up to an irrelevant additive constant, and the dots denote  exponentially suppressed terms.

Thus, the conjectural formula \eqref{f4exact} for the exact $(\nabla \Phi)^4$ coupling in $D=4$ 
predicts that the exact $F^4$ coupling in four dimensions should be given by 
\be 
- \frac{3}{8\pi^2}Ê\log( S_2^{\; k} |\Delta_k(S)|^2)  \delta_{(ab} \delta_{cd)}+F^{(2k-2,6)}_{abcd}(\Phi)  \ , \label{ExactF4} 
\ee
where  for convenience we renamed the indices $\alpha,\beta,\dots$ into $a,b,\dots$ running from $1$ to $2k-2$.  Indeed, it is known that half-maximal supersymmetry in $D=4$ allows for two
types of supersymmetry invariants with four derivatives: the first one is determined in terms of a holomorphic function of $S$, the second depends on the $G_{2k-2,6}$ moduli only, as  described in \eqref{F4allD}, and both contribute to $F^4$ couplings \cite{Bossard:2013rza}. 
The first term in \eqref{ExactF4} corresponds
the first invariant,  which also includes the $\cR^2$ coupling \eqref{fR2}, while the second was considered in \cite{Kiritsis:2000zi},  it is by construction exact at 1-loop and includes a four-derivative scalar couplings studied in \cite{Antoniadis:1997zt}.  

The relative coefficient of the two invariants in \eqref{ExactF4} is in fact fixed by unitarity. Indeed, 
the logarithmic dependence of the one-loop amplitude with respect to the Mandelstam variables ($s_1=s,\, s_2=t,\, s_3=u$) is determined by the 1-loop divergence of the four-photon supergravity amplitude  \cite{Fischler:1979yk}. Because the genus-one string theory amplitude $F^{(2k-2,6)}_{abcd}(\Phi,s_i) $ is finite in the ultra-violet, the corresponding supergravity amplitude pole in dimensional regularisation  $D=4-2\epsilon$ cancels by construction the pole of the coupling $F^{(2k-2,6)}_{abcd}(\Phi,\epsilon) $ regularised according to \eqref{def1looppqFNs} (corresponding formally to $q=6+2\epsilon$). Thus, in the low energy limit $-\ell_s^{\; 2} s_i \ll1$  \footnote{Recall that $2k-2$ is the number of vector multiplets in $D=4$.} 

\bea 
F^{(2k-2,6)}_{abcd}(\Phi,s_i) &\sim&  F^{(2k-2,6)}_{abcd}(\Phi,\epsilon) + \frac{3(2k)}{(4\pi)^2}Ê\Bigl( \frac{1}{\epsilon} - \frac{1}{3} \sum_{i=1}^3 \log(- \ell_s^{\; 2}Ês_i)  \Bigr) \delta_{(ab} \delta_{cd)} Ê \\ 
&\sim&\hat{F}^{(2k-2,6)}_{abcd}(\Phi)  - \frac{3}{8\pi^2} \log(S_2^{\; k}) \delta_{(ab} \delta_{cd)}-  \frac{2k}{(4\pi)^2}Ê \sum_{i=1}^3\log( -\ell_P^{\; 2}Ês_i)  \delta_{(ab} \delta_{cd)} \ , \nn  
\eea
up to a fixed constant, where we used the relation $ \frac{S_2}{2\pi} \ell_P^{\; 2}=  \ell_s^{\; 2}$ between Planck length and string length. Therefore, the relative coefficient of the two invariants in \eqref{ExactF4} is indeed such that the logarithm of $S_2$ in the coupling disappears in string frame,  consistently with the fact that string amplitudes depend analytically on the string coupling constant when formulated in string frame \cite{Green:2010sp}. 

The overal normalisation of the 4-photon amplitude can be determined from the 1-loop divergence as \cite{Fischler:1979yk,Bern:2013qca} (with $t_8 f^4 = f_{\mu\nu} f^{\nu\sigma} f_{\sigma\rho} f^{\rho\mu} - \frac{1}{4}( f_{\mu\nu} f^{\mu\nu} )^2$)
\be  A_4(S,\Phi,s_i) =  \frac{\kappa^4}{8} \biggl(\frac{3}{8\pi^2}ÊÊ\log( S_2^{\; k} |\Delta_k(S)|^2)  \delta_{(ab} \delta_{cd)}- F^{(2k-2,6)}_{abcd}(\Phi,s_i) \biggr) t_8 F^a F^b F^c F^d \ . \ee
More precisely, the 1PI effective action includes the local terms
\bea\label{effAction4D}
S_4 &=&  \hspace{-1mm}\int \de^4 \hspace{-0.2mm}x \sqrt{-g} \biggl( \frac{1}{2\kappa^2}Ê\mathcal{R} - \frac{S_2}{32\pi} ( F^a_{\mu\nu}F_a^{\mu\nu}+F^{\hat{a}}_{\mu\nu}F_{\hat{a}}^{\mu\nu}Ê) + \frac{S_1}{64\pi \scalebox{0.8}{$\sqrt{\mbox{-}g}$}} \varepsilon^{\mu\nu\rho\sigma}Ê ( F^a_{\mu\nu} F_{\rho\sigma\, a}-F^{\hat{a}}_{\mu\nu} F_{\rho\sigma\, \hat{a}} Ê) \biggr . \nn \\
&&\hspace{-5mm}+\frac{\kappa^4}{8}\Bigl( \frac{3}{8\pi^2}Ê\log( S_2^{\; k} |\Delta_k(S)|^2)  \delta_{(ab} \delta_{cd)}- \hat{F}^{(2k-2,6)}_{abcd}(\Phi) \Bigr) t^{\mu\nu\rho\sigma\kappa\lambda\vartheta\tau} \Bigl( \frac{S_2}{8\pi} \Bigr)^2F^a_{\mu\nu}F^{b}_{\rho\sigma}Ê  F^c_{\kappa\lambda}F^{d}_{\vartheta\tau}Ê\nn \\
&&\hspace{-5mm}- \frac{1}{(8\pi)^2}Ê\log( S_2^{\; k} |\Delta_k(S)|^2)  ( \mathcal{R}_{\mu\nu\rho\sigma}\mathcal{R}^{\mu\nu\rho\sigma}-4 \mathcal{R}_{\mu\nu} \mathcal{R}^{\mu\nu} + \mathcal{R}^2) Ê\\
&& \hspace{-5mm}- \frac{\kappa^2}{(8\pi)^2}  \mathcal{R}^{\mu\nu\rho\sigma}   \Bigl(   \cDÊ\log( S_2^{\; k} |\Delta_k(S)|^2)     \frac{S_2}{8\pi}  F_{\mu\nu}^{\hat{a}-} F_{\rho\sigma \hat{a}}^{-} + \overline{\cD}Ê\log( S_2^{\; k} |\Delta_k(S)|^2)    \frac{S_2}{8\pi}  F_{\mu\nu}^{\hat{a}+} F_{\rho\sigma \hat{a}}^{+}  \Bigr) \nn\\
&& \hspace{-5mm} - \frac{\kappa^4}{(8\pi)^2}  \cD^2 Ê\log( S_2^{\; k} |\Delta_k(S)|^2)  \Bigl( \frac{S_2}{8\pi} \Bigr)^2 \Bigl( 2F_{\mu\nu}^{\hat{a}-} F_{\rho\sigma \hat{a}}^{-}  F^{\mu\nu}_{\hat{b}-} F^{\rho\sigma \hat{b}}_{-}    +  F_{\mu\nu}^{\hat{a}-} F^{\mu\nu}_{ \hat{a} -}  F^{\rho\sigma}_{\hat{b}-} F_{\rho\sigma}^{\hat{b}-} \Bigr) \nn \\
&& \hspace{-5mm} - \frac{\kappa^4}{(8\pi)^2}  \overline{\cD}^2 Ê\log( S_2^{\; k} |\Delta_k(S)|^2)  \Bigl( \frac{S_2}{8\pi} \Bigr)^2 \Bigl( 2F_{\mu\nu}^{\hat{a}+} F_{\rho\sigma \hat{a}}^{+}  F^{\mu\nu}_{\hat{b}+} F^{\rho\sigma \hat{b}}_{+}    +  F_{\mu\nu}^{\hat{a}+} F^{\mu\nu}_{ \hat{a} +}  F^{\rho\sigma}_{\hat{b}+} F_{\rho\sigma}^{\hat{b}+} \Bigr) +\dots \biggr) \  , \nn \eea
which includes in particular  the exact $\mathcal{R}^2$ coupling  \eqref{fR2}. The components of \eqref{F34deg0}, \eqref{Z2F34deg0} with $\mu,\nu$ indices correspond to scalar field parametrizing the circle radius $R$, the scalar field $\psi$ dual to  the Kaluza--Klein vector, and the axiodilaton scalar field $S$ in four dimensions. The components involving the derivative of the function of $S$ depend on the complex (anti)selfdual field $F_{\mu\nu}^{\hat{a}\pm} \equiv \frac{1}{2} F_{\mu\nu}^{\hat{a}}  \pm \frac{\I}{4\scalebox{0.8}{$\sqrt{\mbox{-}g}$}}  \varepsilon_{\mu\nu}{}^{\rho\sigma}Ê F_{\rho\sigma}^{\hat{a}}$, with the covariant derivative $\cD$  defined as in Appendix \ref{tensEinsensteinSeries} with $\cD \equiv \cD_0$ and $\cD^2 \equiv \cD_2 \cD_0$. 

Let us now discuss the decompactification limit of the 1PI  effective action to ten dimensions,
focussing for simplicity on the maximal rank case where the lattice decomposes as 
\be \Lambda_{22,6} = D_{16} \oplus \sLambda_{6,6}  , 
\ee
where $D_{16}$ is the weight lattice of $Spin(32)/\IZ_2$.
Identifying $S_2 = \frac{2\pi ( 2\pi R)^6}{g_s^{\; 2}}$, with $g_s$ the heterotic string coupling constant
in 10 dimensions, one obtains for $a,b,c,d$ along $D_{16}$, 
\be  
- \frac{3}{8\pi^2}Ê\log( S_2^{\; k} |\Delta_k(S)|^2)  \delta_{(ab} \delta_{cd)}+\hat{F}^{(2k-2,6)}_{abcd}(\Phi)   = ( 2\pi R)^6 \Bigl( \frac{3}{g_s^{\; 2}} \delta_{(ab} \delta_{cd)}  + \frac{1}{2\pi^5} \delta_{abcd} \Bigr)  +\dots  \ee
up to a threshold contribution and exponentially suppressed terms. Here $\delta_{abcd} = 1$ if all indices are identical, and zero otherwise, and we used 
\bea
\int_{SL(2,\IZ)\backslash \cH} \frac{\de^2\tau}{\tau_2^2}\,\frac{\Gamma_{D_{16}} \left[  P_{abcd} \right]}{\Delta} &=& \int_{SL(2,\IZ)\backslash \cH} \frac{\de^2\tau}{\tau_2^2}\, \biggl[  
\Big( \tfrac{E_4^{\; 3}Ê- 2 \hat{E}_2 E_4 E_6 + \hat{E}_2^{\; 2} E_4^{\; 2}}{48 \Delta } - 24 \Big) \delta_{(ab} \delta_{cd)} + 48  \delta_{abcd} \biggr] 
\nn\\ 
&=& 32 \pi \delta_{abcd}  \ .
\eea
This equation follows from known results about the elliptic genus 
of the heterotic string \cite{Lerche:1987qk}.
Using an orthogonal basis for a Cartan subalgebra of $SO(32)$, one easily computes that this coupling gives the following trace combination in the vector representation of $SO(32)$
\be   \Bigl( \frac{3}{g_s^{\; 2}} \delta_{(ab} \delta_{cd)}  + \frac{1}{2\pi^5} \delta_{abcd} \Bigr)  t_8 F^a F^b F^c F^d =  \frac{( 2\pi R)^6}{4} t_8\Bigl( \frac{3}{g_s^{\; 2}} (\mbox{Tr} F^2 )^2    + \frac{1}{\pi^5} \mbox{Tr}  F^4\Bigr) \ . \ee
Using $\kappa^2 = 4 \alpha^\prime$ and reabsorbing the $ ( 2\pi R)^6 \alpha^{\prime\; 3}$ into the 6-torus volume one obtains in Einstein frame 
\bea
S_{10} &=&  \int \de^{10}x\, \sqrt{-g} \biggl( \frac{1}{8  \alpha^{\prime\; 4}}Ê\mathcal{R} + \frac{1}{8 \alpha^{\prime \; 3}} e^{-\frac{1}{2}\phi} \Bigl( \mbox{Tr} F_{\mu\nu} F^{\mu\nu} + \mathcal{R}_{\mu\nu\rho\sigma}\mathcal{R}^{\mu\nu\rho\sigma}-4 \mathcal{R}_{\mu\nu} \mathcal{R}^{\mu\nu} + \mathcal{R}^2\Bigr) \nn \\
&& \hspace{20mm}Ê- \frac{1}{2 \alpha^\prime } t_8 \Bigl( 3 e^{-\frac{3}{2} \phi} \mbox{Tr} F^2\mbox{Tr} F^2 + \frac{1}{\pi^5} e^{\frac{1}{2} \phi}\mbox{Tr} F^4 \Bigr)+\dots  \biggr)  \ , 
\eea
 which reproduces the tree level $\cR^2$ and $(\Tr F^2)^2$ coupling computed in \cite{Gross:1986mw} upon identifying $\phi = \sqrt{2}\kappa D - 6 \log 2$, and the 1-loop $\Tr F^4$ coupling computed in \cite{Ellis:1987dc,Lerche:1988zy}.

 \subsubsection{BPS dyons}
 
The contributions of non-zero vectors  to the rank-one orbit yield
exponentially suppressed corrections of order $e^{-2\pi \RV \cM(Q,P)}$ \eqref{F34deg}, where $\cM$ is the mass of a 1/2-BPS state of electromagnetic charge $(Q,P)$ in four dimensions.
The phase $e^{2\pi \I (a^1 Q+a^2 P)}$ multiplying \eqref{F34deg} is the expected minimal coupling of a dyonic state with charge $(Q,P)$ to the holonomies of the electric and magnetic gauge fields along the circle. The corresponding instanton is a saddle point of the three-dimensional Euclidean supergravity theory obtained by formal reduction along a time-like Killing vector, in the duality
frame where the axionic scalars $a_1,a_2$ are dualized into vector fields. Following the same
steps as \cite{Denef:2000nb}, one finds that  the classical action 
is then $S_{\rm cl}=2\pi R \cM(Q,P)$.

In the maximal rank case, the summation measure \eqref{muP34} is given by 
\be
\bar c(Q,P) = \sum_{\substack{d\geq 1\\(Q,P)/d\,\in \Lambda_{em}}} c\left(-\tfrac{{\rm gcd}(Q^2,P^2,Q\cdot P)}{2d^2}\right)\ ,
\ee
where $c(m)$ are the Fourier coefficients of $1/\Delta$. For $(Q,P)$ primitive, this agrees with the helicity supertrace \eqref{Om4max} of 1/2-BPS states with charges $(Q,P)$. 
In the case of CHL models, the summation measure is
instead given by  \eqref{electricZNDeg} or \eqref{magneticZNDeg} with 
$q=8$, $\tilde\upsilon=1$, depending whether the dyon is related by  $\Gamma_0(N)$, acting as $\big(\colvec[0.8]{Q\\P}\big)\to\big(\colvec[0.8]{a&b\\c&d}\big)\big(\colvec[0.8]{Q\\P}\big)$, to 
a purely electric or a purely magnetic state. It is interesting to note that
these two  formulas can be combined as follows. We first notice using the decomposition $(Q,P) =(j',p')\hat Q$ and $(Q,P) =(j',p')\hat P$ when $(Q,P)$ belong the electric and magnetic orbit respectively, with $(j',p')=1$, one obtains 
\bea \frac{\hat{Q}}{d} \in \Lambda_m &\Rightarrow& \frac{(Q,P)}{d} \in  \Lambda_m\oplus N \Lambda_m\ ,  \nn\\
 \frac{\hat{P}}{d} \in N\Lambda_e &\Rightarrow& \frac{(Q,P)}{d} \in  N\Lambda_e\oplus N \Lambda_e  \ , \qquad 
 \eea
 such that in both cases $(Q,P)/d\in \Lambda_m \oplus N  \Lambda_e$. Moreover,  if  $(Q,P)/d \in  \Lambda_m \oplus   N\Lambda_e$, then $\hat{Q}/d \in\Lambda_m$ or $\hat{P}/d \in N\Lambda_e$, depending of the orbit to which   $(Q,P)$ belongs to, therefore one  has the equivalence 
 \be \frac{(Q,P)}{d}  \in  \Lambda_m \oplus N  \Lambda_e\hspace{2mm}Ê \Leftrightarrow \hspace{2mm}Ê \frac{\hat{Q}}{d} \in \Lambda_m \hspace{2mm}Ê   {\rm or} \hspace{2mm}Ê   \frac{\hat{P}}{d} \in N\Lambda_e \ , \ee
for $(Q,P)$ conjugate to either an electric charge $\hat{Q}$ or a magnetic charge $\hat{P}$.  Similarly,
 \bea \frac{\hat{Q}}{d} \in \Lambda_e &\Rightarrow& \frac{(Q,P)}{d} \in  \Lambda_e\oplus N \Lambda_e \ ,  \nn\\
 \frac{\hat{P}}{d} \in \Lambda_m &\Rightarrow& \frac{(Q,P)}{d} \in  \Lambda_m\oplus  \Lambda_m \ , \qquad 
 \eea
 such that
  \be \frac{(Q,P)}{d}  \in  \Lambda_e \oplus \Lambda_m\hspace{2mm}Ê \Leftrightarrow \hspace{2mm}Ê \frac{\hat{Q}}{d} \in \Lambda_e \hspace{2mm}Ê   {\rm or} \hspace{2mm}Ê   \frac{\hat{P}}{d} \in \Lambda_m \ , \ee
for $(Q,P)$ conjugate to either an purely electric charge $(\hat{Q},0)$ or a purely magnetic charge $(0,\hat{P})$. Moreover, we have that $\gcd(NQ^2,P^2,Q\cdot P)=N\hat{Q}^2$ for a dyon in the $\Gamma_0(N)$ orbit of a purely electric charge, because then $\gcd(Nj^{\prime 2},p^{\prime 2},j^\prime p^\prime)=N$ since $p^\prime =0$ mod $N$, and $\gcd(NQ^2,P^2,Q\cdot P)=\hat{P}^2$ for a dyon in the $\Gamma_0(N)$ orbit of a purely magnetic charge, because then $\gcd(Nj^{\prime 2},p^{\prime 2},j^\prime p^\prime)=1$ since $p^\prime \ne 0$ mod $N$.  Putting these observations together we conclude that the summation measure  for a general 1/2 BPS dyon 
is given by 
 \be
 \label{dyonicZNDeg}
\bar c_k(Q,P)=  \hspace{-2mm}\sum_{\substack{d\geq 1\\(Q,P)/d\,\in \Lambda_e\oplus \Lambda_m}} 
\hspace{-4mm}c_k\Big(-\tfrac{\gcd(NQ^2,P^2,Q\cdot P)}{d^2}\Big)\hspace{2mm}
+ \hspace{-2mm} \sum_{\substack{d\geq 1\\(Q,P)/d\,\in\Lambda_m\oplus N\Lambda_e}} 
\hspace{-4mm}c_k\Big(-\tfrac{\gcd(NQ^2,P^2,Q\cdot P)}{2N d^2}\Big)\ .
\ee
It is worth noting that $\gcd(NQ^2,P^2,Q\cdot P)$ is invariant under $\Gamma_0(N)$ and Fricke S-duality, so that  each term in \eqref{dyonicZNDeg} is separately invariant under Fricke duality. Further noticing that $\Lambda_m\oplus N\Lambda_e\simeq\Lambda_e[N]\oplus \Lambda_m[N]$, \eqref{dyonicZNDeg} can be rewritten in a more suggestive way as   
 \be\label{dyonicZNDegen}
\bar c_k(Q,P)= \sum_{a|N} \sum_{\substack{d\geq 1\\(Q,P)/d\,\in\Lambda_{em}[a]}} c_k\Big(-\tfrac{\gcd(NQ^2,P^2,Q\cdot P)}{2a\,d^2}\Big)\ .
\ee
Most importantly, \eqref{dyonicZNDegen} agrees with the helicity 
supertrace $\Omega_4(Q,P)$ of a half-BPS dyon with primitive charge $(Q,P)$ which
was determined in \eqref{Om4twi} and \eqref{Om4utwi}.

\subsubsection{Taub-NUT instantons}
  Finally, the rank-two orbit \eqref{FQM1M2}  yields contributions schematically of the form
 \be \label{NUTactionweight} 
 \sum_{M_1\ne 0,M_2,P} \, \bar c(M_1,M_2,P)\, 
 e^{-2\pi \sqrt{\left( {\RV^2}M_1+S_2 \tilde M_2\right)^2+2{\RV^2} {S_2} \tilde P_R^2} + 
 2\pi\I (P\cdot a_2+M_1(\psi-\frac12 a_1 \cdot a_2)+
\tilde M_2 S_1 )}
 \ee
 where the summation measure \eqref{Z2Rank2untwist} is given by 
 \be
 \label{Z2Rank2untwist8}
\bar c_k(M_1,M_2,P)=\sum_{\substack{d|(M_1,M_2)\\P/d\in\Lambda_m}}\,
d\, c_k\Big(\frac{D}{d^2}\Big)
+\sum_{\substack{Nd|(M_1,M_2)\\P/d\in N\Lambda_e}}\,Nd\,
c_k\Big(\frac{D}{Nd^2}\Big)\ ,
\ee
and we denoted $\tilde M_2=M_2-a_1\cdot P+\tfrac12 (a_1\cdot a_1)M_1$, $\tilde P=P-M_1 a_1$, and $D=-\tfrac12P^2+M_1M_2$.
These $\cO(e^{-2\pi R^2 |M_1|})$ contributions are characteristic of an Euclidean Taub-NUT solution of the form ${\rm TN}_{M_1} \times T^6$, where the Taub-NUT space asymptotes to $\IR^3\times S_1(R)$ at spatial infinity \cite{Gibbons:1979xm}. 

The detailed semi-classical interpretation of these effects is complicated by the fact that
in a Taub-NUT background, similarly to the case of NS5-branes, large gauge transformations of the electric and magnetic holonomies  $a_1$ and $a_2$ do not commute, thus cannot be diagonalized simultaneously. The representation \eqref{FQNA} corresponds to the case where translations in $a_2$ and $\psi$ are diagonalized. Accordingly, the argument of the exponential
in \eqref{NUTactionweight} should be interpreted as the classical action in the duality frame in which the fields $\psi, S_1, a_2$ associated to the conserved charges $M_1,M_2$ and $P$ are dualized into vector fields $\omega, B,A$ in three dimensions.
In order to reach a positive definite action after dualization, 
one should first analytically continue the non-linear sigma model 
on $\frac{O(2k,8)}{O(2k)\times O(8)}$ into
 $\frac{O(2k,8)}{O(2k-1,1)\times O(7,1)}$ by taking $\psi, S_1,a_2$ to be purely imaginary. Equivalently, this is the non-linear sigma model obtained by reduction of a Euclidean four-dimensional theory. 
Denoting by $U,\phi,\zeta$  the scalar fields whose asymptotic values are given by 
$\log R, - \frac{1}{2} \log S_2$ and $a_1$,
the Lagrange density in three dimensions is
\be
\begin{split}
\cL =&
|\de U|^2 +\frac{1}{4} e^{4U} | \de \omega|^2 + | \de\phi|^2 + \frac{1}{4} e^{-4\phi}Ê| \de B 
- (  \zeta , \de A) + \tfrac{1}{2} (\zeta,\zeta) \de \omega  |^2\\
&  +\frac{1}{4} e^{2U-2\phi}  g(\de A- \zeta \de \omega, \de A- \zeta \de \omega) + \frac{1}{4} e^{-2U-2\phi}  g(\de \zeta, 
\de\zeta) + P_{a\hat{b}} \star P^{a\hat{b}} \ , 
\end{split}
\ee
where we denote $|f|^2=f \wedge \star f$, $g(F, F)\equiv F_L^a \star F_{L a}+  F^{\hat{a}}_R \star F_{R\hat{a}}$. For simplicity we shall consider only  instantons for which the electromagnetic fields vanish, $\de A=\zeta=0$. One can then write the Lagrangian as a sum of squares
\be 
\cL = \frac{1}{4} e^{4U} \Big| \star \de e^{-2U} \pm \de\omega \Big|^2 
\pm \frac{1}{2} \de ( e^{2U} \de \omega) + \frac{1}{4} e^{-4\phi}  \Big| \star \de e^{2\phi} \pm \de B\Big|^2 \pm 
\frac{1}{2} \de ( e^{-2\phi} \de B ) +  P_{a\hat{b}} \star P^{a\hat{b}} \ . 
\ee
The corresponding 1/2-BPS solutions describe $M_2$ Euclidean NS5-branes on a self-dual Taub-NUT space of charge $M_1$, with $M_1 M_2\ge 0$.\footnote{Solutions with $M_1 M_2\le 0$ exist 
but do not preserve eight supercharges.} For simplicity we consider the NS5-branes at the tip of the Taub-NUT space, with 
\be 
e^{-2U} = \frac{1}{R^2} + \frac{|M_1|}{r} \ , \quad 
e^{2\phi}  = \frac{1}{S_2} +  \frac{|M_2|}{ r} \ , \quad  \omega = -M_1
 \cos\theta\, \de \varphi \ , \quad B = - M_2 \cos\theta \, \de \varphi \ , 
\ee
and the fields $\Phi$ on the Grassmannian $G_{r-6,6}$ are uniform. The  action then reduces to the boundary term $S_{\rm cl} = 2\pi (R^2 |M_1|+S_2 |M_2|) = 2\pi |R^2 M_1+S_2 M_2|$. Note that the measure factor \eqref{Z2Rank2untwist8} vanishes for $P=0$ 
unless $M_1 M_2 \ge -1$. We shall refrain from constructing 1/2-BPS instantons with generic magnetic charge $P$ such that $D\ge 0$, although we expect that their action will reproduce $S_{\rm cl}$ in \eqref{Scl}.

\section{Discussion}

In this work, we have proposed a formula \eqref{f41loop} for the exact $(\nabla\Phi)^4$ coupling
in a class of three-dimensional string vacua obtained as freely acting orbifolds of the heterotic
string on $T^7$ under a $\IZ_N$ action with $N$ prime. Our formula is  manifestly invariant under the U-duality group $G_3(\IZ)$, which unifies the S and T-duality in $D=4$ along with Fricke duality. We derived the supersymmetric Ward identities that the exact coupling function $F_{abcd}(\Phi)$ must satisfy, and showed that the formula \eqref{f41loop} satisfies this constraint.
Furthermore, we analyzed its behavior in the weak coupling regime $g_3\to 0$ and large radius 
regime $R\to \infty$, and found that it correctly reproduces the known  tree-level and one-loop contributions in $D=3$, and the correct non-perturbative $F^4$ couplings in $D=4$.
In addition, we extracted the exponential corrections to these power-like terms in both regimes,
corresponding to non-zero Fourier coefficients with respect to parabolic subgroups $\IR^+ \times G_{2k-1,7} \ltimes \IR^{2k+6}$ and $\IR^+ \times [SL(2)/SO(2)\times G_{2k-2,6} ] \ltimes \IR^{2\times(2k+4)} \times \IR$, 
and found agreement with the expected form of the contributions of NS5-brane, Kaluza--Klein monopoles and H-monopole instantons as $g_3\to 0$, and the contributions of half-BPS dyons
and Taub-NUT instantons as $R\to \infty$. In the case of half-BPS dyons, we found a precise
match between the summation measure $\bar c_k(Q,P)$ and the helicity supertrace 
$\Omega_4(Q,P)$,
at least when the charge vector $(Q,P)$ is primitive. This vindicates the general expectation that
BPS saturated couplings in dimension $D$ encode BPS indices in dimension $D+1$. It would be interesting to determine the helicity supertrace $\Omega_4(Q,P)$ when $(Q,P)$ is
not primitive (which requires a careful treatment of threshold bound states), and compare with
the summation measure $\bar c_k(Q,P)$. 

It is natural to ask whether our formula \eqref{f4exact} is the unique solution to
 the Ward identities \eqref{wardf3} which is invariant under $G_3(\IZ)$, and reproduces the correct power-like terms in the weak coupling and large radius expansions $g_3\to 0$ and 
 $R\to \infty$.
Typically, theorems in the mathematical literature guarantee that smooth automorphic forms on $K\backslash G/G(\IZ)$ which
vanish at all cusps  and have sufficiently sparse Fourier coefficients (in mathematical terms, are attached to a sufficiently small nilpotent orbit) necessarily vanish; so that the only smooth automorphic functions satisfying to \eqref{D2scalar} are necessary Eisenstein series.
However, these theorems are typically concerned with
Chevalley subgroups of reductive groups in the split or quasi-split real form, which is not the case here ($G_3(\IZ)$ is a proper subgroup of the Chevalley group of $O(2k,8)$ for $N>1$), and smoothness away from the cusps is essential. 

As far as the support of Fourier coefficients is concerned, the Ward identities \eqref{D2scalar},  imply that the trace of the modular integral \eqref{f4exactpqtr}  is attached to the  vectorial character of $O(p,q)$,  corresponding to the next-to-minimal orbit. However, the constraints imposed by the differential equations \eqref{LinearF4}, \eqref{D2F4susy} are stronger than \eqref{D2scalar}, {\it e.g.} we show in Appendix \ref{eqDiffDecomp} that the tensor $F_{abcd}$ derived from the scalar Eisenstein series defined in Appendix \ref{sec_eispq} is not a solution to \eqref{D2F4susy}. The general form of 
the Fourier coefficients  is in fact very reminiscent of the one for automorphic forms attached to the minimal orbit of $O(p,q)$: it allows for only two power-like terms at the cusp, rather than three for the next-to-minimal orbit; they involve ordinary Bessel function of one single variable, similarly to $A_1$ Whittaker vectors, rather than more complicated functions of two variables or the typical $2A_1$ Whittaker vectors which appear in the Fourier coefficients of generic vectorial Eisenstein series \cite{Gustafsson:2014iva}. 

However, as we emphasized repeatedly, \eqref{def1looppq} has singularities in the bulk of $G_{p,q}$ on codimension $q$ loci where the projection $P_R^{\hat a}$ of a vector $P$ in $\Lambda_{p,q}$ with
norm 2 (or the projection $Q_R^{\hat a}$ of a vector $Q$ in $\Lambda_{p,q}^*$ with
norm $2/N$) vanishes.  
In order to argue for uniqueness, it is crucial
to ensure that  the modular integral \eqref{f41loop} correctly captures the behavior of the $(\nabla\Phi)^4$ coupling at all singular loci. Since \eqref{f41loop} reproduces correctly the one-loop contribution to  $(\nabla\Phi)^4$, it is clear that it correctly captures
the singular behavior on the loci associated to vectors $P,Q$ in the `perturbative Narain lattice'
$\Lambda_{r-5,7}\subset \Lambda_{r-4,8}$, at least in the weak coupling limit. Presumably,
this suffices to guarantee agreement on all singular loci, but we do not know how to prove
this rigorously.

Let us note finally that, independently of our proposed identification of the $U$-duality group in three dimensions, the general solution to the Ward identities \eqref{LinearF4}, \eqref{D2F4susy} derived in Appendix \ref{eqDiffDecomp} implies that the exact coupling must be of the form \eqref{Fexppert}, up to the determination of the measure factor $\bar c_k(Q)$. The property that we recover the exact coupling in four dimensions implies that the mesure factor is correct for null vectors by  $O(r-5,7,\IZ)$ T-duality. Indeed, for $Q^2=0$, the summation measure in \eqref{measureNS5KK} reproduces the
summation measure for NS5-brane instantons in \eqref{R2weak}. The computation of the BPS index associated to an arbitrary NS5-brane, Kaluza--Klein monopole, H-monopole instanton, would therefore give a direct proof of our result.

Clearly, it would be interesting to generalize our construction to the complete class of heterotic CHL models, whose duality properties and BPS spectrum in 4-dimensions are by now well understood.
It is natural to conjecture that the duality group in $D=3$ will still be given by the automorphism
group of the non-perturbative Narain lattice \eqref{Narainnp}, which naturally incorporates the
S and T-duality symmetries in $D=4$.  
More pressingly however, the present study was a warm-up towards the more challenging problem
of understanding the 1/4-BPS saturated coupling $\nabla^2 (\nabla\Phi)^4$ in four dimensions, which we shall address in forthcoming work.

\vspace*{1cm}


\section*{Acknowledgements}
We are grateful to Herv\'e Partouche and especially Roberto Volpato for useful discussions and correspondence.
CCH and GB  are grateful to CERN Theory Department for hospitality during part of this project.


\appendix

\section{Perturbative spectrum and one-loop $F^4$ couplings in heterotic CHL orbifolds \label{sec_F4CHL}}

In this section, we construct the one-loop vacuum amplitude in CHL models obtained as a freely acting $Z_N$-orbifold of the standard heterotic string on $T^d$ with $N$ prime. From this, we deduce the helicity supertrace for perturbative BPS states, and the one-loop  contribution to the $F^4$ and $(\nabla\Phi)^4$ couplings. We start with the simplest model with $N=2$, and then generalize the
construction to $N=3,5,7$.

\subsection{$\IZ_2$ orbifold\label{Z2PartitionFunction}}

The simplest CHL model is obtained by orbifolding the $E_8\times E_8$ heterotic string
compactified on $T^d$, by an involution $\sigma$ which exchanges the two $E_8$ gauge groups and performs a translation by half a period along one circle in $T^d$ \cite{Chaudhuri:1995bf}. This 
perturbative BPS spectrum in this model was further studied in \cite{Mikhailov:1998si,
Dabholkar:2005dt}.
The symmetry $\sigma$
exists only on a codimension $8d$ space inside the Narain moduli space $G_{d+16,d}$ and 
preserves only a $U(1)^{2d+8}$ subgroup of the original $U(1)^{2d+16}$ gauge symmetry, corresponding to the usual $2d$ Kaluza--Klein and Kalb-Ramond gauge fields, and the Cartan torus of the diagonal combination of the two $E_8$ gauge groups. To implement the quotient by $\sigma$,  it is simplest to work at the point in $G_{d+16,d}$  where the lattice factorizes as 
\be
\Lambda_{d+16,d} = E_8 \oplus E_8 \oplus \sLambda_{d,d}\ .
\ee
The integrand of the one-loop vacuum amplitude of the original heterotic string is then 
\be
\label{Ahet}
\cA = Z_{E_8\times E_8}\times \sPart{d,d} \times \frac{1}{2} \sum_{\alpha,\beta\in\{0,1\}} (-1)^{\alpha\beta+\alpha+\beta} \frac{\overline{\vartheta}^4\ar{\alpha}{\beta}}{\tau_2^4\eta^8 \overline{\eta}^{12}}
\ee
where
\be
Z_{E_8\times E_8} = \left[ \frac{\sum_{Q_1\in E_8} q^{\frac12 Q_1^2}}{\eta^8} \right]\, 
\left[ \frac{\sum_{Q_2\in E_8} q^{\frac12 Q_2^2}}{\eta^8} \right]= \frac{[E_4(\tau)]^2}{\eta^{16}}
\ee
is the partition function of the 16 chiral bosons on the $E_8\times E_8$ root lattice, and the last
factor in \eqref{Ahet} represents  the contribution of the transverse bosonic and fermionic oscillators, while  
the sum over $\alpha,\beta$ implements the GSO projection. As a consequence of space-time supersymmetry, the integral \eqref{Ahet} vanishes pointwise, but it will no longer be so in the presence of vertex operators. Note that the right-moving part in \eqref{Ahet} will not play any role in our case, and will be later replaced by an insertion of the polynomial $P_{abcd}$ \eqref{defP4}.

Following standard rules, the one-loop partition function of the orbifold by $\sigma$ is obtained
by replacing $\cA$ by a sum $\frac12 \sum_{h,g\in\{0,1\}} \cA\ar{h}{g}$,  where $\cA\ar{h}{g}$ is
obtained by twisting the boundary conditions of the fields by $\sigma^g$ along the spatial direction 
of the string, and $\sigma^h$ along the Euclidean time direction, so that $\frac12
(\cA\ar{0}{0}+\cA\ar{0}{1})$ counts $\sigma$-invariant states in the untwisted sector, while
$\frac12( \cA\ar{1}{0}+\cA\ar{1}{1})$ counts $\sigma$-invariant states in the twisted sector.
Modular invariance permutes the three blocks $\ar{0}{1}, \ar{1}{0}, \ar{1}{1}$ according to
\be\label{orbBlocks}
\cA\ar{h}{g}\left(\tfrac{a\tau+b}{c\tau+d}\right) = \cA\ar{a h+c g}{b h+g d}(\tau)
\ee
where $h,g$ are treated modulo 2. In particular, the block $\ar{0}{1}$ is invariant under
the Hecke congruence subgroup $\Gamma_0(2)$, and all other blocks
can be obtained by acting on it with elements of $SL(2,\IZ)/\Gamma_0(2)=\{1,S,ST\}$.

In the case at hand, the involution $\sigma$ exchanges $Q_1\leftrightarrow Q_2$ and the corresponding oscillators, so $\sigma$-invariant states must
have $Q_1=Q_2$ and the same oscillator state on both factors, thus 
\be
Z_{E_8\times E_8}\ar{0}{1}(\tau) = \frac{\sum_{Q\in E_8} q^{Q^2}}{\eta^8(2\tau)} \ .
\ee
The two remaining orbifold blocks are then fixed by modular covariance,
\be
\label{ZE81loop}
\begin{split}
Z_{E_8\times E_8}\ar{0}{0} &= \frac{E_4^2(\tau)}{\eta^{16}(\tau)}\ ,\quad 
Z_{E_8\times E_8}\ar{0}{1} = \frac{E_4(2\tau)}{\eta^{8}(2\tau)}\ ,\quad  \\
Z_{E_8\times E_8}\ar{1}{0} &= \frac{E_4(\tfrac{\tau}{2})}{\eta^{8}(\tfrac{\tau}{2})}\ ,\quad\,
Z_{E_8\times E_8}\ar{1}{1} = \frac{ E_4(\tfrac{\tau+1}{2})}{e^{2\I\pi/3}\eta^{8}(\tfrac{\tau+1}{2})}\ ,\quad 
\end{split}
\ee
As for the action of $\sigma$ on the torus $T^d$, it can be taken into account by replacing the partition function
$\sPart{d,d}$ by 
\be
\label{defGdd}
\sPart{d,d}\ar{h}{g} = \tau_2^{d/2} \sum_{Q \in\sLambda_{d,d}+\tfrac{h}{2}\delta} (-1)^{g\,\delta\cdot Q}\,
q^{\frac12 Q_L^2}\, \bar q^{\frac12 Q_R^2}\ .
\ee
where $\delta$ must be null modulo 2, and depends on the choice of circle $S_1$ inside $T^d$.
The resulting one-loop vacuum amplitude is then the modular integral of 
\be
\label{Ahetorb}
\cA_{\rm orb} =
\frac12 \sum_{h,g\in\{0,1\}} Z_{E_8\times E_8}\ar{h}{g}\,\sPart{d,d}\ar{h}{g}
\times
\frac{1}{2}\sum_{\alpha,\beta\in\{0,1\}} (-1)^{\alpha\beta+\alpha+\beta} \frac{\overline{\vartheta}^4\ar{\alpha}{\beta}}{\tau_2^4 \eta^8 \overline{\eta}^{12}},
\ee
where the one-half factor is explained above \eqref{orbBlocks}. Now, a key observation is that the numerator in the
 blocks $Z_{E_8\times E_8}\ar{h}{g}$ for $(h,g)\neq (0,0)$ 
can be written as a partition functions for the lattice $\Lambda=E_8[2]$ and for its dual $\Lambda^*=E_8[1/2]$, 
\be
\label{ZE81loop2}
\begin{split}
Z_{E_8\times E_8}\ar{0}{1}=&\frac{1}{\eta^{8}(2\tau)} \sum_{Q\in E_8[2]} q^{\tfrac12 Q^2}\\
Z_{E_8\times E_8}\ar{1}{0}=&\frac{1}{\eta^{8}(\tfrac{\tau}{2})} \sum_{Q\in E_8[1/2]} q^{\tfrac12 Q^2}\\
Z_{E_8\times E_8}\ar{1}{1}=&\frac{1}{e^{2\I\pi/3}\eta^{8}(\tfrac{\tau+1}{2})} \sum_{Q\in E_8[1/2]} (-1)^{Q^2} q^{\tfrac12 Q^2}\ .
 \end{split}
\ee
Moreover, the untwisted, unprojected partition function  satisfies
\be
\label{ZE81loop3}
\begin{split}
Z_{E_8\times E_8}\ar{0}{0} =  
&
 \frac{E_4(2\tau)}{\eta^{8}(2\tau)}+  \frac{E_4(\tfrac{\tau}{2})}{\eta^{8}(\tfrac{\tau}{2})}
 + \frac{ E_4(\tfrac{\tau+1}{2})}{e^{2\I\pi/3}\eta^{8}(\tfrac{\tau+1}{2})}
\\
= & Z_{E_8\times E_8}\ar{0}{1} +
Z_{E_8\times E_8}\ar{1}{0}+Z_{E_8\times E_8}\ar{1}{1}\ .
\end{split}
\ee
This relation can be checked using the explicit form of the blocks $Z_{E_8\times E_8}\ar{0}{1}$, but more conceptually, it follows by decomposing $Z_{E_8\times E_8}\ar{0}{0}$, the character of 
the level 1 representation of $\hat E_8 \oplus \hat E_8$, into characters of level 2 representations
of the diagonal $\hat E_8$ \cite{Forgacs:1988iw}.
It follows from \eqref{ZE81loop2}, \eqref{ZE81loop3} that the one-loop amplitude \eqref{Ahetorb} can be written as 
\be
\label{Ahetorb2}
\cA_{\rm orb} =
\frac{1}{2}\sum'_{h,g\in\{0,1\}} \frac{\widetilde\Gamma_{d+8,d}\ar{h}{g}}{\Delta_8\ar{h}{g}}
\times
\frac{1}{2}
\sum_{\alpha,\beta\in\{0,1\}} (-1)^{\alpha\beta+\alpha+\beta} \frac{\overline{\vartheta}^4\ar{\alpha}{\beta}}
{\tau_2^4 \overline{\eta}^{12}}
\ee
where the sum over $(h,g)$ no longer includes $(0,0)$. Here, we defined the eta products
\be
\label{MasonZ2}
\begin{split}
 \Delta_8\ar{0}{1}=&\eta^{8}(\tau) \eta^8(2\tau) = 2^{-4}\eta^{12} \vartheta_2^4 \equiv \Delta_8(\tau) \\
\Delta_8\ar{1}{0}=&\eta^{8}(\tau) \eta^8(\tfrac{\tau}{2})= \eta^{12} \vartheta_4^4 = \Delta_8(\tfrac{\tau}{2}), \\
\Delta_8\ar{1}{1}=&e^{2\I\pi/3}\,\eta^{8}(\tau) \eta^8(\tfrac{\tau+1}{2}) = -\eta^{12} \vartheta_3^4
= \Delta_8(\tfrac{\tau+1}{2}),\ ,
 \end{split}
 \ee
 satisfying
 \be
 \Delta_8\ar{0}{1}(-1/\tau)=2^{-4} \tau^8\, \Delta_8\ar{1}{0}(\tau)\ ,\quad   
 \Delta_8\ar{1}{0}(\tau+1)= \Delta_8\ar{1}{1}(\tau)\ ,
 \ee
 and the partition functions $\widetilde\Gamma_{d+8,d}$ are defined over $\tilde\Lambda_{d+8,d}=E_8[2]\oplus \sLambda_{d,d}$
 and its dual  $\tilde\Lambda_{d+8,d}^*=E_8[1/2]\oplus \sLambda_{d,d}$:
\be
\label{ZE81loop4}
\begin{split}
\widetilde\Gamma_{d+8,d}\ar{0}{1}=&\tau_2^{d/2} \sum_{Q\in \tilde\Lambda_{d+8,d}} 
\big[1+(-1)^{\delta\cdot Q}\big]\, q^{\tfrac12 Q_L^2} \bar q^{\tfrac12 Q_R^2}
 \\
\widetilde\Gamma_{d+8,d}\ar{1}{0}=&\tau_2^{d/2}
\Big[ \sum_{Q\in \tilde\Lambda_{d+8,d}^*}  +\sum_{Q\in \tilde\Lambda_{d+8,d}^*+\frac12\delta} 
\Big]
q^{\tfrac12 Q_L^2} \bar q^{\tfrac12 Q_R^2}
\\
\widetilde\Gamma_{d+8,d}\ar{1}{1}=&\tau_2^{d/2}
\Big[ \sum_{Q\in \tilde\Lambda_{d+8,d}^*}  +\sum_{Q\in \tilde\Lambda_{d+8,d}^*+\frac12\delta} 
\Big] (-1)^{Q^2} q^{\tfrac12 Q_L^2}  \bar q^{\tfrac12 Q_R^2}
 \end{split}
\ee
These relations were derived at the special point where the lattice $\tilde\Lambda_{d+8,d}$ is factorized,
but it is now clear that they hold at arbitrary points on the moduli space $G_{d+8,d}\subset G_{d+16,d}$ where the $\IZ_2$ symmetry exists. 

Choosing $\delta=(0^d;0^{d-1},1)$, so that the involution $\sigma$ acts by 
a translation along the $d$-th circle by a half period, this can be further written as
\be
\label{ZE81loop5}
\begin{split}
\Part{d+8,d} &\equiv 
\tau_2^{d/2} \sum_{Q\in \Lambda_{d+8,d}}  q^{\tfrac12 Q_L^2} \bar q^{\tfrac12 Q_R^2} =\frac12
\widetilde\Gamma_{d+8,d}\ar{0}{1}
 \\
(2^5/\tau^4)\Part{d+8,d}(-1/\tau)=\PartD{d+8,d} &\equiv
\tau_2^{d/2} \sum_{Q\in \Lambda_{d+8,d}^*}  q^{\tfrac12 Q_L^2} \bar q^{\tfrac12 Q_R^2}
= \widetilde\Gamma_{d+8,d}\ar{1}{0}
\\
\PartD{d+8,d}\big[(-1)^{Q^2}\big] &\equiv 
\tau_2^{d/2}  \sum_{Q\in \Lambda_{d+8,d}^*}  (-1)^{Q^2} q^{\tfrac12 Q_L^2} \bar q^{\tfrac12 Q_R^2} =\widetilde\Gamma_{d+8,d}\ar{1}{1}
 \end{split}
\ee
where $\Lambda_{d+8,d}$ is related to $\tilde\Lambda_{d+8,d}$ by rescaling a $\sLambda_{1,1}$ 
summand\footnote{Note that this rescaling  implies an extra volume factor upon Poisson resummation, 
namely $\PartD{d+8,d}(\tau)=(2^5/\tau^4)\Part{d+8,d}(-1/\tau)$.},
\be
\Lambda_{d+8,d}= E_8[2]\oplus \sLambda_{1,1}[2]\oplus \sLambda_{d-1,d-1} \ .
\ee
Here $\sLambda_{1,1}[2]$ is the usual sum over momentum $m_d$ and winding $n_d$, with
$m_d$ running only over even integers. The dual lattice is 
\be
\label{Lambdae}
\Lambda_{d+8,d}^* = E_8[1/2]\oplus \sLambda_{1,1}[1/2]\oplus \sLambda_{d-1,d-1} \ ,
\ee
where $\sLambda_{1,1}[1/2]$ is the usual sum over momentum $m_d$ and winding $n_d$, with
$n_d$ running over $\IZ/2$. For $d=6$, since $\Lambda_{14,6}\subset  \Lambda_{14,6}^*$, we 
see that the electric charges carried by excitations of the heterotic string lie in the lattice 
$\Lambda_e=\Lambda_{14,6}^*$, in agreement with the result stated 
in Table \ref{TableauCHL}. Moreover, it is apparent
that the degeneracy of perturbative BPS states with charge $Q\in\Lambda_{d+8,d}^*,\, Q\notin \Lambda_{d+8,d}$ in the twisted sector is given by the coefficient of 
$q^{-Q^2/2}$ in $1/\Delta_8\ar{1}{0}=1/\Delta_8(\tau/2)$, or equivalently the coefficient of 
$q^{-Q^2}$ in $1/\Delta_8$, while the degeneracy of perturbative BPS states with
charge $Q\in\Lambda_{d+8,d} \subset  \Lambda_{d+8,d}^*$ has an additional 
contribution from the coefficient of 
$q^{-Q^2/2}$ in $1/\Delta_8$, in agreement with \eqref{om4pert_tw} and
\eqref{om4pert_utw}, and the analysis in \cite{Mikhailov:1998si,Dabholkar:2005dt}.

At last, we can turn to the one-loop $F^4$ amplitude in this model. As is the case in the usual heterotic string, the insertion of four vertex operators replaces the right-moving contribution in the vacuum
amplitude \eqref{Ahetorb2} by an insertion of the polynomial $P_{abcd}$ in \eqref{defP4}. Thus, we
get
\be
F_{abcd}^{\mbox{\tiny (1-loop)}} = 
\RN
\int_{SL(2,\IZ)\backslash\cH}\! \!\!\frac {\de\tau_1\de\tau_2}{\tau_2^{\, 2}} \sum_{\gamma\in \Gamma_0(2)\backslash SL(2,\IZ)}\left. \frac{\Part{d+8,d}[P_{abcd}]}
{\Delta_8}\right|_{\gamma}\ ,
\ee
where $\Part{d+8,d}[P_{abcd}]$ denotes the lattice partition function 
$\Part{d+8,d}\ar{h}{g}$ with an insertion of the polynomial $P$ as in \eqref{defZpq1}.
Equivalently, we can unfold the integral over a fundamental domain $\Gamma_0(2)\backslash\cH$
for the action of $ \Gamma_0(2)$ on $\cH$, at the expense of keeping only the identity in the 
sum over cosets,
\be
F_{abcd}^{\mbox{\tiny (1-loop)}} =
\RN
\int_{\Gamma_0(2)\backslash\cH}\! \!\!\frac {\de\tau_1\de\tau_2}{\tau_2^2}
\frac{\Part{d+8,d}[P_{abcd}]}
{\Delta_8}\ ,
\ee
which demonstrates \eqref{f41loop} in this case. 

\subsection{$\IZ_N$ orbifold with $N=3,5,7$ \label{ZNPartitionFunction}}

The construction detailed in the previous section can be easily generalized to $\IZ_N$ orbifolds,
provided one can find a point in the moduli space $G_{d+16,d}$ where $\IZ_N$ acts on 
the lattice $\Lambda_{d+16,d}$ by a permutation with cycle shape $1^k N^k$.  It turns out
that for $N=3,5,7$, such a lattice can be obtained by applying a Wick rotation on the
Niemeier lattices $D_6^4$, $D_4^6$ and $D_3^8$, respectively. Indeed, recall that
given an even self-dual Euclidean lattice 
\be
\Lambda=\cup_{(\lambda,\lambda')\in \cG} (D_k+\lambda) \oplus (\Lambda'+\lambda')
\ee
of dimension $n$, where the glue code $\cG$ is a given sublattice of  $D_k^*/D_k \oplus \Lambda'^*/\Lambda'$, one can obtain an even self-dual lattice of dimension $n-8$, by replacing $D_k$ by $D_{k-8}$, while keeping the same glue code $\cG$, using the fact that $\cG_k= D_k^*/D_k$ 
is invariant under $k\mapsto k-8$\footnote{Indeed, $\cG_k=\IZ_2\oplus \IZ_2$ is $k$ is
even, or $\IZ_4$ is $k$ is odd, with the 4 elements in one-to-one correspondence with the 
highest weights $0,s,v,c$ of the adjoint, spinor, vector
and conjugate spinor representations.},
\be
\hat\Lambda=\cup_{(\lambda,\lambda')\in \cG} (D_{k-8}+\lambda) \oplus (\Lambda'+\lambda')\ .
\ee
If $1\leq k<8$, then $D_{k-8}$ should be understood as $D_{8-k}[-1]$, so that
the new lattice is a Lorentzian
lattice with signature $(n-k, 8-k)$ \cite[\S A.4]{Lerche:1998nx}. In this way, starting from
the Niemeier lattice $\Lambda=D_k^{N+1}$ for $N=3,5,7$, which is symmetric under 
cyclic permutations of the $N+1$ $D_k$ factors,  we obtain an even self-dual lattice 
$\hat\Lambda=D_k^N \oplus D_{8-k}[-1]$ of signature $(Nk,8-k)$ with a $\IZ_N$ symmetry $\sigma$ acting by cyclic permutations of the $N$ $D_k$ factors. Using the explicit description of the
glue code for Niemeier lattices given in \cite[Table 16.1]{conway2013sphere}, it is possible to check that the only elements
$(\lambda_1,\dots \lambda_{N+1})$ in the
glue code $\cG\subset \cG_k^{N+1}$ which are invariant under $\IZ_N$ are those
of the form $(\lambda,\dots, \lambda)$ with $\lambda$ running over $\cG_k$. The partition
function of the lattice 
$\hat\Lambda$ with an insertion of the element $\sigma^g$ with $g\neq 0 \mod N$ is thus
\be
\label{ZNiemWick}
\begin{split}
Z_{\hat\Lambda}\ar{0}{g} =&  
\tfrac{\vartheta_3^k+\vartheta_4^k}{2\eta^k}(N\tau) 
\overline{\tfrac{\vartheta_3^{8-k}+\vartheta_4^{8-k}}{2\eta^{8-k}}}
+ \tfrac{\vartheta_3^k-\vartheta_4^k}{2\eta^k}(N\tau) \overline{\tfrac{\vartheta_3^{8-k}-\vartheta_4^{8-k}}{2\eta^{8-k}}} 
\\ & 
+\tfrac{\vartheta_2^k+\vartheta_1^k}{2\eta^k}(N\tau) \overline{\tfrac{\vartheta_2^{8-k}+\vartheta_1^{8-k}}{2\eta^{8-k}}}
+ \tfrac{\vartheta_2^k-\vartheta_1^k}{2\eta^k}(N\tau) \overline{\tfrac{\vartheta_2^{8-k}-\vartheta_1^{8-k}}{2\eta^{8-k}}}  \ .
\end{split}
\ee
The other blocks are obtained by modular covariance, leading for $h\neq 0 \mod N$ to
\be
\begin{split}
Z_{\hat\Lambda}\ar{h}{0} =&  
\tfrac{\vartheta_3^k+\vartheta_2^k}{2\eta^k}\big(\frac{\tau}{N}\big) 
\overline{\tfrac{\vartheta_3^{8-k}+\vartheta_2^{8-k}}{2\eta^{8-k}}}
+ \tfrac{\vartheta_3^k-\vartheta_2^k}{2\eta^k}\big(\frac{\tau}{N}\big) 
\overline{\tfrac{\vartheta_3^{8-k}-\vartheta_2^{8-k}}{2\eta^{8-k}}}
 \\ & 
+\tfrac{\vartheta_4^k+\vartheta_1^k}{2\eta^k}\big(\frac{\tau}{N}\big) 
\overline{\tfrac{\vartheta_4^{8-k}+\vartheta_1^{8-k}}{2\eta^{8-k}}}
+ \tfrac{\vartheta_4^k-\vartheta_1^k}{2\eta^k}\big(\frac{\tau}{N}\big) 
\overline{\tfrac{\vartheta_4^{8-k}-\vartheta_1^{8-k}}{2\eta^{8-k}}}\ ,
\end{split}
\ee
while the remaining blocks with $g\neq 0 \mod N$ follow by acting with $\tau\to\tau+1$,
\be
Z_{\hat \Lambda}\ar{h}{g}(\tau) =  Z_{\hat \Lambda}\ar{h}{0}(\tau+gh^{-1})
\ee
where $h^{-1}$ is the inverse of $h$ in the multiplicative group $\IZ_N$. The untwisted, unprojected
block is then 
\be
\label{Z00sum}
\begin{split}
Z_{\hat\Lambda}\ar{0}{0} =
    Z_{\hat\Lambda}\ar{0}{1} + \sum_{g=0}^{N-1} Z_{\hat\Lambda}\ar{1}{g}\ ,
\end{split}
\ee
{\it i.e.} a sum over images of  $Z_{\hat\Lambda}\ar{0}{1}$
under $\Gamma_0(N) \backslash SL(2,\IZ)=\{1,S,TS,\dots, T^{N-1} S\}$. 
 As a consistency check, one can verify that the analogous sum
for the Euclidean lattice $\Lambda$ reproduces the partition function of the Niemeier lattice,
\be
\begin{split}
\frac{\Theta_{D_k^{N+1}}}{\eta^{24}} & =
  Z_{\Lambda}\ar{0}{1} + \sum_{g=0}^{N-1} Z_{\Lambda}\ar{1}{g}
= \frac{E_4^3}{\eta^{24}} + 48 k - 768\ ,
\end{split}
\ee
where $Z_{\Lambda}\ar{0}{1}$ is obtained by replacing $
\overline{\vartheta_i^{8-k}/\eta^{8-k}}$
by $(\vartheta_i/\eta)^k$ in  \eqref{ZNiemWick}. 

The integrand of the one-loop vacuum amplitude follows in the same way as in the previous
subsection, by combining the orbifold blocks $Z_{\hat\Lambda}\ar{h}{g}(\tau)$ for the lattice
$\hat\Lambda$ with the shifted partition function for the remaining $d-8+k$ compact directions (where $d$ is assumed to be greater that $8-k$)
\be
\label{defGdd2}
\Gamma_{d-8+k,d-8+k}\ar{h}{g} = \tau_2^{\tfrac{d-8+k}{2}} \sum_{Q \in\Lambda_{d-8+k,d-8+k}+\tfrac{h}{N}\delta} (-1)^{\frac{2}{N}g\,\delta\cdot Q}\,
q^{\tfrac12 Q_L^2}\, \bar q^{\tfrac12 Q_R^2}\ .
\ee
After eliminating $Z_{\hat\Lambda}\ar{0}{0}$ using \eqref{Z00sum}, grouping
terms into an orbit of $\Gamma_0(N)\backslash SL(2,\IZ)$, and rescaling a $\sLambda_{1,1}$ factor in $\Lambda_{d+2k-8,d}$ as\footnote{Note that this rescaling implies $\PartD{d+2k-8,d}(\tau)=(N^{\frac{k}{2}+1}/\tau^{k-4})\Part{d+2k-8,d}(-1/\tau)$.}
\be\label{ZNlattice}
\Lambda_{d+2k-8,d}= D_k[N]\oplus D_{8-k}[-1] \oplus 
\sLambda_{1,1}[N] \oplus \sLambda_{d+k-9,d+k-9}  \ ,
\ee
with a glue code $\{(0,0), (s,s), (v,v), (c,c)\}$ for the first two factors, we find
\be
\label{Ahetorb2N}
\cA_{\rm orb} =
\left[ \frac{\Part{d+2k-8,d}}{\Delta_k\ar{0}{1}} + \frac{1}{N}\sum_{g=0}^{N-1} 
\frac{\PartD{d+2k-8,d}[(-1)^{gQ^2}]}{\Delta_k\ar{1}{g}}
\right]
\times
\frac{1}{2}
\sum_{\alpha,\beta\in\{0,1\}} (-1)^{\alpha\beta+\alpha+\beta} \frac{\overline{\vartheta}^4\ar{\alpha}{\beta}}{\tau_2^4\overline{\eta}^{12}}
\ee
where we defined the eta products 
\be\label{etaProducts}
\Delta_k\ar{0}{1} = \eta(\tau)^{k}\, \eta(N\tau)^{k}\ ,\quad 
\Delta_k\ar{1}{g} = e^{\frac{\I \pi gk}{12} } \,  \eta(\tau)^{k}\, \eta\big(\tfrac{\tau+g}{N}\big)^{k}\,
\ee
and
\be\label{partitionFunctionRepresentative}
\begin{split}
\Part{d+2k-8,d}
&= 
\tau_2^{\frac{d}{2}} \sum_{Q\in \Lambda_{d+2k-8,d}}  q^{\tfrac12 Q_L^2} \bar q^{\tfrac12 Q_R^2}
\\
\PartD{d+2k-8,d}\big[(-1)^{gQ^2}\big] 
&= 
\tau_2^{\tfrac{d}{2}} \sum_{Q\in \Lambda_{d+2k-8,d}^*}  (-1)^{gQ^2} \, 
q^{\tfrac12 Q_L^2} \bar q^{\tfrac12 Q_R^2} \ .
\end{split}
\ee

From this description, it is apparent
that the degeneracy of twisted perturbative BPS states with charge $Q\in\Lambda_{d+2k-8,d}^*,\, Q\notin\Lambda_{d+2k-8,d}$  is given by the coefficient of 
$q^{-Q^2/2}$ in $1/\Delta_k\ar{1}{0}=1/\Delta_k(\tau/N)$, or equivalently the coefficient of 
$q^{-NQ^2/2}$ in $1/\Delta_k$, while the degeneracy of perturbative BPS states with
charge $Q\in  \Lambda_{d+2k-8,d} \subset \Lambda_{d+2k-8,d}^*$ has an additional 
contribution from the coefficient of 
$q^{-Q^2/2}$ in $1/\Delta_k$, in agreement with \eqref{om4pert_tw} and
\eqref{om4pert_utw}. For four-dimensional vacua $(d=6)$, we see that
 the electric charges  carried by perturbative BPS states lie in the lattice $\Lambda_e=\Lambda_m^*$ where 
\be
\begin{split}
N=3:  &\quad \Lambda_m = D_6[3] \oplus D_2[-1] \oplus \sLambda_{1,1}[3]\oplus \sLambda_{3,3}
\\
N=5: &\quad \Lambda_m =  D_4[5] \oplus D_4[-1] \oplus \sLambda_{1,1}[5]\oplus\sLambda_{1,1}
\\
N=7: &\quad \Lambda_m = D_3[7] \oplus D_5[-1]\oplus \sLambda_{1,1}[7]
\end{split}
\ee
This is in fact in agreement with the results stated in Table \ref{TableauCHL}, thanks to 
the isomorphisms
\be
\label{latiso}
\begin{split}
D_6[3] \oplus D_2[-1] &\simeq A_2\oplus A_2 \oplus \sLambda_{2,2}[3] \\
D_4[5]\oplus D_4[-1] &\simeq \sLambda_{2,2}[5] \oplus \sLambda_{2,2} 
\\
D_3[7]\oplus D_5[-1] &\simeq {\scriptsize \begin{pmatrix} -4 & -1 \\ -1 & -2 \end{pmatrix}} \oplus \sLambda_{1,1}[7] \oplus \sLambda_{2,2}  
\end{split}
\ee
Indeed, both lattices on each line have the same genus, in particular the same discriminant
group $L^*/L=\IZ_N^k$. For $N=2$ (hence $k=8$), Eq. \eqref{ZNlattice} continues to hold with the understanding that 
$D_8[2]\oplus D_0[-1]\equiv E_8[2]$.

Finally, we can obtain the one-loop $F^4$ amplitude by replacing the last factor 
in \eqref{Ahetorb2N} by   an insertion of the polynomial $P_{abcd}$ in \eqref{defP4}, and integrating over the fundamental domain $ \cH / SL(2,\IZ)$. As before, the integral can be unfolded onto a fundamental domain $\Gamma_0(N)\backslash\cH$ for the action of $ \Gamma_0(N)$ on $\cH$, at the expense of keeping only the block $\ar{0}{1}$,
\be
F_{abcd}^{\mbox{\tiny (1-loop)}} = \RN
\int_{\Gamma_0(N)\backslash\cH}\! \!\!\frac {\de\tau_1\de\tau_2}{\tau_2^{2}}  
\frac{\Part{d+2k-8,d}[P_{abcd}]}
{\Delta_k}\ ,
\ee
where $\Delta_k\equiv \Delta_k\ar{0}{1}$, 
thus establishing  \eqref{f41loop} for this class of models.

\section{Ward identity in the degeneration $O(p,q)\to O(p-1,q-1)$}
\label{eqDiffDecomp}

In section \ref{modularIntegral}, we proved that the differential equations \eqref{LinearF4} and  \eqref{EqD2F} are satisfied by the one-loop modular integral $F_{abcd}$ defined in \eqref{def1looppq}. 
Here, we verify explicitly that the differential equation in \eqref{EqD2F}  is verified by each Fourier mode
in the degeneration limit  $O(p,q)\to O(p-1,q-1)$, and that the solution is uniquely determined
up to a moduli-independent summation measure.  

Using the decomposition \eqref{sopqDec1} and changing variable $R=e^{-\phi}$ for the non-compact Cartan generator of $O(p,q)$, the metric on moduli space reads

\be 
2 P_{a\hat{b}} P^{a\hat{b}} = 2 \de \phi^2+ 2 P_{\alpha\hat{\beta}} P^{\alpha\hat{\beta}}+ e^{2\phi} \bigl(  p_{L\,\alpha\,I}p_{L}\_^{\alpha}\__J+p_{R\,\hat\alpha\,I}p_{R}\_^{\hat\alpha}\__J \bigr)\,\de a^I\,\de a^J 
\ee
with 
\be 
P_{0\hat{0}} = -  \de \phi \ , \qquad P_{0\hat{\alpha}} = \frac{1}{\sqrt{2}} e^{\phi}p_{R\,\hat\alpha\,I}\,\de a^I ,\, \qquad P_{\alpha\hat{0}} = \frac{1}{\sqrt{2}} e^{\phi}p_{L\,\alpha\,I}\,\de a^I \  .  
\ee 
Beware that in this section we use the same notations $p_L$ and $p_R$ for both $O(p,q)$ and $O(p-1,q-1)$, so $p_{L\,\alpha\,I}Q^I$ is not  $p_{L\,a\,\cI}Q^\cI$ for $a=\alpha$. 

One can compute the covariant derivative in tangent frame such that 
\be 
\de Z_a = 2 P^{b\hat{c}}  \partial_{b\hat{c}} Z_a = 2 P^{b\hat{c}}  ( \cD_{b\hat{c}} Z_a - B_{b\hat{c}\, a}{}^d Z_d ),  
\label{spinConnEquation}
\ee
 and similarly for hatted indices. This way one computes that, for any tensor $F_a=(F_0,\,F_\alpha,\,F_{\hat \alpha},\,F_{\hat 0})$, $F_b=(F_0,\,F_\beta,\,F_{\hat \beta},\,F_{\hat 0})$, ...
\begin{eqnarray} \cD_{0\hat{0}} F_a &=& - \frac{1}{2} \frac{\partial\, }{\partial \phi} F_a  \ , 
 \nonumber\\
 \cD_{\alpha\hat{0}} F_a &=& \frac{1}{\sqrt{2}}e^{- \phi} v^{\mbox{\tiny-1} I}{}_\alpha \frac{\partial\, }{\partial a^I} F_a  + \frac{1}{2}\left( F_\alpha, - \delta_{\alpha\beta} F_0,0,0\right) 
 \nonumber \\
\cD_{0\hat{\alpha}} F_a &=& \frac{1}{\sqrt{2}} e^{- \phi}v^{\mbox{\tiny-1} I}{}_{\hat{\alpha}}  \frac{\partial\, }{\partial a^I} F_b  + \frac{1}{2}\left( 0,0, - \delta_{\alpha\beta} F_{\hat{0}} , F_{\hat{\alpha}} \right)\ ,
\end{eqnarray} 
and finally the operator $\cD_{\alpha\beta}$ will only be acting on the moduli fields through the projectors $p^I_{L\,\gamma}$, $p^I_{R\,\hat\gamma}$:
\be
\cD_{\alpha\hat{\beta}} p_L^I\__\gamma= \tfrac{1}{2}\delta_{\alpha\gamma}p_R^I\__{\hat\beta},\qquad\cD_{\alpha\hat{\beta}} p_L^I\__{\hat\gamma}= \tfrac{1}{2}\delta_{\hat\beta\hat\gamma}p_R^I\__{\alpha}\ .
\ee

Recall the differential equation \eqref{wardf3}
\be 
\mathcal{D}_{(e}^{\hat{c}} \mathcal{D}_{f)\,\hat{c}} F_{abcd} = \frac{2-q}{4} \delta_{ef} F_{abcd} + (4-q) \delta_{a)(e} F_{f)(bcd} + 3 \delta_{(ab} F_{cd)ef} \  .  
\ee
For brevity we define the vector $\vec{F}$ 
\be 
\vec{F} = \left(F_{1111},F_{111\alpha},F_{11\alpha\beta},F_{1\alpha\beta\gamma},F_{\alpha\beta\gamma\delta}\right)^\intercal   
\ee
and $\vec{F}_Q$ such that  $\vec{F} = \sum_Q \vec{F}_Q e^{2\pi\I Q \cdot a} $.  
The first component $(e,f)=(0,0)$gives 
\be 
\label{firstDiffEq} 4  \mathcal{D}_0{}^{\hat{c}} \mathcal{D}_{0\hat{c}}\vec{F}_Q =(\partial_\phi ( \partial_\phi +q-1) -8\pi^2 e^{-2\phi} Q_R^2 )  \vec{F}_Q = -\left( \begin{array}{c} 5(q-6)F_{1111} \\
4(q-5)F_{111\alpha}\\
3(q-4)F_{11\alpha\beta}-2 \delta_{\alpha\beta}F_{1111}  \\
2(q-3)F_{1\alpha\beta\gamma}-6 \delta_{(\alpha\beta}F_{111\gamma)}\\
(q-2)F_{\alpha\beta\gamma\delta}-12\delta_{(\alpha\beta}F_{11\gamma\delta)}  \end{array}\right) \  .  
\ee
Then the action of the differential operator 
\begin{eqnarray}  2  \mathcal{D}_0{}^{\hat{c}} \mathcal{D}_{\eta\hat{c}}\vec{F}_Q+2  \mathcal{D}_\eta{}^{\hat{c}} \mathcal{D}_{0\hat{c}}\vec{F}_Q &=&- 2\pi\I \sqrt{2} e^{-\phi} (Q_{L\,\eta} ( \partial_\phi + q-2) + 2 Q_{R\,\hat\alpha} {\cal D}_\eta{}^{\hat{\alpha}} ) \vec{F}_Q
\nonumber\\
\,&&- (\partial_\phi + \frac{q-2}{2}) \left( \begin{array}{c} 4F_{111\eta} \\
3F_{11\alpha\eta}- \delta_{\eta\alpha} F_{1111} \\
2F_{1\alpha\beta\eta}- 2\delta_{\eta(\alpha} F_{111\beta)} \\
F_{\alpha\beta\gamma\eta}- 3\delta_{\eta(\alpha} F_{11\beta\gamma)}\\
- 4\delta_{\eta(\alpha} F_{1\beta\gamma\delta)} \end{array}\right) \  ,\qquad   \end{eqnarray} 
allows to obtain the second component $(e,f)=(0,\alpha)$ of the differential equation 
\begin{eqnarray} 
\label{secondDiffEq}&& -2\pi\I \sqrt{2} e^{-\phi} (Q_{L\,\eta} ( \partial_\phi + q-2) + 2 Q_{R\,\hat\alpha}  {\cal D}_\eta{}^{\hat{\alpha}} ) \vec{F}_Q 
\nn\\
&=&  \left( \begin{array}{c} 4( \partial_\phi + 4) F_{111\eta} \\
3( \partial_\phi + 3)F_{11\alpha\eta}- \delta_{\eta\alpha}( \partial_\phi +q-3)  F_{1111} \\
2( \partial_\phi + 2)F_{1\alpha\beta\eta}- 2\delta_{\eta(\alpha}( \partial_\phi +q-3)   F_{111\beta)}+ 2 \delta_{\alpha\beta} F_{111\eta}  \\
( \partial_\phi + 1)F_{\alpha\beta\gamma\eta}- 3\delta_{\eta(\alpha} ( \partial_\phi +q-3)  F_{11\beta\gamma)}+ 6 \delta_{(\alpha\beta} F_{11\gamma)\eta} \\
- 4\delta_{\eta(\alpha} ( \partial_\phi +q-3)  F_{1\beta\gamma\delta)}+ 12 \delta_{(\alpha\beta} F_{1\gamma\delta)\eta} \ \end{array}\right) \  . \quad \end{eqnarray} 
The final differential operator for $(e,f)=(\eta,\vartheta)$
\begin{eqnarray}  
4 \mathcal{D}_{(\eta}{}^{\hat{c}} \mathcal{D}_{\vartheta)\hat{c}}\vec{F}_Q&=& (4 \mathcal{D}_{(\eta}{}^{\hat{\gamma}} \mathcal{D}_{\vartheta)\hat{\gamma}} + \delta_{\eta\vartheta} \partial_\phi - 8\pi^2 e^{-2\phi} Q_{L\,\eta}  Q_{L\,\vartheta}   )\vec{F}_Q 
\nn\\
&&+ 4\pi\I \sqrt{2} e^{-\phi} Q_{L\,(\eta}  \left( \begin{array}{c} 4F_{111\vartheta)} \\
3F_{11\alpha|\vartheta)}- \delta_{\vartheta)\alpha} F_{1111} \\
2F_{1\alpha\beta|\vartheta)}- 2\delta_{\vartheta)(\alpha} F_{111\beta)} \\
F_{\alpha\beta\gamma|\vartheta)}- 3\delta_{\vartheta)(\alpha} F_{11\beta\gamma)}\\
- 4\delta_{\vartheta)(\alpha} F_{1\beta\gamma\delta)} \end{array}\right) 
\\
&& + \left( \begin{array}{c} 12F_{11\eta\vartheta}-4 \delta_{\eta\vartheta} F_{1111}  \\
6F_{1\alpha\eta\vartheta}-3\delta_{\eta\vartheta} F_{111\alpha} - 7 \delta_{\alpha(\eta} F_{111\vartheta)}   \\
2F_{\alpha\beta\eta\vartheta}-2\delta_{\eta\vartheta} F_{11\alpha\beta} - 10 \delta_{\alpha)(\eta} F_{11\vartheta)(\beta}+2 \delta_{\alpha)(\eta} \delta_{\vartheta)(\beta} F_{1111} \\
-\delta_{\eta\vartheta} F_{1\alpha\beta\gamma} - 9 \delta_{\alpha)(\eta} F_{1\vartheta)(\beta\gamma}+6 \delta_{\alpha)(\eta} \delta_{\vartheta)(\beta} F_{111\gamma}\\
- 4 \delta_{\alpha)(\eta} F_{\vartheta)(\beta\gamma\delta}+12 \delta_{\alpha)(\eta} \delta_{\vartheta)(\beta} F_{11\gamma\delta}\end{array}\right)\nn \  ,    
\end{eqnarray} 
gives a third differential equation 
\begin{multline} \label{thirdDiffEq}(4 \mathcal{D}_{(\eta}{}^{\hat{\gamma}} \mathcal{D}_{\vartheta)\hat{\gamma}} + \delta_{\eta\vartheta} \partial_\phi - 8\pi^2 e^{-2\phi}Q_{L\,\eta}  Q_{L\,\vartheta}  )\vec{F}_Q + 4\pi\I \sqrt{2} e^{-\phi} Q_{L\,(\eta}  \left( \begin{array}{c} 4F_{111\vartheta)} 
\\
3F_{11\alpha|\vartheta)}- \delta_{\vartheta)\alpha} F_{1111} \\
2F_{1\alpha\beta|\vartheta)}- 2\delta_{\vartheta)(\alpha} F_{111\beta)} \\
F_{\alpha\beta\gamma|\vartheta)}- 3\delta_{\vartheta)(\alpha} F_{11\beta\gamma)}\\
- 4\delta_{\vartheta)(\alpha} F_{1\beta\gamma\delta)} \end{array}\right) 
\\
= - \left( \begin{array}{c} (q-6) \delta_{\eta\vartheta} F_{1111}  \\
 (q-5) \delta_{\eta\vartheta} F_{111\alpha} +(q-11) \delta_{\alpha(\eta} F_{111\vartheta)}   \\
 (q-4) \delta_{\eta\vartheta} F_{11\alpha\beta} - 2 \delta_{\alpha\beta} F_{11\eta\vartheta} +2(q-9) \delta_{\alpha)(\eta} F_{11\vartheta)(\beta}+2 \delta_{\alpha)(\eta} \delta_{\vartheta)(\beta} F_{1111} \\
 (q-3) \delta_{\eta\vartheta} F_{1\alpha\beta\gamma}- 6\delta_{(\alpha\beta} F_{1\gamma)\eta\vartheta} +3(q-7) \delta_{\alpha)(\eta} F_{1\vartheta)(\beta\gamma}+6 \delta_{\alpha)(\eta} \delta_{\vartheta)(\beta} F_{111\gamma}\\
 (q-2) \delta_{\eta\vartheta} F_{\alpha\beta\gamma\delta}- 12\delta_{(\alpha\beta} F_{\gamma\delta)\eta\vartheta} +4(q-5)  \delta_{\alpha)(\eta} F_{\vartheta)(\beta\gamma\delta}+12 \delta_{\alpha)(\eta} \delta_{\vartheta)(\beta} F_{11\gamma\delta}\end{array}\right)   \end{multline} 

One can then check that the only exponentially suppressed solution to the three equations \eqref{firstDiffEq}, \eqref{secondDiffEq} and \eqref{thirdDiffEq} is given, up to a moduli-independent
prefactor, by 
\be 
\vec{F}_Q = \left( \begin{array}{c} F^{(4)}_1 \\ Q_{L\,\alpha} F^{(3)}_1\\ Q_{L\,\alpha}Q_{L\,\beta}  F^{(2)}_1+\delta_{\alpha\beta} F^{(2)}_2 \\ Q_{L\,\alpha}Q_{L\,\beta}Q_{L\,\gamma}  F^{(1)}_1+\delta_{(\alpha\beta} Q_{L\,\gamma)}(Q) F^{(1)}_2\\ Q_{L\,\alpha}Q_{L\,\beta}Q_{L\,\gamma}Q_{L\,\delta} F^{(0)}_1+\delta_{(\alpha\beta} Q_{L\,\gamma}Q_{L\,\delta)} F^{(0)}_2+\delta_{(\alpha\beta} \delta_{\gamma\delta)} F^{(0)}_3\end{array}\right) \ , 
\ee
\begin{eqnarray} 
F^{(k)}_1 &=& \Big(\frac{\I}{\sqrt 2}\Big)^k 2^{\frac{q-2}{2}}(2\pi)^{\frac{q-3-2k}{2}}  R^{\frac{q-1}{2}} \sqrt{\textstyle 2 |Q_R|^2}^{\frac{2k+3-q}{2}}  K_{\frac{2k+3-q}{2}}(2\pi R \sqrt{\textstyle 2|Q_R|^2}) 
\nonumber\\
F^{(k)}_2 &=&-\Big(\frac{\I}{\sqrt 2}\Big)^k 2^{\frac{q-4}{2}}\frac{(4-k)(3-k)}{2} (2\pi)^{\frac{q-5-2k}{2}} R^{\frac{q-3}{2}}\sqrt{\textstyle 2|Q_R|^2}^{\frac{2k+5-q}{2}}  \hspace{-2mm} K_{\frac{2k+5-q}{2}}(2\pi R\sqrt{\textstyle 2 |Q_R|^2})
 \nonumber\\
F^{(0)}_3 &=& 3 \times 2^{\frac{q-6}{2}}(2\pi)^{\frac{q-7}{2}}R^{\frac{q-5}{2}}\sqrt{2|Q_R|^2}^{\frac{7-q}{2}}   K_{\frac{7-q}{2}}(2\pi R\sqrt{\textstyle 2|Q_R|^2}) \ ,   \end{eqnarray} 
In particular, the tensorial part of the function $\vec F_Q$ is polynomial in $Q_{L\alpha},\,\ldots$, and the rest only depends on the moduli through $Q_R^2$ and $R=e^{-\phi}$. We conclude that the Fourier coefficient $\vec F_Q$ for a fixed $Q$ is uniquely determined by the differential equations \eqref{LinearF4} and  \eqref{EqD2F} up to an overal constant corresponding to the measure factor. 

The power-low terms satisfy to the same equations for $Q=0$. One easily computes that the only two solutions are such that 
\be \vec{F} = \left(\colvec[1.0]{ (7-q)(9-q) c_0 R^{q-6} \\ 0 \\ (7-q)c_0 R^{q-6} \delta_{\alpha\beta} \\0\\3 c_0 R^{q-6}  \delta_{(\alpha\beta}\delta_{\gamma\delta)}+R F^{p-1,q-1}_{\alpha\beta\gamma\delta} }\right)   \ , \ee
for an arbitrary constant $c_0$ and a solution $F^{p-1,q-1}_{\alpha\beta\gamma\delta}$  to \eqref{LinearF4} and  \eqref{EqD2F} on $G_{p-1,q-1}$.

\section{Polynomials appearing in Fourier modes}
\label{DecompPolynomials}

In the degeneration limit $O(p,q)\to O(p-1,q-1)$ studied in \S\ref{pertLimit}, the monomials \linebreak $\tilde P^{(\ell)}_{\alpha_{h+1}\ldots \alpha_{4}}(Q)$  with $\ell\geq0$ are of degree $4-2\ell-h$ in $Q$, and defined by
\be\label{polynomialsPertLimit}
\begin{split}
&\sum_{\ell\geq0}\tilde P^{(\ell)}_{\alpha\beta\gamma\delta}(Q)=Q_{L,\alpha}Q_{L,\beta}Q_{L,\gamma}Q_{L,\delta}-\frac{3}{2\pi}\delta_{(\alpha\beta}Q_{L,\gamma}Q_{L,\delta)}+\frac{3}{16\pi^2}\delta_{(\alpha\beta}\delta_{\gamma\delta)},
\\
&\sum_{\ell\geq0}\tilde P^{(\ell)}_{\alpha\beta\gamma}(Q)=Q_{L,\alpha}Q_{L,\beta}Q_{L,\gamma}-\frac{3}{4\pi}Q_{L,(\alpha}\delta_{\beta\gamma)},
\\
&\sum_{\ell\geq0}\tilde P^{(\ell)}_{\alpha\beta}(Q)=Q_{L,\alpha}Q_{L,\beta}-\frac{1}{4\pi}\delta_{\alpha\beta},
\\
&\sum_{\ell\geq0}\tilde P^{(\ell)}_{\alpha}(Q)=Q_{L,\alpha},
\\
&\sum_{\ell\geq0}\tilde P^{(\ell)}(Q)=1.
\end{split}
\ee

\noindent In the degeneration limit $O(p,q)\to O(p-2,q-2)$ studied in \S\ref{decompLimit}, the monomials \linebreak $\cP^{(\ell)}_{\mu_1\ldots\mu_h\,\alpha_{h+1}\ldots\alpha_4}(Q'^i,S)$ with $\ell\geq0$ are of degree $4-2\ell-h$ in $Q'^i$, and defined by
\be\label{polynomialsDecompLimit}
\begin{split}
&\sum_{\ell\geq0}\cP^{(\ell)}_{\alpha\beta\gamma\delta}(Q'^i,S)=Q'^i_{L,(\alpha}Q'^j_{L,\beta}Q'^k_{L,\gamma}Q'^l_{L,\delta)}M_{ij}M_{kl}-\frac{3}{2\pi}\delta_{(\alpha\beta}Q'^i_{L,\gamma}Q'^j_{L,\delta)}M_{ij}+\frac{3}{16\pi^2}\delta_{(\alpha\beta}\delta_{\gamma\delta)},
\\
&\sum_{\ell\geq0}\cP^{(\ell)}_{\mu\alpha\beta\gamma}(Q'^i,S)=Q'_{L,\mu(\alpha}Q'^i_{L,\beta}Q'^j_{L,\gamma)}M_{ij}-\frac{3}{4\pi}Q'_{L,\mu(\alpha}\delta_{\beta\gamma)},
\\
&\sum_{\ell\geq0}\cP^{(\ell)}_{\mu\nu\alpha\beta}(Q'^i,S)=Q'_{L,\mu\alpha}Q'_{L,\nu\beta}-\frac{1}{4\pi}\delta_{\alpha\beta}\frac{Q'_\mu\cdot Q'_\nu}{Q'_\tau\cdot Q'^\tau},
\\
&\sum_{\ell\geq0}\cP^{(\ell)}_{\mu\nu\rho\alpha}(Q'^i,S)=Q'_{L,\mu\alpha}\frac{Q'_\nu\cdot Q'_\rho}{Q'_\tau\cdot Q'^\tau},
\\
&\sum_{\ell\geq0}\cP^{(\ell)}_{\mu\nu\rho\sigma}(Q'^i,S)=\frac{Q'_{(\mu}\cdot Q'_\nu\,Q'_\mu\cdot Q'_{\sigma)}}{(Q'_\tau\cdot Q'^\tau)^2},
\end{split}
\ee
where $M_{ij}=v_{i\mu}v^\mu\__j$ is the torus metric \eqref{torusVielbein}, and $Q'_{\mu}\cdot Q'_\nu=\frac{1}{S_2}\Big(\colvec[0.7]{(Q+S_1P)^2&(Q+S_1 P)S_2P\\(Q+S_1 P)S_2P&S_2^2P^2}\Big)$.

\section{Tensorial Eisenstein series}
\label{tensEinsensteinSeries}

In the degeneration limit $O(p,q)\to O(p-2,q-2)$ studied in \S\ref{decompLimit}, the power-like terms in  \eqref{Z2F34deg0} involve tensorial Eisenstein series that we rewrote as tensorial derivatives of real analytic Eisenstein series, using $\cD_{\mu\nu}$ the traceless differential operator on $SL(2,\IR)/O(2)$. Here we exhibit these relations, and show how this operator can be rewritten in terms of lowering and raising operators $\cD_w$ and $\overline{\cD}_w$.

The non-holomorphic Eisenstein series 
\be
\cE_{s,w}(S)=\frac{1}{2\zeta(2s)}\sum_{(c,d)\in\IZ^2\smallsetminus \{0,0\}}\frac{S_2^s}{(c+ d S)^{s+\frac{w}{2}}(c+d\bar S )^{s-\frac{w}{2}}}
\ee
has modular weight $(\tfrac{w}{2},-\tfrac{w}{2})$ under $SL(2,\IZ)$. The raising and lowering operators, $\cD_w=2\I S_2\partial_S +\frac{w}{2}$ and $\overline{\cD}_w=-2\I S_2\partial_{\bar S}-\frac{w}{2}$ act on  $\cE_{s,w}(S)$ according to
\be
\cD_w \, \cE_{s,w}=\Big(s+\frac{w}{2}\Big)\cE_{s,w+2},\qquad \overline{\cD}_w\, \cE_{s,w}=\Big(s-\frac{w}{2}\Big)\cE_{s,w-2}.
\ee
Non-holomorphic Eisenstein series are thus eigenmodes of the laplacian $\Delta_w=\bar \cD_{w+2} \cD_w$ with eigenvalue $\big(s+\tfrac{w}{2}\big)\big(s-\tfrac{w}{2}-1\big)$.

Alternatively, one can denote the momenta and winding along a torus as $z_\mu=m_i v^i_\mu$ 
with $(m_1,m_2)=(c,d)$, $v_\mu\_^i$ is the vielbein defined in \eqref{torusVielbein}, such that $z_\mu z^\mu=\frac{1}{S_2}|c+dS|^2$ is invariant under $SL(2,\IZ)$. The traceless differential operator $\cD_{\mu\nu}$ acts as  
\be
\cD_{\mu\nu}z_\rho=\frac{1}{2}\delta_{\rho(\mu}z_{\nu)}-\frac{1}{4}\delta_{\mu\nu}z_\rho.
\ee
 One can show that they are related to the lowering and raising operator through 
 \be
 \cD_{\mu\nu}=-\frac{1}{2}\sigma^+_{\mu\nu}\,\cD_w-\frac{1}{2}\sigma^-_{\mu\nu}\,\bar \cD_w
 \ee
where $\sigma^\pm=\frac{1}{2}(\sigma_3\pm\I\sigma_1)$ and $\sigma_i$ are the Pauli matrices. By acting on non-holomorphic Eisenstein series of weight 0 with $\cD_{\mu\nu}$ and $\cD_{(\mu\nu}\cD_{\rho\sigma)}$, one obtains the relations
\be\label{EseriesRelations}
\begin{split}
&\frac{s}{2}\,\sigma^+_{\mu\nu}\cE_{s,2}+\frac{s}{2}\,\sigma^-_{\mu\nu}\cE_{s,-2}=\,\frac{s}{2\zeta(2s)}\sum'_{(j,p)}\frac{1}{(z_\tau z^\tau)^s}\left(\frac{z_\mu z_\nu}{z_\tau z^{\tau}}-\frac{1}{2}\delta_{\mu\nu}\right)
\\
&\frac{s(s+1)}{4}\sigma^+_{(\mu\nu}\sigma^+_{\rho\sigma)}\cE_{s,4}+\frac{s(s-1)}{4}\sigma^-_{(\mu\nu}\sigma^-_{\rho\sigma)}\cE_{s,-4}+s(s-1)\left(\sigma^+_{(\mu\nu}\sigma^-_{\rho\sigma)}-\frac{1}{8}\delta_{(\mu\nu}\delta_{\rho\sigma)}\right)\cE_{s,0}=
\\&\qquad\qquad\qquad\qquad\qquad \frac{s(s+1)}{2\zeta(2s)}\sum'_{(j,p)}\frac{1}{(z_\tau z^\tau)^s}\left(\frac{z_\mu z_\nu z_\rho z_\sigma}{(z_\tau z^{\tau})^2}-\frac{\delta_{(\mu\nu}z_\rho z_{\sigma)}}{z_\tau z^\tau}+\frac{1}{8}\delta_{(\mu\nu}\delta_{\rho\sigma)}\right)
\end{split}
\ee
where the second line is traceless. 

Now, the components $F^{(p,q),1,0}_{\alpha\beta\mu\nu}$ and $F^{(p,q),1,0}_{\mu\nu\rho\sigma}$  in \eqref{F34deg0} were obtained originally as
\be 
\begin{split}
F^{(p,q),1,0}_{\alpha\beta\mu\nu} &= R^{q-6}\frac{c(0)}{4\pi^2}\Big(\frac{8-q}{2}\Big)\frac{1}{2\zeta(8-q)}\sum'_{(j,p)}\frac{1}{(z_\tau z^\tau)^{\frac{8-q}{2}}}\frac{z_\mu z_\nu}{z_\tau z^{\tau}},
\\
F^{(p,q),1,0}_{\mu\nu\rho\sigma} &= R^{q-6}\frac{c(0)}{2\pi^2}\Big(\frac{8-q}{2}\Big)\Big(\frac{10-q}{2}\Big)\frac{1}{2\zeta(8-q)}\sum'_{(j,p)}\frac{1}{(z_\tau z^\tau)^{\frac{8-q}{2}}}\frac{z_\mu z_\nu z_\rho z_\sigma}{(z_\tau z^{\tau})^2}
\end{split}
\ee
They can be written as in \eqref{F34deg0} by rewritting the relations above, for $s\neq-1$
\be \label{EseriesRelations2}
\begin{split}
\frac{s}{2\zeta(2s)}\sum'_{(j,p)}\frac{1}{(z_\tau z^\tau)^s}\frac{z_\mu z_\nu}{z_\tau z^{\tau}}&=\frac{s}{2}\left(\delta_{\mu\nu}\cE_{s,0}+\sigma^+_{\mu\nu}\cE_{s,2}+\sigma^-_{\mu\nu}\cE_{s,-2}\right),
\\
\frac{s(s+1)}{2\zeta(2s)}\sum'_{(j,p)}\frac{1}{(z_\tau z^\tau)^s}\frac{z_\mu z_\nu z_\rho z_\sigma}{(z_\tau z^{\tau})^2}&=
\frac{s(s+1)}{4}\sigma^+_{(\mu\nu}\sigma^+_{\rho\sigma)}\cE_{s,4}+\frac{s(s-1)}{4}\sigma^-_{(\mu\nu}\sigma^-_{\rho\sigma)}\cE_{s,-4}
\\
&+\frac{s(s+1)}{2}\big(\delta_{(\mu\nu}\sigma^+_{\rho\sigma)}\cE_{s,2}+\delta_{(\mu\nu}\sigma^-_{\rho\sigma)}\cE_{s,-2}\big)
\\
&+\frac{s^2}{2}\big(\sigma^+_{(\mu\nu}\sigma^-_{\rho\sigma)}-\frac{1}{4}\delta_{(\mu\nu}\delta_{\rho\sigma)}\big)\cE_{s,0}+\frac{3s(s+1)}{8}\delta_{(\mu\nu}\delta_{\rho\sigma)}\cE_{s,0}
\end{split}
\ee
In other words, all the tensorial series in \eqref{Z2F34deg0} appearing as low-energy propagators on the torus can be rewritten a combination of $\cE_{s,0},\,\cD\cE_{s,0},\,\overline{\cD}\cE_{s,0},\,\cD^2\cE_{s,0}$ and $\overline{\cD}^2\cE_{s,0}$. This is used extensively to rewrite the 1-PI effective action in four dimensions \eqref{effAction4D}.

Similarly, they can also be rewritten using traceless differential operators $\cD_{\mu\nu}$ and $\cD^2_{\mu\nu\rho\sigma}=\cD_{(\mu\nu}\cD_{\rho\sigma)}-\tfrac{1}{4}\delta_{(\mu\nu}\delta_{\rho\sigma)}\cD_{\tau\kappa}\cD^{\tau\kappa}$ 
\be \label{EseriesRelationsBis}
\begin{split}
\frac{s}{2\zeta(2s)}\sum'_{(j,p)}\frac{1}{(z_\tau z^\tau)^s}\frac{z_\mu z_\nu}{z_\tau z^{\tau}}&=\Big(\frac{s}{2}\delta_{\mu\nu}-\cD_{\mu\nu}\Big)\cE_{s,0}
\\
\frac{s(s+1)}{2\zeta(2s)}\sum'_{(j,p)}\frac{1}{(z_\tau z^\tau)^s}\frac{z_\mu z_\nu z_\rho z_\sigma}{(z_\tau z^{\tau})^2}&=
\Big(\cD^2_{\mu\nu\rho\sigma}-(s+1)\delta_{(\mu\nu}\cD_{\rho\sigma)}+\frac{3}{8}s(s+1)\delta_{(\mu\nu}\delta_{\rho\sigma)}\Big)\cE_{s,0}
\end{split}
\ee

\section{Poincar\'e series and Eisenstein series for $O(p,q,\IZ)$ \label{sec_AFP}}

In this section, we evaluate the modular integrals \eqref{def1looppq} and \eqref{f4exactpqtr} 
using the method developed in \cite{Angelantonj:2012gw,Angelantonj:2013eja}, which keeps
invariance under the automorphism group $O(p,q,\IZ)$ of the lattice $\Lambda_{p,q}$ manifest.
The result is expressed as a sum over lattice vectors with fixed norm, which is a special type 
of Poincar\'e series for $O(p,q,\IZ)$. In \S\ref{sec_eispq}, we use a similar method to
construct Eisenstein series for $O(p,q,\IZ)$.

\subsection{Poincar\'e series representation of $F^{p,q}$}
The method developed in \cite{Angelantonj:2012gw,Angelantonj:2013eja} relies on 
expressing the factor multiplying the lattice sum in the integrand in terms of a special type of 
Poincar\'e series for $\Gamma_0(N)$, known as the  Niebur-Poincar\'e series of 
weight $w\in2\IZ$,
 \be
\label{defNP}
\cF_N(s,\kappa,w;\tau) = \frac12\sum_{\gamma\in \Gamma_{\infty}\backslash \Gamma_0(N)}
\cM_{s,w}(-\kappa\tau_2)\, e^{-2\pi\I\kappa\tau_1}\vert_w \gamma \ ,
\ee
where $\cM_{s,w}(y)$ is the Whittaker function defined 
in \cite[Eq. (2.7)]{Angelantonj:2012gw}, and $\vert_w \gamma$ is the Petersson
slash operator, $[f\vert_w \gamma](\tau)=(c\tau+d)^{-k} f(\tfrac{a\tau+b}{c\tau+d})$
for $\gamma=\big(\colvec[0.7]{a &b\\c &d }\big)$. 
The series converges
absolutely for $\Re(s)>1$, grows as $\frac{\Gamma(2s)}{\Gamma(s+\tfrac{w}{2})}q^{-\kappa}$ near the
cusp $\tau\to\I\infty$ and is regular at the cusp $\tau=0$.
It transforms under the Maass raising and lower operators according to
\be
\label{barDFNP}
\begin{split}
D \cF_N(s,\kappa,w)=& 2\kappa \left(s+\tfrac{w}{2}\right)\,\cF_N(s,\kappa,w+2)\ ,
\\
\bar D \cF_N(s,\kappa,w)=& \frac{1}{8\kappa}\left(s-\tfrac{w}{2}\right)\cF_N(s,\kappa,w-2)\ ,
\end{split}
\ee
which implies that it is  an eigenmode of the
weight $w$ Laplacian on $\cH$ with eigenvalue $(s-\tfrac{w}{2})(s-1+\tfrac{w}{2})$. In particular, 
for $w<0$ and $s=1-\tfrac{w}{2}$, $\cF_N(s,\kappa,w)$ is a harmonic Maass form of weight $w$. In cases where there exists no cusp form of weight $2-w$, it is actually a
weakly holomorphic modular form of weight $w$ \cite{1004.11021}. The Fourier expansion
of $\cF_N(s,\kappa,w)\equiv \cF_{\infty}(s,\kappa,w;\tau)$ around the cusps at $\infty$ and at 0 is
given in \cite[Eq. (5.8-10)]{Angelantonj:2013eja}, in terms of the Kloosterman sums $\cZ_{\infty\infty}(m,n;s)$ and 
$\cZ_{0\infty}(m,n;s)$ defined in Eq. A.3 and A.4 of loc. cit.

For $N=1$, one has, by matching the residue of the pole at $\tau=\I\infty$,
\be
\frac{1}{\Delta(\tau)} = \lim_{s\to 7} \frac{\cF_1(s,1,-12;\tau)}{\Gamma(2s)}\ .
\ee
For $N=2,3,5,7$, using the fact that $\Delta_k$ is invariant under the Fricke involution, one has instead
\be
\frac{1}{\Delta_k(\tau)} = \lim_{s\to 1+\frac{k}{2}} 
 \frac{ \left[\cF_N(s,1,-k;\tau) +
 \hat\cF_N (s,1,-k;\tau)\right]}{\Gamma(2s)}\ ,
\ee
where $\hat\cF_N (s,\kappa,w;\tau)$ is the image of 
$\cF_N (s,\kappa,w;\tau)$ under the  Fricke involution.\footnote{
For $N=7$, $1/\Delta_3$ is a modular form
of odd weight with character $\chi=( \tfrac{\cdot}{7})$, so the Petersson slash
operator $\vert_w \gamma$ in  \eqref{defNP} involves an additional factor of $\chi(d)^{-1}$.
This results in additional factors of $\chi(d)^{-1}$ and $\chi(c)^{-1}$ in the Kloosterman sums 
$\cZ_{\infty\infty}(m,n;s)$ and $\cZ_{0\infty}(m,n;s)$, respectively.}

We shall compute the family of integrals
\be
\begin{split}
\label{defpqtenss}
F^{(p,q)}(\Phi,s,\kappa) =& \frac{1}{\Gamma(2s)}  \int_{ \Gamma_0(N)\backslash\cH}
\frac{\de\tau_1 \de\tau_2}{\tau_2^2}\,\Part{p,q}\,
\cF_N(s,\kappa,-\tfrac{p-q}{2};\tau) \ ,
\\
F^{(p,q)}_{abcd}(\Phi,s,\kappa) =& \frac{1}{\Gamma(2s)} \int_{ \Gamma_0(N)\backslash\cH}
\frac{\de\tau_1 \de\tau_2}{\tau_2^2}\,\Part{p,q}[P_{abcd}] \,
\cF_N(s,\kappa,-\tfrac{p-q}{2}-4;\tau) \ ,
\end{split}
\ee
which converges absolutely for $\Re(s)>\tfrac{p+q}{4}$. 
Here, $\Part{p,q}[P_{abcd}]$
is the partition function of a $N$-modular lattice $\Lambda_{p,q}$ of signature $(p,q)$.
It follows from the $N$-modularity property that $|\Lambda_{p,q}^*/\Lambda_{p,q}|=N^{(p+q)/2}$
and that $\Part{p,q}[P_{abcd}]$ satisfies
\be
\Part{p,q}[P_{abcd}](\Phi,\tau) = 
\left(-\I \tau\sqrt{N}\right)^{-4-\frac{p-q}{2}}\, \Part{p,q}[P_{abcd}]
\left(\sigma\cdot\Phi,-\frac{1}{N\tau}\right)
\ee
where $\sigma$ is the $O(p,q,\IR)$ transformation realizing the isomorphism $\Lambda_{p,q}^*\simeq \Lambda_{p,q}[1/N]$. The desired integrals \eqref{f4pq} are then obtain by taking a limit
\be
\label{AFPlim}
\begin{split}
F^{(p,q)}(\Phi)=& \frac18 \lim_{s\to 1+\frac{k}{2}}
\left[ F^{p,q}(\Phi,s,1) + 
F^{(p,q)}(\sigma\cdot\Phi,s,1) \right]
\\
F^{(p,q)}_{abcd}(\Phi)=& \lim_{s\to 1+\frac{k}{2}}
\left[ F^{(p,q)}_{abcd}(\Phi,s,1) +  F^{(p,q)}_{abcd}(\sigma\cdot\Phi,s,1) \right]\ .
\end{split}
\ee
By unfolding the integration domain against the sum over $\gamma$, 
one obtains, for $\Re(s)>\tfrac{p+q}{4}$,
\be
\begin{split}
F^{(p,q)}_{abcd}(s,\kappa) =& \frac{1}{\Gamma(2s)} 
	\sum_{Q\in\Lambda_{p,q}}\int_{\cS}\de\tau_1 \de\tau_2\,\tau_2^{q/2-2}\, P_{abcd}
	\,e^{\I\pi(\tau p_L^2-\bar\tau p_R^2)} 
	\cM_{s,w}(-\kappa\tau_2)\,
	e^{-2\pi \I \tau_1\kappa},
	\end{split}
\ee
where $\cS$ denotes the strip $-\tfrac12 < \tau_1<\tfrac12, \tau_2>0$.  The integral
over $\tau_1$ enforces the BPS condition $Q^2 = Q_L^2-Q_R^2 =2\kappa$. Decomposing 
 \be
 P_{abcd}(Q,\tau_2) = \sum_{0\leq \ell\leq 2} \tilde P_{abcd,\ell}(Q)\, \tau_2^{-\ell}\ ,
 \ee
 where $ \tilde P_{abcd,\ell}$ is a polynomial of degree $4-2\ell$ in $Q$, 
 and integrating over $\tau_2$, we get
 \be
 \label{BPShyper}
 \begin{split} 
 F^{(p,q)}_{abcd}(s,\kappa) =& \frac{1}{\Gamma(2s)} \sum_{0\leq \ell\leq 2} \
	(4\pi\kappa)^{\ell+1-\frac{q}{2}}\sum_{\stackrel{Q\in\Lambda_{p,q}}{Q^2=2\kappa}} \,
	\tilde P_{abcd,\ell}(Q) \left(\frac{Q_L^2}{2\kappa}\right)^{\ell+1-s-\tfrac{q-w}{2}}
	\\
	&\times
	\Gamma\left(s+\tfrac{q-w}{2}-\ell-1\right)\,_2F_1\left(s+\tfrac{w}{2},s+\tfrac{q-w}{2}-\ell-1;2s;\frac{2\kappa}{Q_L^2}\right)
	\\
	=& \frac{1}{\Gamma(2s)} \sum_{1\leq k\leq 3} \
	(4\pi\kappa)^{\ell+1-\frac{q}{2}}\sum_{\stackrel{Q\in\Lambda_{p,q}}{Q^2=2\kappa}} \,
	\tilde P_{abcd,\ell}(Q) \left(\frac{Q_R^2}{2\kappa}\right)^{\ell+1-s-\tfrac{q-w}{2}}
	\\
	&\times
	\Gamma\left(s+\tfrac{q-w}{2}-\ell-1\right)\,_2F_1\left(s-\tfrac{w}{2},s+\tfrac{q-w}{2}-\ell-1;2s;-\frac{2\kappa}{Q_R^2}\right)
	\end{split}
\ee
where in the second line, we used Pfaff's equality $_2F_1(a,b;c;z)=(1-z)^{-b} {}_2F_1(b,c-a;c;\tfrac{z}{z-1})$. 
Similarly, for the scalar integral we get
\be
 \label{BPShyperpot}
 \begin{split} 
 F^{(p,q)}(s,\kappa) = & \frac{(4\pi\kappa)^{1-\frac{q}{2}}}{\Gamma(2s)} \
	\sum_{\stackrel{Q\in\Lambda_{p,q}}{Q^2=2\kappa}} \,
	\left(\frac{Q_R^2}{2\kappa}\right)^{1-s-\frac{p+q}{4}} \, 
_2F_1\left(s+\tfrac{q}{4},s+\tfrac{p+q}{4}-1;2s;-\frac{2\kappa}{Q_R^2}\right)
\end{split}
\ee
For $q<6$, the series \eqref{BPShyper} and \eqref{BPShyperpot}
are absolutely convergent at $s=1+\tfrac{k}{2}$, so the limit \eqref{AFPlim}
can be taken term by term. For $q\geq 6$, the limit must
be taken after analytically continuing the sum, and subtracting
the pole when $q=6$. In either case, the series \eqref{BPShyper}
and \eqref{BPShyperpot} correctly encode the singular behavior
of the integral at codimension-$q$ singularities in $G_{p,q}$ where $P_R^2\to 0$ for 
a norm $2\kappa$  in $\Lambda_{p,q}$ or $Q_R^2\to 0$ for 
a norm $2\kappa/N$ vector in $\Lambda_{p,q}^*$. Near these loci, 
the leading singular behavior of \eqref{BPShyper} is given, for $\kappa=1$,  by 
\be
\label{Fpqsing}
F^{(p,q)}_{abcd}\sim\frac{\Gamma\big(\tfrac{q-2}{2}\big)}{(2\pi)^{\frac{q-2}{2}}}\left[\frac{Q_{L,a}Q_{L,b}Q_{L,c}Q_{L,d}}{(Q_R^2)^{\frac{q-2}{2}}}-\frac{6}{q-4}\frac{\delta_{(ab}Q_{L,c}Q_{L,d)}}{(Q_R^2)^{\frac{q-4}{2}}}-\frac{3}{(q-6)(q-4)}\frac{\delta_{(ab}\delta_{cd)}}{(Q_R^2)^{\frac{q-6}{2}}}\right]
\ee
and similarly for $F^{(p,q)}$.

Using the same argument as in \eqref{polyEq} and making use of \eqref{barDFNP}, it is easy to show that the integrals \eqref{defpqtenss} satisfy the differential equation
\be
\begin{split}
\cD^2_{ef} \, F^{(p,q)}_{abcd}(s)  = & 
(2-q)\,\delta_{ef}\,F^{(p,q)}_{abcd}(s)
+(16-4q) \,\delta_{e)(a}\,F^{(p,q)}_{bcd)(f}(s) 
+12\,\delta_{(ab}\,F_{cd)ef}(s) \\
& 
+\int_{\Gamma_0(N)\backslash\cH}\frac{\de\tau_1\de\tau_2}{\tau_2^2} \frac{2(2s+k)}{2\kappa\Gamma(2s)}\,
\cF_N(s,\kappa,-k-2)\, \Part{p,q}[P_{abcdef}]
\end{split}
\ee
The modular integral on the second line can again by evaluating by the unfolding trick, as a sum over vectors $Q\in\Lambda_{p,q}$ with $Q^2=2\kappa$ . For the relevant value $s=1+\tfrac{k}{2}$ with small $|k|$, such that $\cF_N(s,\kappa,-k)$ is weakly holomorphic, $\cF_N(s,\kappa,-k-2)$
vanishes so the sum over $Q$ must vanish. We have checked that this is indeed the case
in the Euclidean case $q=0, N=1$, such that only a finite number of vectors $Q$ contribute.

\subsection{Eisenstein series for $O(p,q,\IZ)$\label{sec_eispq}}

While the modular integrals  \eqref{defpqtenss} result into
automorphic forms with singularities on $G_{p,q}$, due to the pole of order $\kappa$
in the Niebur-Poincar\'e series $\cF_N(s,\kappa,w;\tau)$, it is useful to consider the
analogue 
\be
\label{defEpqs}
E^{(p,q)}(\Phi,s) =  \int_{\Gamma_0(N)\backslash\cH}
\frac{\de\tau_1 \de\tau_2}{\tau_2^2}\,\Part{p,q}\,
E_N(s,-\tfrac{p-q}{2};\tau) \ ,
\ee
where $\cF_N(s,\kappa,w;\tau)$ is replaced by the non-holomorphic
Eisenstein series 
for $\Gamma_0(N)$, 
\be
E_N(s,w;\tau) = \frac12\sum_{\gamma\in \Gamma_{\infty}\backslash \Gamma_0(N)}
\tau_2^{s-\tfrac{w}{2}}\vert_w \gamma \ ,
\ee
which can be obtained formally by taking the limit $\kappa\to 0$ in \eqref{defNP}. 
The integral converges for $\Re(s)>\tfrac{p+q-2}{2}$, and can be computed using the unfolding 
trick, leading to a standard vectorial Eisenstein series for $O(p,q,\IZ)$, the automorphism
group of $\Lambda_{p,q}$,
\be
\label{EpqEins}
E^{(p,q)}(\Phi,s) =  \pi^{-s'} \, \Gamma(s')\, 
\sum_{\substack{P\in \Lambda_{p,q} \smallsetminus \{0\}\\ P^2 = 0}} \frac{1}{(P_L^2+P_R^2)^{s'}}
\ee
with $s^\prime=s+\tfrac{p+q}{4}-1$. Another Eisenstein series for the same group is obtained by 
replacing $E_N(s,w;\tau)$ by its image under the Fricke involution, which amounts to changing
$\Phi\mapsto \sigma\cdot \Phi$ in \eqref{EpqEins}. Unlike \eqref{defpqtenss}, both Eisenstein series are smooth automorphic forms on $G_{p,q}$.
Their behavior in the degeneration limits  $O(p,q)\to O(p-1,q-1)$ and $O(p,q)\to O(p-2,q-2)$
is easily obtained by applying the same methods as in \S\ref{pertLimit} and \S\ref{decompLimit}. 
In particular, the constant terms
proportional to $\tau_2^{s-\tfrac{w}{2}}$ and to 
$\tau_2^{1-s-\tfrac{w}{2}}$ in the Fourier expansion of $E_N(s,w;\tau)$ lead to power-like terms 
proportional to $R^{2s^\prime}$ and $R^{p+q-2-2s^\prime}$ in the degeneration limit  $O(p,q)\to O(p-1,q-1)$.

By direct computation, or using the fact that $E_N(s,w;\tau)$ is an  eigenmode of the weight $w$ Laplacian on $\cH$ with eigenvalue $(s-\tfrac{w}{2})(s-1+\tfrac{w}{2})$, one sees  that
\be
\label{DeltaEpqs}
\Delta_{G_{p,q}}\, E^{(p,q)}(\Phi,s) = s'(2s'-p-q+2)\, E^{(p,q)}(\Phi,s)\ .
\ee
For  $s'=\frac{p+4}{2}$, corresponding to $s=3+\tfrac{p-q}{2}$,  
the eigenvalue 
coincides with the eigenvalue of $F^{(p,q)}$ in \eqref{D2scalar} (the other
value $s'=\frac{q-6}{2}$, $s=-2-\tfrac{p-q}{2}$ lies outside the fundamental domain, and is related
to the former by the functional equation $s\mapsto 1-s$). Moreover, 
using the same methods as in \S\ref{modularIntegral} it is easy to check that  $E^{(p,q)}(\Phi,s)$
satisfies the second constraint in \eqref{D2scalar}. It is thus natural to ask if the exact 
$(\nabla \Phi)^4$ coupling could involve an extra term proportional to $E^{(p,q)}(\Phi,3+\tfrac{p-q}{2})$
in addition to the proposed formula \eqref{f4exact}. However, it turns out that the latter
contains terms of order $R^{p+4}$ and $R^{q-6}$
in the degeneration limit $O(p,q)\to O(p-1,q-1)$ with a non-zero coefficient, respectively, and the first term $R^{p+4}$ is 
ruled out by the differential equation \eqref{EqD2F}.

%

\end{document}